\documentclass[11pt,tightenlines,eqsecnum,floats,aps,amssymb,nofootinbib,prd,shownopacs,floatfix,superscriptaddress]{revtex4-2}

\usepackage{graphicx}
\usepackage{epstopdf}
\usepackage{latexsym}
\usepackage{amssymb}
\usepackage{amsmath}
\usepackage{color}
\usepackage{mathrsfs}
\usepackage{xparse}
\usepackage[dvipsnames]{xcolor}
\usepackage{xparse}
\usepackage[normalem]{ulem}
\usepackage{dsfont}
\usepackage{mathtools}
\usepackage[toc,title,page]{appendix}
\usepackage[sort&compress]{natbib}
\usepackage{bbm}
\usepackage{etoolbox}
\usepackage{tensor}
\usepackage{braket}
\usepackage[colorlinks=false, pdfborder={0 0 0}]{hyperref}

\renewcommand{\thesection}{\Roman{section}}
\renewcommand{\thesubsection}{\thesection.\Alph{subsection}}
\renewcommand{\thesubsubsection}{\thesubsection.\arabic{subsubsection}}
\makeatletter
\renewcommand{\p@subsection}{}
\renewcommand{\p@subsubsection}{}
\makeatother
\usepackage{nomencl}
\makenomenclature
\begin{document}
\begin{abstract}
We formulate a model of gravitationally induced decoherence for photons starting from Maxwell theory coupled to linearised gravity, expressed in terms of Ashtekar–Barbero variables and treated as an open quantum field theoretic system. In contrast to quantum mechanical models, the interaction between the system (Maxwell field) and the environment (gravitational field) is not postulated phenomenologically, but is instead dictated by the underlying action in a post-Minkowskian approximation.
This framework extends earlier models for a scalar field and enables a more detailed analysis of the role of dynamical reference fields (clocks) within the relational formalism. We show that, for a suitable choice of geometrical clocks together with a U(1)-Gauß clock, and by employing an appropriate combination of the observable map and its dual, the resulting Dirac observables are given directly by the transverse components of the photon field as well as the symmetric-transverse-traceless degrees of freedom of gravitational waves on the linearised phase space of the coupled system. In addition we also compare different choices of Dirac observables and their dynamics. Upon applying a Fock quantisation to the reduced system, we derive the time-convolutionless (TCL) master equation, truncated at second order, and analyse its structural properties. These results provide a foundation for further investigations of the decoherence model, including its renormalisation and a detailed study of its one-particle sector, and are found to be structurally consistent with former master equations for photons derived using ADM variables and a specific gauge fixing.
\end{abstract}
\title{A gravitationally induced decoherence model for photons in the context of the relational formalism}
\author{Max Joseph Fahn}
\email{maxjoseph.fahn@unibo.it}
\affiliation{Dipartimento di Fisica e Astronomia, Università di Bologna, Via Irnerio 46, 40126 Bologna, Italy.}
\affiliation{INFN, Sezione di Bologna, Viale C. Berti Pichat, 6/2, 40127 Bologna, Italy.}

\author{Kristina Giesel}
\email{kristina.giesel@gravity.fau.de}
\affiliation{Institute for Quantum Gravity, Theoretical Physics III, Department of Physics,  Friedrich-Alexander-Universit\"at Erlangen-N\"urnberg, Staudtstr. 7, 91058 Erlangen, Germany.}

\author{Roman Kemper}
\email{roman.kemper@fau.de}
\affiliation{Institute for Quantum Gravity, Theoretical Physics III, Department of Physics,  Friedrich-Alexander-Universit\"at Erlangen-N\"urnberg, Staudtstr. 7, 91058 Erlangen, Germany.}

\maketitle
\newpage
\tableofcontents
%So funktionierts: Im Header des main.tex folgende zwei Zeilen einfügen:
% \usepackage{nomencl}
% \makenomenclature

% An der Stelle im Dokument, an die die Notationsliste soll, dann einfügen (wenn diese Datei "notation.tex" heißt):
% \input{notation}

% Für arxiv: Zusätzlich .nls-Datei mithochladen und den gleichen Namen wie der Hauptdatei geben.

% ========================================================
\setlength{\nomitemsep}{-0.1\baselineskip}
\renewcommand{\nomgroup}[1]{%
  \vspace{1\baselineskip}%
  \item[\bfseries
  \ifstrequal{#1}{A}{Indices}{%
  \ifstrequal{#1}{B}{General Symbols}{%
  \ifstrequal{#1}{C}{Canonical Variables}{%
  \ifstrequal{#1}{D}{Constraints}{%
  \ifstrequal{#1}{E}{Clocks, Reference Fields and (Dual) Observables}{}}}}}
]}
\renewcommand{\nomname}{List of Symbols}

\nomenclature[A]{\(\mu,\nu,...\)}{Spacetime indices}
\nomenclature[A]{\(a,b,...\)}{Spatial indices}
\nomenclature[A]{\(i,j,...\)}{Internal Lie(SU(2)) indices}
\nomenclature[A]{\(I,J,...\)}{Specific collections, e.g. $C_I$ denotes all constraints}

\nomenclature[B]{\(g_{\mu\nu}\)}{Spacetime metric (mostly plus signature)}
\nomenclature[B]{\(\eta_{\mu\nu}\)}{Minkowski metric (mostly plus signature)}
\nomenclature[B]{\(\kappa\)}{Gravity-matter coupling constant $\kappa=16\pi G_N$ with Newton constant $G_N$}
\nomenclature[B]{\(\delta h_{\mu\nu}\)}{Linearised metric ($g_{\mu\nu} = \eta_{\mu\nu} + \delta h_{\mu\nu}$)}
\nomenclature[B]{\(\epsilon_{abc}\)}{3d Levi-Civita symbol}
\nomenclature[B]{\(\beta\)}{Barbero-Immirzi parameter}
\nomenclature[B]{\(G^\Delta\)}{Greens function of $-\Delta$ (negative Laplacian), see eq. \eqref{Greens function}}
\nomenclature[B]{\(G^{\Delta\Delta}\)}{Greens function of the squared Laplacian, see eq. \eqref{eq: greens fct Laplacian squared}}
\nomenclature[B]{\(\ast\)}{3d Convolution: $(f\ast g)(\vec x,t) = \int_{\mathbb{R}^3} d^3y \; f(\vec y,t) \: g(\vec x-\vec y,t)$ for two functions $f(\vec x,t)$ and $g(\vec x,t)$ }
\nomenclature[B]{\(\partial_x^a; \partial^a\)}{Partial derivative $\frac{\partial}{\partial x_a}$; subscript $_x$ omitted where clear}
\nomenclature[B]{\(\tilde{\kappa}\)}{Abbreviation; equal to $1$ if in combination with $\mathcal{G}^{U(1)}$; equal to $\kappa$ if in combination with $\mathcal{G}^I_{\rm geo}$}
\nomenclature[B]{\(P_b^a\)}{Projector onto the transverse subspace, $P_b^a X = \delta_b^a X + \partial_b \partial^a (X\ast G^\Delta)$}
\nomenclature[B]{\(P_{aj}^{ib}\)}{Projector onto the symmetric transverse traceless (STT) subspace, see eq. \eqref{eq: STT projector position space}}
\nomenclature[B]{\(X_{[\mu\nu]}\)}{Antisymmetrisation: $X_{[\mu\nu]}= X_{\mu\nu} - X_{\nu\mu}$}
\nomenclature[B]{\(:\widehat{X}:\)}{Normal ordering of an operator $\widehat{X}$}
\nomenclature[B]{\(\widehat{\widetilde{X}}\)}{Interaction picture of operator $\widehat{X}$}

\nomenclature[C]{\(\mathcal{A}_a^i(\vec{x},t)\)}{Ashtekar connection (configuration variable for gravity in Ashtekar-Barbero variables)}
\nomenclature[C]{\(\mathcal{E}^a_i(\vec{x},t)\)}{Densitised triad (canonically conjugated momentum to Ashtekar connection $\mathcal{A}_a^i(\vec{x},t)$)}
\nomenclature[C]{\(\delta \mathcal{A}_a^i(\vec{x},t)\)}{Linearised Ashtekar connection}
\nomenclature[C]{\(\delta \mathcal{E}^a_i(\vec{x},t)\)}{Linearised densitised triad}
\nomenclature[C]{\(A_a(\vec{x},t)\)}{Maxwell photon field (configuration variable for electromagnetic field)}
\nomenclature[C]{\(E^a(\vec{x},t)\)}{Electric field (canonically conjugated momentum to (minus) photon field $-A_a(\vec{x},t)$)}
\nomenclature[C]{\(F_{ab}^i(\vec{x},t)\)}{Gravitational Ashtekar-curvature $F_{ab}^i= \partial_{[a}\mathcal{A}^i_{b]} + \tensor{\epsilon}{^i_j_k} \mathcal{A}_a^j \mathcal{A}^k_b$}
\nomenclature[C]{\(T^{\mu\nu}(\vec{x},t)\)}{Energy-momentum tensor of the Maxwell field}

\nomenclature[D]{\(C_I\)}{General constraint}
\nomenclature[D]{\(C_I^{\rm geo}\)}{Geometric part of constraint}
\nomenclature[D]{\(C_I^{\rm ph}\)}{Part of constraint with electromagnetic degrees of freedom}
\nomenclature[D]{\(C(\vec{x},t)\)}{Hamiltonian (scalar) constraint}
\nomenclature[D]{\(C_a(\vec{x},t)\)}{Spatial diffeomorphism (vector) constraint}
\nomenclature[D]{\(G_i^{\rm geo}(\vec x,t)\)}{Geometric Gauß constraint}
\nomenclature[D]{\(G^{U(1)}(\vec x,t)\)}{U(1) Gauß constraint}
\nomenclature[D]{\(N(\vec x,t)\)}{Lapse function (Lagrange multiplier for $C(\vec x,t)$)}
\nomenclature[D]{\(N^a(\vec x,t)\)}{Shift vector function (Lagrange multiplier for $C_a(\vec x,t)$)}
\nomenclature[D]{\(\Lambda^i(\vec x,t)\)}{Lagrange multiplier for $G_i^{\rm geo}(\vec x,t)$}
\nomenclature[D]{\(\phi(\vec x,t)\)}{Lagrange multiplier for $G^{U(1)}(\vec x,t)$}
\nomenclature[D]{\(C_I'\)}{Modified set of constraints, canonically conjugated to exactly one reference field and commutes with all other reference fields; introduced in eq. \eqref{eq:primedConstraints}}
\nomenclature[D]{\(C_I[N_I](t)\)}{Smeared constraints (mainly used in Poisson brackets): $C_I[N_I](t) = \int_{\mathbb{R}^3} d^3x \: C_I(\vec x,t) \: N_I(\vec x,t)$}

\nomenclature[E]{\(T^I\)}{General clock}
\nomenclature[E]{\(\mathcal{G}^I\)}{General shifted clock, $\mathcal{G}^I = T^I-\tau^I$}
\nomenclature[E]{\(\tau^I\)}{General clock parameter for clock $T^I$}
\nomenclature[E]{\(\delta\mathcal{G}\)}{Shifted clock for the Hamiltonian constraint, $\delta\mathcal{G} =\delta T-\tau$}
\nomenclature[E]{\(\delta T\)}{Clock for the Hamiltonian constraint}
\nomenclature[E]{\(\tau\)}{Clock parameter for $T$}
\nomenclature[E]{\(\delta\mathcal{G}^a\)}{Shifted clock for the spatial diffeomorphism constraint, $\delta\mathcal{G}^a = \delta T^a-\sigma^a$}
\nomenclature[E]{\(\delta T^a\)}{Clock for the spatial diffeomorphism constraint}
\nomenclature[E]{\(\sigma^a\)}{Clock parameter for $T^a$}
\nomenclature[E]{\(\delta\mathcal{G}^j_{\rm geo}\)}{Shifted clock for the geometric Gauß constraint, $\delta\mathcal{G}^j_{\rm geo} = \delta\Xi^j - \xi^j$}
\nomenclature[E]{\(\delta \Xi^j\)}{Clock for the geometric Gauß constraint}
\nomenclature[E]{\(\xi^j\)}{Clock parameter for $\Xi^j$}
\nomenclature[E]{\(\delta\mathcal{G}^{U(1)}\)}{Shifted clock for the U(1) Gauß constraint, $\delta\mathcal{G}^{U(1)} = \delta T^{U(1)}-\gamma$}
\nomenclature[E]{\(\delta T^{U(1)}\)}{Clock for the U(1) Gauß constraint}
\nomenclature[E]{\(\gamma\)}{Clock parameter for $T^{U(1)}$}
\nomenclature[E]{\(\delta\mathcal{G}^I_{\rm geo}\)}{Collection of $\delta \mathcal{G}, \delta \mathcal{G}^a, \delta \mathcal{G}^j_{\rm geo}$}
\nomenclature[E]{\(\mathcal{O}_{F,\{\mathcal{G}^I\}}\)}{Observable of a phase space function $F$ that (weakly) Poisson commutes with all constraints}
\nomenclature[E]{\(\mathcal{O}^{\rm dual}_{F,\{C_I\}}\)}{Dual observable of a phase space function $F$ that (weakly) Poisson commutes with all clocks}
\nomenclature[E]{\(\mathcal{O}^{\rm dual,vac(i)}_{F,\{C_I\}}\)}{i-th order of the dual observable of a phase space function $F$ that (weakly) Poisson commutes with all clocks, built by only taking into account the geometric constraints in the dual observable map}

%Template: \nomenclature[]{\(\)}{}

\clearpage
\markboth{}{} % Ansonsten erscheint "NOTATION" im header
\printnomenclature
\markboth{}{} % Ansonsten erscheint "NOTATION" im header der folgenden Seiten
\clearpage
\newpage
\section{Introduction}
\label{sec:Intro}
Open quantum systems play a role in various areas of physics, ranging from applications in non-relativistic quantum mechanics \cite{Zurek:1991vd}, quantum optics \cite{Plenio:1997ep,Landi:2023ktg} to condensed matter systems \cite{Daley:2014fha,Minganti:2024uic} and relativistic quantum field theory models \cite{Breuer:2007juk} that include gravity \cite{Bassi:2017szd,Anastopoulos:2021jdz}, whereby most of the references listed above refer to textbooks or reviews. What all these applications have in common is that quantum models are formulated taking into account the fact that physical systems are generally not completely shielded from their environment and therefore cannot be treated as isolated. To this end, a quantum system consisting of the system of interest and its environment is considered, as well as the corresponding dynamics, which includes the free evolution of both the system and the environment, along with a certain interaction between the two. Given this setting, a master equation can be derived for the system that describes the dynamics of the system of interest's density matrix under the effective influence of the environment. This master equation is obtained by tracing out the degrees of freedom of the environment in the von-Neumann equation, assuming a choice for the density matrix of the environment. Compared to the dynamics of a closed system, a master equation includes the unitary evolution that is also present when we treat the system as closed, as well as an additional dissipator that is not unitary and can encode effects such as dissipation, dephasing, and decoherence, see for instance \cite{Breuer:2007juk, Breuer:2003avm, Rivas:2012ugu} for textbooks on this topic. A well-known example of such a master equation is the Lindblad equation \cite{Lindblad:1975ef}, in which the dissipator is typically characterised by a selection of Lindblad operators describing the coupling between the system and the environment, together with a set of decoherence or, more generally, damping parameters encoding properties of the environment. The aspect of open quantum systems that we will focus on in this work is the decoherence induced by the environment. 

There has been recent interest in the investigation in environmentally induced decoherence from both the experimental and theoretical side. The focus we will have in this work is gravitationally induced decoherence corresponding to the choice of a gravitational environment, in particular linearised gravity, see \cite{Bassi:2017szd,Anastopoulos:2021jdz} on reviews on gravitational decoherence in a broader context. Models for such kind of decoherence have been formulated at the phenomenological level in the context of neutrino oscillations, see for examples \cite{Calatayud-Cadenillas:2024wdw,ESSnuSB:2024yji,Guzzo:2014jbp,Coloma:2018idr,DeRomeri:2023dht,Lessing:2023uxb, KM3NeT:2024jji,IceCube:2023bwd,Xu:2020lhc} as well as the microscopic level, see for instance \cite{Blencowe:2012mp,Anastopoulos:2013zya,Oniga:2015lro,Lagouvardos:2020laf,Fahn:2022zql,Fogedby:2022wbz,Fahn:2024fgc,Kading:2025cwg}. In the first case often a Lindblad equation is taken as a starting point, whereas for the latter an underlying quantum model is chosen and then a Lindblad or more general master equation is derived. For both approaches models have been investigated at the quantum mechanical and quantum field theoretical level. In this work we will work in the field theoretical setting and the environment that we consider is linearised gravity and the system of interest is a photon field. Similar models for a scalar field have been formulated in \cite{Anastopoulos:2013zya,Blencowe:2012mp,Fahn:2022zql}, for photons in \cite{Lagouvardos:2020laf} and for general bosonic fields in \cite{Oniga:2015lro}.
~\\
~\\
The new aspects that we will consider in this work are the following: 
First we will formulate the classical model in terms of Ashtekar-Barbero variables \cite{Ashtekar:1986yd,BarberoG:1994eia} which opens the possibility, in principle, to apply an LQG-inspired quantisation to linearised gravity \cite{Ashtekar:1991mz,Varadarajan:2002ht}. As an initial step in this work we will still apply a Fock quantisation also in order to be in future work able to compare the similarities and differences that potentially arise from the two different quantisation procedures. An open quantum model in the context of polymer quantum mechanics (that uses an LQG inspired quantisation) can be found in \cite{Giesel:2022pzh} and in \cite{Feller:2016zuk} a toy model for surface state decoherence in LQG is presented. 
~\\
~\\
Secondly, with linearised gravity as the chosen environment independently of the chosen system of interest the underlying microscopic system has constraints. Therefore, when quantising the microscopic QFT model we need to choose whether we do this via a Dirac or reduced phase space quantisation. The first approach was chosen in \cite{Oniga:2015lro}, while \cite{Blencowe:2012mp,Anastopoulos:2013zya,Lagouvardos:2020laf} apply a gauge-fixing before quantising and \cite{Fahn:2022zql} uses the relational formalism \cite{Rovelli:1990ph,Rovelli:2001bz} to extract the physical phase space. This involves the choice of a dynamical physical reference field, often also denotes as clocks in this context, by means of which gauge invariant Dirac observables can be constructed \cite{Vytheeswaran:1994np,Dittrich:2004cb,Dittrich:2005kc,Pons:2009cz,Pons:2010ad} and a gauge invariant formulation of the dynamics can be obtained in terms of a physical Hamiltonian \cite{Thiemann:2004wk} that does not vanish in the physical sector in contrast to the canonical Hamiltonian in linearised gravity. The advantage of this is that the formulation of the model is then valid in any gauge. 
~\\
~\\
While in \cite{Fahn:2022zql} the system of interest was a scalar field without internal gauge degrees of freedom, we generalise this approach here to the case of photons, which involve an additional $U(1)$ gauge symmetry. At the level of the relational formalism, this requires the introduction of an additional dynamical reference field for the corresponding constraint. In this approach one often distinguishes matter and geometrical clocks, which are constructed from only matter or only geometrical degrees of freedom respectively. Applications in the framework of LQG of matter reference frames can be for instance found in \cite{Giesel:2007wn,Domagala:2010bm,Husain:2011tk,Giesel:2012rb,Giesel:2016gxq} in full LQG. Geometrical clocks have been so far mainly used in perturbative approaches such as for instance in linearised gravity \cite{Arnowitt:1962hi,Dittrich:2006ee,Giesel:2024xtb} or cosmological perturbation theory \cite{Giesel:2017roz,Giesel:2018opa}.  We will show that one can choose a suitable set of reference fields consisting of geometrical clocks constructed solely from geometrical degrees of freedom for the gauge symmetries related to linearised gravity, and an $U(1)$-Gauß clock constructed from the degrees of freedom of the photon field, so that the corresponding Dirac observables directly correspond to the two polarizations of the gravitational wave and the two physical polarizations of the photon field. Interestingly, it turns out that this can only be achieved by combining two methods. One is the observable map \cite{Dittrich:2004cb,Vytheeswaran:1994np}, which maps a given gauge-variant phase space function to its gauge-invariant extension, which can be constructed within the relational formalism once a set of dynamical reference frames has been selected. The second method is called dual observable map, since in this map the roles of the dynamical reference fields (clocks) and the constraints are interchanged and it was introduced in \cite{Fahn:2022zql}. It can be used, for example, to obtain a set of (weakly) mutually commuting dynamical reference fields or a (weakly) abelianised set of constraints but can be also be applied to other phase space functions. In this sense it allows for a broader range of applications than the weak abelianisation procedure introduced in \cite{Dittrich:2004cb}, which is restricted to the set of constraints there. In our application in this work such maps yield not only weak but strong equalities due to choice of clocks that we consider. While in \cite{Fahn:2022zql} we only applied the observable map, but not its dual, to construct a set of Dirac observables in the gravitational sector, and then manually applied the projector to the symmetric transverse traceless degrees of freedom, in contrast in this work no projector is introduced manually, but it is obtained directly as a result after applying a combination of the observable map and its dual.
~\\
Furthermore, the kinematical phase space of the model exhibits a Kucha\v{r} decomposition due to such a choice of dynamical reference fields, which means that the kinematical phase space can be clearly separated into gauge and physical degrees of freedom, which is less straightforward to obtain when a gauge-fixing is obtained. 
~\\
~\\
Thirdly, although we will consider this choice as the classical starting point for the open quantum model, we will also discuss different choices for Dirac observables by using an alternative form of the dual observable map and compare their properties and dynamics with the chosen set. Interestingly, our results show that the complexity of both the form of the physical Hamiltonian and the form of the algebra of Dirac observables is strongly related to the choice of the chosen set of Dirac observables, and that our choice is also strongly guided by the requirement that a simple form of the algebra of Dirac observables is very convenient for canonical quantisation.
~\\
~\\
With the classical physical phase space given and the corresponding physical Hamiltonian we apply a Fock quantisation to the system and then derive the TCL master equation in the projection operator formalism, generalising the results from \cite{Fahn:2022zql} from scalar fields to the case of Maxwell vector fields. Generalising \cite{Fahn:2022zql}, in the present work the matter sector is in the quantum theory entirely formulated in terms of components of the energy-momentum tensor, which might facilitate the application to other matter fields. While in \cite{Lagouvardos:2020laf} the final master equation presented involves already the Born- and Markov approximation and then the application of the rotating wave approximation is discussed, here we only apply the Born approximation when deriving the final form of the master equation. This puts the model obtained in this work in a form such that in future work renormalisation techniques used in \cite{Fahn:2024fgc} can be extended and applied to it. Furthermore, it allows to investigate more in detail under which conditions the Markov and rotating wave approximation can be applied in the field theoretical model.
~\\
~\\
The article is structured as follows: After the introduction in section \ref{sec:Intro}, we give a brief overview of the underlying classical formulation of Maxwell's theory  coupled to linearised gravity using Ashtekar-Barbero variables in a post-Minkowski approximation \cite{Will:2016sgx} in section \ref{sec:CassicalSetup}. Section \ref{sec:ConstrObs} presents the choice of geometrical and $U(1)$-Gauß clocks in the context of the relational formalism, as well as the construction of the corresponding Dirac observables for the gravitational and photon field degrees of freedom. Furthermore, the corresponding physical Hamiltonian  is derived and the algebra of observables is calculated. In addition to the final choice of Dirac observables used in this article, other choices and their properties are discussed in comparison to the selected set of Dirac observables. The reduced phase space obtained in section \ref{sec:ConstrObs} is then taken as the classical starting point for the canonical quantisation discussed in section \ref{sec:Meq}. First, the entire reduced system is quantised using Fock quantisation and then treated as an open quantum system in which the physical photon degrees of freedom are the system of interest and the gravitational waves form the environment. To derive the final master equation for photons, we assume a thermal state for the environment, calculate the corresponding environment correlation functions, and formulate the master equation as a TCL equation truncated to second order using the projection operator formalism. Finally, in section \ref{sec:Conclusions}, we discuss our results, compare them with those available in the literature, and provide an outlook on possible future work based on the results obtained in this article. The appendices contain more detailed calculations and derivations of the results presented in the main parts of the article.
\section{Maxwell theory coupled to linearised gravity using Ashtekar-Barbero variables}
\label{sec:CassicalSetup}
We briefly review the classical model that we take as a starting point for the decoherence model, which is Maxwell's theory coupled to gravity, given by the sum of the Einstein-Hilbert action and the Maxwell action in a generic curved spacetime
\begin{align}\label{eq:Action}
    S[g_{\mu\nu},A_{\mu}] = S_\textrm{EH}+ S_{\textrm{Maxwell}}=\frac{1}{\kappa}\int\limits_{\cal M} d^4x \sqrt{-g}R-\frac{\mu_0}{4}\int
\limits_{\cal M}d^4x \sqrt{-g}F_{\mu\nu}F^{\mu\nu}\,,
\end{align}
with the Ricci scalar $R$ and the gravitational coupling constant $\kappa=16\pi G_N$, where $G_N$ denotes Newton's constant and we denote the Lorentzian metric by $g_{\mu\nu}$ with $\mu,\nu=0,\dots, 3$ with signature $(-,+,+,+)$. For Maxwell theory we consider a massless 1-form $A_\mu$ as the dynamical field variable with $\mu_0$ as the coupling constant. Without loss of generality, we will set $\mu_0 = 1$ in the following. The electromagnetic field strength tensor on a curved spacetime takes the form
\begin{align}\label{eq:NoCovDerInF}
    F_{\mu\nu}=\nabla_{[\mu}A_{\nu]}=\partial_{[\mu}A_{\nu]}-\Gamma_{[\mu\nu]}^\tau A_\tau = \partial_{[\mu}A_{\nu]}, 
\end{align}
where the last equality holds because of the metric compatible and torsion-free Levi-Civita connection for which we have $\Gamma_{\mu\nu}^\tau=\Gamma_{\nu\mu}^\tau$ and we will use in the entire article the following notation for anti-symmetrisation 
\begin{equation*}
T_{[\mu\nu]}:=T_{\mu\nu}-T_{\nu\mu}.    
\end{equation*}
Since we want to apply canonical quantisation later on and further aim at formulating the model in such a way that quantisation techniques from loop quantum gravity can  be used in future work, we formulate the action in terms of Ashtekar-Barbero variables \cite{Ashtekar:1986yd, Ashtekar:1987zz, Ashtekar:1991hf, Ashtekar:1987gu}. For this purpose we consider a 3+1 decomposition of the action\footnote{For this, we use a foliation of the spacetime $\mathcal{M}$ into spacelike hypersurfaces $\sigma$ labeled by a time parameter $t \in \mathbb{R}$.} in \eqref{eq:Action} and perform a Legendre transformation. As a first step we go to the partially reduced phase space in which the primary constraint corresponding to the momenta of the lapse function and the shift vector as well as the one of $A_0=-\phi$ are solved. We denote the Ashtekar-Barbero connection and its conjugate momenta by $({\cal A}_a^i(x),{\cal E}^a_i(x))$ and the canonical variables of the Maxwell field as $(A_a(x),\Pi^a(x):=-E^a(x))$. In terms of self-dual Ashtekar-Barbero variables the action can be found in \cite{Ashtekar:1991hf} and for real Ashtekar-Barbero variables in \cite{Thiemann:2001gmi}. In the latter case  
in terms of Ashtekar-Barbero variables the corresponding action reads
\begin{equation}
\begin{aligned}\label{eq: action with constraints}
S=\int_{\mathbb{R}} d t \int_\sigma d^3 x & \left(\frac{1}{\kappa \beta} \dot{\cal A}_a^i(\vec{x}, t) {\cal E}_i^a(\vec{x}, t)-\dot{A}_a(\vec{x}, t) E^a(\vec{x}, t)\right. \\
& \left.-\frac{1}{\kappa}\left(\Lambda^i(\vec{x}, t) G_i^{\textrm{geo}}(\vec{x}, t)+N^a(\vec{x}, t) C_a(\vec{x}, t)+N(\vec{x}, t) C(\vec{x}, t)\right)-\phi(\vec{x}, t) G^{\textrm{U(1)}}(\vec{x}, t)\right)
\end{aligned}
\end{equation}
with the canonical Hamiltonian
\begin{equation}
    H^{\textrm{can}}= \int\limits_{\sigma}d^3x 
    \left(\frac{1}{\kappa}\left( \Lambda^i(\vec{x}, t) G_i^{\textrm{geo}}(\vec{x}, t)+N^a(\vec{x}, t) C_a(\vec{x}, t)+N(\vec{x}, t) C(\vec{x}, t)\right)+\phi(\vec{x}, t) G^{\textrm{U(1)}}(\vec{x}, t)\right)
\end{equation}
that consists of secondary constraints only which in Ashtekar-Barbero variables take the form
\begin{align}\label{eq:constraintsAshtekar1}
    &\begin{aligned}[t]
    &G^{\textrm{U(1)}} = -E_{\:\:,a}^{a}, &
    &G_i^{\textrm{geo}}= \frac{1}{2\beta}\left(\mathcal{E}^a_{i,a}+\epsilon_{ij}^{\:\:\:k}\mathcal{A}^j_a\mathcal{E}^a_k\right),\\
    &C = C^{\rm geo} + C^{\rm ph}, & 
    &C_a = C_a^{\rm geo}+ C_a^{\rm ph}  \nonumber \\
    \end{aligned}
  \\
  &\begin{aligned}[t]
    \text{with} \hspace{0.2in} &C^{\rm geo} =  \left(F_{ab}^i-\frac{\beta^2+1}{\beta^2}\epsilon_{ilm}\left(\mathcal{A}_a^l-\Gamma_a^l\right)\left(\mathcal{A}_b^m-\Gamma_b^m\right)\right)\frac{\epsilon^{ijk}\mathcal{E}_j^a\mathcal{E}_k^b}{2\sqrt{\mathcal{E}}}\\
    &C^{\rm ph}=\frac{\kappa E^{a} E^d \mathcal{E}_a^{i} \mathcal{E}_d^{i}\sqrt{\mathcal{E}}}{2}+\frac{\kappa}{4\sqrt{\mathcal{E}}^3}\mathcal{E}^{a}_i \mathcal{E}^{b}_i \mathcal{E}^{c}_j \mathcal{E}^{d}_j F_{ca} F_{db},
    \\
    &C_a^{\rm geo}= \frac{1}{\beta}F_{ab}^i\mathcal{E}_i^b  \hspace{0.2in}\text{ and }\hspace{0.2in} C_a^{\rm ph} =\kappa E^c F_{ca}.
  \end{aligned}
\end{align}
Here $ G^{\textrm{U(1)}}$ denotes the electromagnetic U(1)-Gau\ss{} constraint, whereas $G^{\textrm{geo}}$ the SU(2) geometrical Gau\ss{} constraint that results from the extension of the ADM phase space to a phase space in terms of Ashtekar-Barbero variables. $F_{ab}^i=\mathcal{A}^i_{[b,a]}+\epsilon^i_{\:\:jk}\mathcal{A}_a^j\mathcal{A}_b^k$ denotes the curvature associated with the Ashtekar-Barbero connection $\mathcal{A}_a^i$. Furthermore, $C$ and $C_a$ denote the Hamiltonian and spatial diffeomorphism constraint, where we neglected the terms proportional to $G^{\textrm{geo}}$ in both expressions since this set of constraints defines an equivalent constraint hypersurface. 
~\\
~\\
Given the canonical form of the action, we can read off the form of the non-vanishing Poisson brackets from the symplectic potential. These are given by
\begin{align}\label{eq:elementrayPBgeo}
    \{\mathcal{A}_a^i(\vec{x},t),\mathcal{E}_j^b(\vec{y},t)\}&=\kappa\beta\delta_j^i\delta_a^b\delta(\vec{x},\vec{y}),\\
    \label{eq:elementrayPBelectro}
    \{A_a(\vec{x},t),\Pi^b(\vec{y},t)\}&=\delta_a^b\delta(\vec{x},\vec{y}),\quad{\rm with}\quad \Pi^b(\vec{y},t)=-E^b(\vec{y},t)\,.
\end{align}
All remaining combinations vanish.\\
The constraint algebra takes the following form 
\begin{align}\label{ConstraindAlgebraADM}
    \{C[N],C[N']\}& =-\kappa\vec{C}[\vec{f}(N,N',q)],\\
    \{\vec{C}[\Vec{N}],\Vec{C}[\Vec{N}']\}&=-\kappa\vec{C}[\mathcal{L}_{\Vec{N}}\Vec{N}'],\\
    \{\vec{C}[\Vec{N}],C[N']\}&=-\kappa C[\mathcal{L}_{\Vec{N}}N'],\\
    \{G[\Lambda],G[\Lambda']\}&=-\kappa G[\Lambda,\Lambda'],\\
    \{G^{\textrm{U(1)}}[\phi],C[N]\}&=\int_\sigma d^3x\; \partial_{[b}\partial_{d]}\left(\frac{\sqrt{q}}{2}q^{ab}q^{cd}A_{[a,c]}N\phi\right)(\vec{x},t)=0,\\
    \{G^{\textrm{U(1)}}[\phi],\vec{C}[\vec{N}]\}&=-\int_\sigma d^3 x\; \partial_{[b}\partial_{a]}\left(\phi N^a E ^b\right)(\Vec{x},t)=0,\\
    \{G^{\textrm{U(1)}}[\phi],G^{\textrm{geo}}[\Lambda]\}&=0,\\
    \{G^{\textrm{U(1)}}[\phi],G^{\textrm{U(1)}}[\phi']\}&=0,
\end{align}
where $f^a(N,N',q) = q^{ab}\left(N N_{,b}'- N'N_{,b}\right)$ are so called structure functions with $q^{ab}({\cal E})=\frac{{\cal E}^a_i{\cal E}^b_j\delta^{ij}}{\det(\cal E)}$ to be understood as a function of the densitised triads ${\cal E}^a_i$ and $\mathcal{L}_{\vec{N}}$ is the Lie-derivative with respect to $\vec{N}$. The first three equations are only valid on the $G^{\textrm{geo}}\approx 0,\:\:G^{\textrm{U(1)}}\approx 0$ hypersurface. The reason for this is that when rewriting the constraints originally given in ADM variables in terms of Ashtekar-Barbero variables, both the spatial diffeomorphism and the Hamiltonian constraints contain contributions that involve the geometrical SU(2) Gaussian constraint. If we omit these from the definition of these constraints,  as done in \eqref{eq:constraintsAshtekar1},
we obtain additional terms in the resulting Poisson brackets of the constraint algebra that are proportional to the SU(2) Gau\ss{} constraint. Moreover, as discussed, for example, in \cite{Pons:1999xu}, we also obtain for Yang-Mills theories additional contributions of the corresponding Gau\ss{} constraints on the right-hand side of the Poisson brackets in the constraint algebra. It is convenient to work in the present work with the set of constraints defined in \eqref{eq:constraintsAshtekar1}, since this set of constraints defines an equivalent constraint hypersurface and, moreover, these additional contributions vanish strongly in the order we are interested in for the linearised case, that we will consider later. The fact that the U(1)-Gau\ss{} constraint $G^{\textrm{U(1)}}$ strongly Poisson commutes with itself as well as with $C^{\rm geo}$ will become important later when the corresponding Dirac observables will be constructed.
A similar model in ADM variables can be found in \cite{Lagouvardos:2020laf}, while a model using Ashtekar-Barbero variables but involving a scalar field as the matter part can be found in \cite{Fahn:2022zql}.
\subsection{Post-Minkowski approximation for Maxwell theory coupled to linearised gravity}
\label{sec:PostMinkowski}
In this section, we will linearise the geometrical degrees of freedom and discuss how we can implement the matter contributions consistently in this perturbative treatment in the context of a post-Minkowski approximation, which is based on the Landau-Lifshitz formulation of general relativity \cite{Will:2016sgx}.
A particular focus of the discussion will be on the linearised constraint algebra because this provides the basis for the construction of the Dirac observables for the model in section \ref{sec:ConstrObs}.
We denote the perturbation around the flat Minkowski metric by $\delta h_{\mu\nu}$, then we have
\begin{align}\label{eq:perturbgeneralmetric}
    g_{\mu\nu}&=\eta_{\mu\nu}+\kappa\delta h_{\mu\nu},
\end{align}
where the gravitational coupling constant $\kappa$ plays the role of the perturbation parameter. For the Ashtekar-Barbero variables in the geometrical sector the linearisation takes the following form\footnote{When considering linearised quantities, we sometime use internal indices on the same footage as spatial ones, as the linearisation of $\mathcal{E}^a_i$ implies that the soldering form mapping between them is in lowest order a Kronecker delta, so e.g. $\delta \mathcal{E}^a_b := \delta^i_b \, \delta \mathcal{E}^a_i$.} \cite{Ashtekar:1991mz}
\begin{align}\label{eq:AllLinearisedVariables}
    \mathcal{E}^{a}_i &= \delta^{a}_i + \kappa \delta \mathcal{E}_i^{a}, 
    \:\: \mathcal{A}^{i}_a= \kappa \delta \mathcal{A}^{i}_a,  \:\:
    N = 1+\kappa \delta N,\:\: N^{a} = \kappa \delta N^{a},\:\: \Lambda^{i} = \kappa \delta\Lambda^{i},
\end{align}
where the elementary phase space variables of the linearised geometric sector are the perturbations $\delta\mathcal{A}_a^i$ and $\delta\mathcal{E}^a_i$. 
They have the following non-vanishing Poisson bracket
\begin{equation}
\label{eq:linearisedPBGeo}
\{\delta\mathcal{A}_a^i(\vec{x},t), \delta\mathcal{E}^b_j(\vec{y},t)\} =\frac{\beta}{\kappa}\delta^b_a\delta^i_j \delta^{(3)}(\vec{x},\vec{y}).
\end{equation}
We have also introduced the perturbation of the Lagrange multipliers encoded in $\delta N,\:\delta N^a,\:\delta \Lambda$.\\
To consistently include matter in this perturbation expansion, we use the post-Minkowski approximation, see for instance \cite{Fahn:2022zql} for a brief introduction in the context of linearised gravity. From the post-Minkowski approximation we can conclude that the interaction Hamiltonian of the perturbed model is given by
\begin{align}\label{eq:interactionHamiltonian}
    \begin{split}
    \delta H_{I} &= -\frac{\kappa}{2}\delta h_{\mu\nu}T^{\mu\nu}(A,\eta)\\
    &=\kappa\delta N T^{00}(A,\eta)-\kappa\delta N_{a}T^{0a}(A,\eta)-\frac{\kappa}{2}W_{abc}^i \delta\mathcal{E}_i^c T^{ab}(A,\eta),
    \end{split}
\end{align}
where we used
\begin{align}
    \begin{split}
    \delta h_{00}&= -2\delta N,\:\:\:\: \delta h_{0a}=\delta N_a,\\
    \delta h_{ab}&= W_{abc}^i\delta\mathcal{E}_i^c \quad{\rm with}\quad
   W_{abc}^i:= -\delta_b^i\delta_{ac}-\delta_a^i\delta_{bc}+\delta_{ab}\delta_c^i.
    \end{split}
\end{align}
Here, $T^{\mu\nu}(A,\eta)$ denotes the energy-momentum tensor of the field $A_\mu$ evaluated on Minkowski spacetime $\eta$. As for the order we are considering in this work $T^{\mu\nu}$ is always evaluated only on Minkowski background, we drop the argument $\eta$ from $T^{\mu\nu}$ to simplify the notation. In the lowest order of the post-Minkowski approximation, $T^{\mu\nu}$ provides the source for the metric perturbations and is for the Maxwell photon field
\begin{align}\label{eq:energymomentumonMinkowski}
     T^{\sigma \gamma}&=F^{\sigma\nu}F^{\gamma\tau}\eta_{\nu\tau}-\frac{1}{4}\eta^{\sigma \gamma} F^{\mu\nu}F^{\rho\tau}\eta_{\rho\mu}\eta_{\tau\nu}
&=A^{[\nu,\sigma]}A^{[\tau,\gamma]}\eta_{\nu\tau}+\eta^{\sigma\gamma}\left(\frac{E_a E^a}{2}-\frac{1}{4}A^{[c,d]}A_{[c,d]}\right).
\end{align}
In terms of the 3+1 decomposition we obtain
\begin{align}\label{eq:EnergyMomentumTensor3+1}
    T^{00}&=\frac{E^a E_a}{2}+\frac{1}{4}A_{[a,b]}A^{[a,b]},\\
    T^{0a}&=-E^b A^{[c,a]}\delta_{bc},\\
    T^{ab}&=-E^a E^b+A^{[c,a]}A^{[d,b]}\delta_{cd}+\delta^{ab}\left(\frac{E_c E^c}{2}-\frac{1}{4}A_{[a,b]}A^{[a,b]}\right).
\end{align}
Using this and \eqref{eq:interactionHamiltonian}, the interaction Hamiltonian of the model reads
\begin{align}\label{eq:postminkowski interaction Hamiltonian}
    \begin{split}
    \delta H_{I}=&\kappa\delta N\left(\frac{E^a E_a}{2}+\frac{1}{4}A_{[a,b]}A^{[a,b]}\right)+\kappa \delta N^a E^b A_{[a,b]}\\
    &+\kappa A_{[d,c]}A^{[d,i]}\delta\mathcal{E}_i^c-\kappa E_c E^i\delta\mathcal{E}_i^c+\kappa\delta_c^i\delta\mathcal{E}^c_i\left(\frac{E^a E_a}{4}-\frac{3}{8}A^{[a,b]}A_{[a,b]}\right).
    \end{split}
\end{align}
We want to generate equations of motion up to the first order in\footnote{While $\kappa$ is the expansion parameter for the gravitational degrees of freedom, with $\delta$ we mean here the expansion order of the appearing gravitational quantities. This means that an expansion ``up to order $\delta^n$" includes the highest order expansion term of order $\delta^n$ as well as products of lower order expansion terms, such as $n$ times order $\delta$. The reason for these two perturbation parameters $\kappa$ and $\delta$ roots in the fact that the expansion in $\kappa$ is carried out asymmetrically in geometric and matter variables: while in the expansion of the former the $\kappa$ and $\delta$ orders agree, the latter are not expanded (this is they are of order $\delta^0$, but come along with $\kappa$ which manifests e.g. in the form of the expanded constraints).} $\kappa,\:\delta$. The Poisson brackets for the perturbations of the gravitational phase space variables reduce the  order in $\kappa$ by one and the order in $\delta$ by two as can be seen from  \eqref{eq:linearisedPBGeo}. This means that we have to consider the canonical Hamiltonian up to second order in $\kappa$ and $\delta$ respectively, to get first order equations of motion. To do this we first expand the constraints up to second order using \eqref{eq:AllLinearisedVariables}. Starting with the geometrical Gauß constraint
\begin{align}\label{eq:lingeogaußcosntraint}
   \delta G_i^{\textrm{geo}}=\frac{\kappa}{2\beta}\left(\delta\mathcal{E}^a_{i,a}+\epsilon_{ij}^{\:\:\:k}\delta\mathcal{A}^j_a\delta_k^a\right),\:\:\:\:\delta^2 G_i^{\textrm{geo}}=\frac{\kappa^2}{2\beta}\epsilon_{ij}^{\:\:\:k}\delta\mathcal{A}^j_a\delta\mathcal{E}^a_k.
\end{align}
For the Hamiltonian constraint we have
\begin{align}\label{eq:linHamiltonConstraint}
    \begin{split}
    \delta C=& \kappa \epsilon_{i}^{\:\:jk}\delta_j^a\delta_k^b\delta\mathcal{A}_{b,a}^i+\kappa\frac{1}{2}E^aE^d\delta_{ad}+\kappa\frac{1}{4}A_{[a,b]}A^{[a,b]}\\
    =& \kappa \epsilon_{i}^{\:\:jk}\delta_j^a\delta_k^b\delta\mathcal{A}_{b,a}^i+\kappa T^{00},
    \end{split}
    \\
    \begin{split}
    \delta^2 C =& \frac{\kappa^2}{2}\Big[\left(\delta_l^a\delta_m^b-\delta_l^b\delta^a_m\right)\delta\mathcal{A}_a^l\delta\mathcal{A}_b^m-\frac{\beta^2+1}{\beta^2}\left(\delta_l^a\delta_m^b-\delta_l^b\delta^a_m\right)\left(\delta\mathcal{A}_a^l-\delta\Gamma_a^l\right)\left(\delta\mathcal{A}_b^m-\delta\Gamma_b^m\right)\\ &+\delta\mathcal{A}^i_{[b,a]}\epsilon_i^{\:\:jk}\left(\delta_j^a\delta\mathcal{E}_k^b+\delta_k^b\delta\mathcal{E}_j^a-\frac{1}{2}\delta_j^a\delta_k^b\delta_d^n\delta\mathcal{E}_n^d\right)\\
    &+\frac{1}{2}E^aE^a\delta_{ad}\delta_f^n\delta\mathcal{E}_n^f-2 E_bE^j\delta\mathcal{E}_j^b+\frac{1}{2}A_{[a,c]}A_{[b,d]}\left(4\delta\mathcal{E}_i^a\delta_i^b\delta^{cd}-\frac{3}{2}\delta^{ab}\delta^{cd}\delta_f^n\delta\mathcal{E}_n^f\right)\Big]\\
    =& \frac{\kappa^2}{2}\Big[\left(\delta_l^a\delta_m^b-\delta_l^b\delta^a_m\right)\delta\mathcal{A}_a^l\delta\mathcal{A}_b^m-\frac{\beta^2+1}{\beta^2}\left(\delta_l^a\delta_m^b-\delta_l^b\delta^a_m\right)\left(\delta\mathcal{A}_a^l-\delta\Gamma_a^l\right)\left(\delta\mathcal{A}_b^m-\delta\Gamma_b^m\right)\\
    &+\delta\mathcal{A}^i_{[b,a]}\epsilon_i^{\:\:jk}\left(\delta_j^a\delta\mathcal{E}_k^b+\delta_k^b\delta\mathcal{E}_j^a-\frac{1}{2}\delta_j^a\delta_k^b\delta_d^n\delta\mathcal{E}_n^d\right)-W_{abc}^i\delta\mathcal{E}_c^i T^{ab}\Big].
    \end{split}
\end{align}
with the linearised spin connection \cite{Fahn:2022zql}
\begin{align}\label{LinearisedSpinconnectionintermsoftriads1}
    \begin{split}
        \delta\Gamma_a^i\coloneqq-\frac{1}{2}\epsilon^{ijk}\delta_k^b\left(-\delta_a^l\delta_c^j\partial_b+\delta_c^j\delta_b^l\partial_a-\delta_{ac}\delta^l_j\partial_b+\delta_a^j\delta_c^l\partial_b\right)\delta\mathcal{E}_l^c.
    \end{split}
\end{align}
The derivation can for instance be found in \cite{Fahn:2022zql}. For the diffeomorphism constraint, we obtain the following expression
\begin{align}\label{LinDiffeoConstraint}
        \delta C_a &= \frac{\kappa}{\beta}\delta\mathcal{A}^i_{[i,a]}+\kappa E^c A_{[a,c]}
        =\frac{\kappa}{\beta}\delta\mathcal{A}^i_{[i,a]}+\kappa T_{0a},\\
    \delta^2 C_a &= \frac{\kappa^2}{\beta}\left(\epsilon^b_{\:\:jk}\delta\mathcal{A}_a^j\delta\mathcal{A}_b^k+\delta\mathcal{A}^i_{[b,a]}\delta\mathcal{E}_i^b\right).
\end{align}
Given the perturbed constraints we can write down the perturbed interaction Hamiltonian by means of the first and second order perturbations of the spatial diffeomorphism and Hamiltonian constraint.
~\\
~\\
 For further investigation, we decompose the total Hamiltonian into three contributions
 \begin{equation}\label{eq: perturbed Hamiltonian 1}
\delta H = \delta H_{\textrm{geo}}+H_{\textrm{Maxwell}}+\delta H_I,   
\end{equation}
where
\begin{align}\label{LingravHamiltonian}
    \begin{split}
    \delta H_\textrm{geo}=&\int\limits_{\mathbb{R}^3} d^3x \Big( -\frac{\kappa}{2\beta^2}\delta_{[m}^a\delta_{n]}^b\left(\delta\mathcal{A}_a^m\delta\mathcal{A}_b^n+(\beta^2+1)\delta\Gamma_a^m\delta\Gamma_b^n-2\delta\mathcal{A}_b^n\delta\Gamma_a  ^m\right)\\
    &-\kappa\delta N_{,a} \epsilon_{i}^{\:\:ab}\delta\mathcal{A}_{b}^i+ \frac{\kappa}{\beta}\delta N^a\delta\mathcal{A}^i_{[i,a]}+\delta\Lambda^i\frac{\kappa}{2\beta}\left(\delta\mathcal{E}^a_{i,a}+\epsilon_{ij}^{\:\:\:a}\delta\mathcal{A}^j_a\right)\Big)
    \end{split}
\end{align}
describes the gravitational system up to second order in the perturbations,
\begin{align}
    H_{\textrm{Maxwell}}= \int\limits_{\mathbb{R}^3} d^3x \left(T^{00} +\phi G^{\textrm{U(1)}}\right)=\int\limits_{\mathbb{R}^3} d^3x \left(\frac{1}{2}E^a E_a+\frac{1}{4}A_{[a,b]}A^{[a,b]} -\phi E_{\:\:,a}^a\right)
\end{align}
is the Hamiltonian for Maxwell theory in  Minkowski spacetime. The interaction Hamiltonian $\delta H_I$ is given in  \eqref{eq:postminkowski interaction Hamiltonian}.
~\\
~\\
The final step in this section is to study the constraint algebra in the linearised phase space.  Here we are only interested in the results up to linear order in $\kappa$ and $\delta$ respectively. A rather long and tedious computation yields the following result:
\begin{align}\label{LinConstraintAlgebra}
    \begin{split}
    \{G^{\textrm{U(1)}}[\phi](t),G^{\textrm{geo}}[\delta\Lambda](t)\}=0,&\quad  \{G^{\textrm{U(1)}}[\phi](t),\vec{C}[\delta\vec{N}](t)\}=0,\\ 
    \{G^{\textrm{U(1)}}[\phi](t), C[\delta N](t)\}=0,&\quad \{G^{\textrm{U(1)}}[\phi](t),G^{\textrm{U(1)}}[\phi](t)\}=0,\\    
    \{G^{\textrm{geo}}[\delta\Lambda](t),\vec{C}[\delta\vec{N}](t)\}= 0+O(\kappa^2,\delta^2),&
    \quad \{G^{\textrm{geo}}[\delta\Lambda](t),C[\delta N](t)\}=0+ O(\kappa^2,\delta^2),\\
    \{\vec{C}[\delta\vec{N}](t), C[\delta N](t)\}=0+O(\kappa^2,\delta^2),&\quad 
    \{C[\delta N](t), C[\delta N](t)\}= 0+O(\kappa^2,\delta^2),\\
    \{\vec{C}[\delta\vec{N}](t), \vec{C}[\delta \vec{N}](t)\}= 0+O(\kappa^2,\delta^2), &\quad
    \{G^{\textrm{geo}}[\delta\Lambda](t),G^{\textrm{geo}}[\delta\Lambda](t)\}=0+O(\kappa^2,\delta^2).
    \end{split}
\end{align}
The algebra for the $U(1)$-Gauß constraint $G^{\rm U(1)}$ does not get modified. This was to be expected, since we have only perturbed the gravitational degrees of freedom. For the algebra of the geometrical constraints we only get contributions that are at least second order in either $\kappa$ and/or $\delta$. The detailed calculation for the algebra of the purely geometrical constraints can be found in \cite{Ashtekar:1991mz} and also in \cite{Fahn:2022zql}. The contributions of the photon degrees of freedom to the diffeomorphism and Hamiltonian constraint are, according to the post-Minkowski approximation, already linear in $\kappa$. The Poisson bracket with respect to the photon degrees of freedom neither reduces the order in $\kappa$ nor in $\delta$. Therefore, all the terms stemming from Poisson brackets in the constraint algebra where the contributions of the photon degrees of freedom to  $C$ and $\vec{C}$ are involved are at least of second order in $\kappa$. Hence, at the order of the post-Minkwoski approximation we consider, the algebra can be treated as being Abelian. Note that here this is true always and not only on the  $\delta G_i^{\textrm{geo}}\approx 0$ hypersurface as it was the case for the constraint algebra in the full theory. Given the kinematical linearised phase space now in the next section we will construct Dirac observables in a way that we perform a so-called Kucha\v{r} decomposition of this phase space. This means that the transition to a set of Dirac observables can be formulated in terms of a canonical transformation in which, next to the elementary Dirac observables, the constraints and their associated reference fields, also often denoted as clocks, become part of the new set of canonical variables.
\section{Dirac observables in the linearised model using geometrical clocks and an U(1)-Gauß clock}
\label{sec:ConstrObs}
In this section we will construct the Dirac observables for the model discussed in section \ref{sec:CassicalSetup}. The partially reduced phase space with respect to the primary constraints consists of the following 24 phase space degrees of freedom
\begin{equation}
\label{eq:OrigPhaseSpace}
 \big({\cal A}_a^i(\vec{x},t), {\cal E}^a_i(\vec{x},t)\big),\,\, \big(A_a(\vec{x},t),E^a(\vec{x},t)\big).   
\end{equation}
In addition, the system has eight first class constraints
\begin{equation}
 G^{\textrm{U(1)}}(\vec{x},t),\,\, G_i^{\rm geo}(\vec{x},t),\,\, C(\vec{x},t), \,\,C_a(\vec{x},t).  
\end{equation}
The first-class nature of the constraints implies that the physical phase space will consist of eight independent degrees of freedom. In addition to constructing the Dirac observables that encode these eight physical degrees of freedom, we want to perform a Kucha\v{r} decomposition of the original phase space that we started with. By this we mean that we can formulate the set of Dirac observables together with the constraints and a suitable choice of reference fields, also called clocks, resulting in a new set of canonical variables that we can obtain by a canonical transformation from the original phase space variables in \eqref{eq:OrigPhaseSpace}. A similar transformation has been constructed in \cite{Fahn:2022zql} for a model in which linearised gravity is coupled to a scalar field, which has next to gravity no additional gauge constraints, unlike the Maxwell system with its U(1)-Gau\ss{} constraint.
~\\
~\\
Following the idea of the relational formalism \cite{Rovelli:1990ph, Rovelli:1990pi,Rovelli:2001bz,Dittrich:2004cb,Dittrich:2005kc,Thiemann:2004wk,Vytheeswaran:1994np} we choose one reference field (clock) for each constraint present and then construct Dirac observables for the remaining degrees of freedom by means of the chosen clocks. As for systems involving gravity we also need to choose clocks for the spatial diffeomorphism and Hamiltonian constraints, we can interpret those clocks as a dynamical physical reference frames that provide a notion of physical temporal and spatial coordinates. Furthermore, it allows to describe the dynamics of the Dirac observables with respect to the temporal clock and in the model considered here the dynamics are generated by a physical Hamiltonian. As we will discuss in the next subsection, there exists a choice of geometrical clocks and U(1)-Gau\ss{} clock such that the application of the observable map and its dual counter part yield directly the two transverse degrees of freedom of the photon as well as the two polarisations degrees of freedom of the gravitational wave. All remaining degrees of freedom in the kinematical phase space can be identified with either clocks or their conjugate momenta, where the latter are the constraints.
\subsection{A choice of a set of geometrical clocks and U(1)-Gauß clock}\label{sec: A choice of a set of geometrical clocks and U(1)-Gauß clock}
As a first step towards the construction of Dirac observables, we will choose a suitable set of phase space functions on the linearised phase space which we denote as clocks, that can be chosen as a dynamical physical reference system. We use the following notation
\begin{align}\label{defconstraints}
    C_I(\vec{x},t)\coloneqq& \left(C (\vec{x},t),C_a (\vec{x},t), G^{\textrm{geo}}_i(\vec{x},t), G^{\textrm{U(1)}} (\vec{x},t)\right),\\
    \delta C_I(\vec{x},t)\coloneqq&\left(\delta C (\vec{x},t),\delta C_a (\vec{x},t), \delta G^{\textrm{geo}}_i (\vec{x},t)\right),\\
    \delta^2 C_I(\vec{x},t)\coloneqq&\left(\delta^2 C (\vec{x},t),\delta^2 C_a (\vec{x},t), \delta^2 G^{\textrm{geo}}_i (\vec{x},t)\right),\\
    \begin{split}\label{eq: Hamilton constraint up to second order}
        C \coloneqq & \delta C^{\textrm{geo}}+\delta C^{\textrm{ph}}+\delta^2 C^{\textrm{geo}}+\delta^2 C^{\textrm{ph}}\\
        =& \kappa \epsilon_{i}^{\:\:jk}\delta_j^a\delta_k^b\delta\mathcal{A}_{b,a}^i+\kappa T^{00}(-E^a,A_a)+\delta^2 C^{\textrm{geo}}-\frac{\kappa^2}{2}W_{abc}^i\delta\mathcal{E}_i^c T^{ab}(-E^a,A_a),
    \end{split}\\
    \begin{split}\label{eq: diffeo constraint up to second order}
        C_a \coloneqq & \delta C_a^{\textrm{geo}}+\delta C_a^{\textrm{ph}}+\delta^2 C_a^{\textrm{geo}}\\
        =&\frac{\kappa}{\beta}\delta_l^b\delta^d_{[b}\partial_{a]}\delta\mathcal{A}^l_{d}+\kappa T_{0a}(-E^a,A_a)+\delta^2 C_a^{\textrm{geo}},
    \end{split}\\
    \begin{split}\label{eq: geometric Gauß constraint up to second order}
         G^{\textrm{geo}}_i \coloneqq & \delta G_i^{\textrm{geo}}+\delta^2 G_i^{\textrm{geo}}\\
         =& \frac{\kappa}{2\beta}\left(\delta\mathcal{E}^a_{i,a}+\epsilon_{ij}^{\:\:\:k}\delta\mathcal{A}^j_a\delta_k^a\right)+\delta^2 G_i^{\textrm{geo}},
    \end{split}\\
    \label{eq: electromagnetic Gauß constraint}
         G^{\textrm{U(1)}}=& -E^a_{\:\:,a}.
\end{align}
Here the notation for the $\delta$ in front of the different constraint orders is chosen such that it numbers the order of the term. Since we have two perturbation parameters, always the highest order in $\kappa$ or $\delta$ appearing is counted. Thus $\delta C^{\textrm{ph}}=\kappa T^{00}(-E^a,A_a)$ comes with a delta since it is first order in $\kappa$ and therefore a first order term. On the other hand $\delta^2 C^{\textrm{ph}}=\frac{\kappa^2}{2}W_{abc}^i\delta\mathcal{E}_i^c T^{ab}(-E^a,A_a)$ is second order in $\kappa$ and first order in $\delta\mathcal{E}^c_i$, thus a second order term, denoted by a $\delta^2$ in front.
~\\
~\\
Next we introduce for each first order constraint one reference field, also denoted as clock that have the property that they consist of elementary linear perturbations and their derivatives only. For the linearised Hamiltonian, spatial diffeomorphism and geometrical Gau\ss{} constraint we chose so called geometrical clocks that are build from geometrical degrees of freedom only. Geometrical clocks have been introduced already in early works of the ADM formalism in \cite{Arnowitt:1962hi}. These have been generalised in the context of linearised gravity in terms of Ashtekar-Barbero variables in \cite{Dittrich:2006ee}. Here we will use the geometrical clocks used in 
\cite{Fahn:2022zql}, that agree in vacuum with the ones in \cite{Dittrich:2006ee}, but slightly deviate from them in case a non-vanishing energy-momentum tensor is present. We will briefly review their construction in \cite{Fahn:2022zql} below. In addition to the geoemetrical clocks, in contrast to \cite{Fahn:2022zql} where a scalar field is coupled to linearised gravity, here we also need to choose a clock associated with  the Maxwell U(1)-Gau\ss{} constraint, for which we construct the clock from the electromagnetic degrees of freedom only. In order to follow the notation for the geometrical clocks used in \cite{Fahn:2022zql}, we introduce the following notation. At the full non-linear level, we introduce for each constraint $C_I$ gauge fixing conditions $\mathcal{G}^I:=T^I-\tau^I$, here $T^I$ denote the in general full non-linear clocks and $\tau^I$ are spacetime functions that determine the specific gauge fixing condition. At the level of perturbation theory we define 
\begin{equation}
\delta\mathcal{G}^I:=\mathcal{G}^I-\overline{\mathcal{G}}^I=\mathcal{G}^I=T^I-\tau^I,
\end{equation}
where we used in the second step that the gauge fixing conditions in the background are satisfied, this $\overline{\mathcal{G}}^I=0$. Now in order to, in principle, choose different gauge-fixing condition at the various orders we write $\tau^I(t,\vec{x})=\overline{\tau}^I(t,\vec{x})+\kappa\: (\tau^{(1)})^I+\frac{1}{2}\kappa^2\:(\tau^{(2)})^I+O(\kappa^3)$, where the specific choices for $\tau^I$ that we will use in this work will be discussed in section \ref{sec: Coordinate gauge fixing}. Taking further into account that for the $U(1)$-Gau\ss{} clock no perturbations are considered and that for the geometrical clocks by assumption all higher than linear order terms vanish, we use the following notation:
\begin{align}\label{defclokcs}
    \begin{split}
    \mathcal{G}^I&\coloneqq\left(\overline{\mathcal{G}}+\delta \mathcal{G} ,\overline{\mathcal{G}}_a+\delta \mathcal{G}^a, \overline{\mathcal{G}}_{\textrm{geo}}^j+\delta \mathcal{G}_{\textrm{geo}}^j,\mathcal{G}^{\textrm{U(1)}}\right),\\
    \delta\mathcal{G}^I_{\rm geo}&\coloneqq \mathcal{G}^I -\overline{\mathcal{G}}^I=\left(\delta \mathcal{G},\delta \mathcal{G}^a, \delta \mathcal{G}_{\textrm{geo}}^j\right),\\
    \delta \mathcal{G}&\coloneqq (\delta T-\tau),\\
    \delta\mathcal{G}^a &\coloneqq (\delta T^a-\sigma^a),\\
    \delta\mathcal{G}^j_{\textrm{geo}} &\coloneqq (\delta\Xi^j-\xi^j),\\
    \mathcal{G}^{\textrm{U(1)}}&\coloneqq (T^{\textrm{U(1)}}-\gamma).
    \end{split}
\end{align}
Here we denoted the individual clock parameters $\tau^I=(\tau,\sigma^a,\xi^j,\gamma)$ for the 
linearised Hamiltonian, spatial diffeomorphism and geometrical Gau\ss{} and $U(1)$-Gau\ss{} constraint respectively, so that  $(\delta T,\:\tau), (\delta T^a,\:\sigma^a)$, $(\delta\Xi^j,\:\xi^j)$ and $(T^{\textrm{U(1)}},\:\gamma)$ list the chosen clocks with their corresponding parameters. For the case of $\delta\mathcal{G}^I_{\rm geo}$ we consider the expansion of $\tau$ up to linear order only. The clock parameters $\tau$ and $\sigma^a$ will be later chosen in such a way that these geometrical clocks provide a physical reference system for a temporal clock and spatial rods, whereas the clock parameters associated with the Gau\ss{} constraints will be chosen to vanish. The Dirac observables are, by construction, gauge invariant, and therefore gauge invariant for all choices of $(\tau,\sigma^a,\xi^i,\gamma)$.
~\\
~\\
In order to formulate the choice of clocks in terms of a Kucha\v{r} decomposition of the linearised phase space, that is that we can clearly separate the gauge and physical degrees of freedom, we need to ensure that 
\begin{itemize}
\item[(i)]\label{it: clocks commute} $\{\mathcal{G}^I(\vec{x},t), \mathcal{G}^J(\vec{y},t)\}=0$
\item[(ii)]\label{it: clocks and constrains canonical conjugated} $\{\delta \mathcal{G}^I(\vec{x},t), \delta C_J(\vec{y},t)\}= \frac{1}{\kappa}\delta^I_J\delta^{(3)}(\vec{x},\vec{y})$ and $\{\mathcal{G}^{\textrm{U(1)}}(\vec{x},t), G^{\textrm{U(1)}}(\vec{y},t)\}= \delta^{(3)}(\vec{x},\vec{y})$
\item[(iii)]\label{it: clocks and second order constraints commute} $\{\delta\mathcal{G}^I(\vec{x},t),\delta^2 C_J(\vec{y},t)\}=0+O(\kappa^2,\delta^2)$\,.
\end{itemize}
The factor $\frac{1}{\kappa}$ is included because it is also included in the Poisson brackets of the linearised elementary gravitational variables. 
We present the final set of clocks and weakly equivalent constraints in section \ref{sec:FinalSetCons+Clocks} for the reader that would like to skip the more detailed discussion on how this set can be constructed.
~\\
~\\
Likewise to the assumptions used in \cite{Fahn:2022zql}, we consider geometrical clocks which are first order in $\delta$ (thus $\delta^2\mathcal{G}=0$). Given this, if we compute the Poisson brackets between the geometrical clocks and constraints up to linear order in $\kappa$ and $\delta$ we obtain
\begin{align}
    \{\mathcal{G}^I_{\rm geo}(\vec{x},t),  C_J(\vec{y},t)\}= 
    \{\delta{\cal G}^I_{\rm geo}(\vec{x},t),\delta C_J(\vec{y},t)\}+
    \{\delta{\cal G}^I_{\rm geo}(\vec{x},t),\delta^2 C_J(\vec{y},t)\} +O(\kappa^2,\delta^2).
\end{align}
For the U(1)-Gau\ss{} clock the analogue equation reads
\begin{align}
    \{\mathcal{G}^{\rm ph}(\vec{x},t), C_J(\vec{y},t)\}= \{{\cal G}^{\rm ph}(\vec{x},t),\delta C_J(\vec{y},t)\}+
    \{{\cal G}^{\rm ph}(\vec{x},t),\delta^2 C_J(\vec{y},t)\} +O(\kappa^2,\delta^2).
\end{align}
To discuss the explicit form of these clocks, as a first step we choose the following set of geometrical clocks introduced in \cite{Fahn:2022zql}
\begin{align}\label{geometrical Gauß constraints}
    \delta \Xi^{i}(\vec{x},t)\coloneqq\frac{2}{\kappa}\partial^{a}\left(\delta \mathcal{A}_a^{i}\ast G^{\Delta}\right)(\vec{x},t)=\frac{2}{\kappa}\partial^a_x\int\limits_{\mathbb{R}^3}d^3y\:\delta\mathcal{A}^i_a(\vec{y},t)G^\Delta(\vec{x}-\vec{y}),
\end{align}
\begin{align}\label{Hamilton constraint clock}
    \delta T(\vec{x},t)\coloneqq-\frac{1}{\kappa\beta}\Big[\frac{1}{2}\epsilon_a^{\:cb}\delta_c^{i}\partial_b(\delta \mathcal{E}_i^{a}\ast G^\Delta)(\vec{x},t)+\delta_i^{a}(\delta \mathcal{A}^i_{a}\ast G^\Delta)(\vec{x},t)\Big],
\end{align}
\begin{align}\label{Diffeo clock}
    \begin{split}
    \delta \widetilde{T}^{a}(\vec{x},t) \coloneqq&\frac{1}{\kappa}\left(\delta_b^{a}\delta_c^{i}\partial^c-\frac{1}{2}\delta_b^{i}\partial^{a}+\delta^{ac}\delta_c^{i}\partial_b\right)(\delta \mathcal{E}_i^{b}\ast G^\Delta)(\vec{x},t).
    \end{split}
\end{align}
and for the U(1)-Gau\ss{} constraint the following clock
\begin{align}\label{Electromagnetic clock}
    T^{\textrm{U(1)}}(\vec{x},t)\coloneqq\partial^{a}( A_a\ast G^\Delta)(\vec{x},t),
\end{align}
where $\ast$ is the convolution and $G^\Delta(\vec{x}-\vec{y})$ is the Green's function for $-\Delta$:
\begin{align}\label{Greens function}
    G^{\Delta}(\vec{x}-\vec{y})\coloneqq\frac{1}{4\pi}\frac{1}{||\vec{x}-\vec{y}||}.
\end{align}
The choice for $T^{\textrm{U(1)}}$ is motivated from the Coulomb gauge $A_a^{\:\:,a}=0$. Since the primary constraint $\pi_0=0$ holds, we can choose the gauge  $A_0=0$ and then the combination of these two gauges also satisfies a Lorentz gauge condition $\partial^\mu A_\mu=0$ in our case if we set $T^{\textrm{U(1)}}=0$. The  Dirac observables however that we will construct below by means of these clocks provide an expression that is invariant under U(1)-Gau\ss{} transformation  also for $T^{\textrm{U(1)}}\not=0$. 
~\\
~\\
The set $( \delta T, \delta\widetilde{T}^a,\delta \Xi^{i},T^{\textrm{U(1)}})$ defined above has the property that if we consider the Poisson bracket with the linearised constraints then each clock is canonically conjugate to the linearised constraints with respect to which it is a reference field but might not commute with other constraints yet. The latter is for instance happening for $T^{\textrm{U(1)}}$ that does not commute with the electromagnetic contribution of the spatial diffeomorphism $\delta C^{\textrm{ph}}_a$ and Hamiltonian constraint $\delta C^{\textrm{ph}}+\delta^2 C^{\textrm{ph}}$. 
~\\
As a first step we want to implement (i), i.e. that all clocks also commute with each other in the required order $O(\kappa^2,\delta^2)$. For this purpose, we use the dual observable map, in which - in contrast to the (non-dual/original) observable map \cite{Dittrich:2004cb,Dittrich:2005kc, Vytheeswaran:1994np} introduced below - the role of the role of the constraints and clocks are interchanged, for those clocks that do not commute with all clocks. This allows to modify the set of clocks in such a way that the requirement that they mutually commute can be achieved. The dual observable formula for a phase space function $F[A_a,E^a,\delta\mathcal{A}_a^i,\delta\mathcal{E}^a_i]=f[A_a,E^a]+\delta f[A_a,E^a,\delta\mathcal{A}_a^i,\delta\mathcal{E}^a_i]+\delta^2 f[A_a,E^a,\delta\mathcal{A}_a^i,\delta\mathcal{E}^a_i]$ up to second order is given by \cite{Fahn:2022zql}
\begin{align}\label{eq:linearised dual observable map}
    \begin{split}
    \mathcal{O}^{\textrm{dual}}_{F,\{C_I\}}\coloneqq&[\exp{(-\Tilde{\kappa}\int\limits_{\mathbb{R}^3} d^3y \beta_I(\vec{y})\{\delta \mathcal{G}^I(\vec{y}),.\})F}]\Big|_{\beta_I=C_I}\\
    =&\mathcal{O}^{\textrm{dual}}_{F,\{G^{\textrm{U(1)}}\}}-\kappa\int\limits_{\mathbb{R}^3} d^3y \delta C_I(\vec{y})\{\delta \mathcal{G}^I(\vec{y}),\delta f\}\\
    &-\kappa\int\limits_{\mathbb{R}^3} d^3y \delta^2 C_I(\vec{y})\{\delta \mathcal{G}^I(\vec{y}),\delta f\}
    -\kappa\int\limits_{\mathbb{R}^3} d^3y \delta C^I(\vec{y})\{\delta \mathcal{G}^I(\vec{y}),\delta^2 f\}\\
    &+\frac{\kappa^2}{2}\int\limits_{\mathbb{R}^3} d^3 y\int\limits_{\mathbb{R}^3} d^3 z \delta C_I(\vec{y})\delta C_J(\vec{z})\{\delta \mathcal{G}^J(\vec{z}),\{\delta \mathcal{G}^I(\vec{y}),\delta^2 f\}\}\\
    &+O(\delta^3,\kappa^3)\,,
    \end{split}
\end{align}
where we defined
\begin{align}\label{eq:definition kappa tilde}
    \Tilde{\kappa} = \begin{cases}
        1 & \mathcal{G}^I = \mathcal{G}^{\textrm{U(1)}}\\
        \kappa & \mathcal{G}^I = \mathcal{G}_{\textrm{geo}}^I \\
    \end{cases}.
\end{align}
By construction, the dual observable map projects a phase space function F to its, weakly equivalent, corresponding phase space function $\mathcal{O}^{\textrm{dual}}_{F,\{C_I\}}$ which weakly Poisson commutes with all clocks.\footnote{Note that at this stage of our construction this is only given up to linear order in perturbation theory, since the chosen clocks are only canonically conjugate to the linearised constraints but potential additional contributions can come from the second order perturbations of the constraints. In a later step we will use the dual observable map to obtain a set of constraints which is indeed canonically conjugate to the set of clocks up to the order of perturbation theory we are interested in circumvented this. Also we will see that, in our case, the dual observable map will produce phase space functions which Poisson commute strongly with all clocks.} Although we have defined a dual observable map for all involved constraints in \eqref{eq:linearised dual observable map}, some caution is required when using this notation. For the Maxwell-Gau\ss{} constraint and the associated clock $(T^{\textrm{U(1)}},G^{\textrm{U(1)}})$, the dual observable map cannot be written as a power series in $\kappa$, since in the post-Minkowski approximation the electromagnetic sector is not perturbed. Therefore, in general, the entire infinite power series for the dual observable map must be considered in \eqref{eq:linearised dual observable map} which is denoted as $\mathcal{O}_{F,\{G^{\textrm{U(1)}}\}}$. However, we will only apply the dual observable map to the electromagnetic elementary phase space variables for which the series terminates after linear order also in the case of $(T^{\textrm{U(1)}},G^{\textrm{U(1)}})$.
\subsubsection{Constructing mutually commuting clocks via the dual vacuum observable map}
We will now use the dual observable map to transform the reference fields such that (i) is fulfilled. The $U(1)$-Gau\ss{} clock commutes with all geometrical clocks because they both consist solely of geometric or electromagnetic degrees of freedom, which mutually commute. We explicitly constructed them in this manner. Therefore, we will employ a modified dual observable map here, containing only the geometric part of the constraints, ensuring that after the application of the dual observable map the geometrical clocks do not acquire any electromagnetic degrees of freedom. This imposition is not restrictive, as the geometrical clocks already commute with the photon part of the first-order geometrical constraints, as well as with the $U(1)$-Gau\ss{} clock. The definition of the  dual vacuum map, which is formulated exclusively in terms of geometric degrees of freedom, applied to $F[\delta\mathcal{A}^i_a,\delta\mathcal{E}_i^a]=\delta f+\delta^2 f$ is given by
\begin{align}\label{eq: vacuum dual map}
    \begin{split}
    \mathcal{O}^{\textrm{dual, vac}}_{F,\{C_I^{\textrm{geo}}\}}\coloneqq & F[\delta\mathcal{A}^i_a,\delta\mathcal{E}_i^a]-\kappa\int\limits_{\mathbb{R}^3} d^3y \delta C_I^{\textrm{geo}}(\vec{y})\{\delta \mathcal{G}^I(\vec{y}),\delta f\}\\
    &-\kappa\int\limits_{\mathbb{R}^3} d^3y \delta^2 C_I^{\textrm{geo}}(\vec{y})\{\delta \mathcal{G}^I(\vec{y}),\delta f\}
    -\kappa\int\limits_{\mathbb{R}^3} d^3y \delta C_I^{\textrm{geo}}(\vec{y})\{\delta \mathcal{G}^I(\vec{y}),\delta^2 f\}\\
    &+\frac{\kappa^2}{2}\int\limits_{\mathbb{R}^3} d^3 y\int\limits_{\mathbb{R}^3} d^3 z \delta C_I^{\textrm{geo}}(\vec{y})\delta C_J^{\textrm{geo}}(\vec{z})\{\delta \mathcal{G}^J(\vec{z}),\{\delta \mathcal{G}^I(\vec{y}),\delta^2 f\}\}\\
    &+O(\delta^3,\kappa^3).
    \end{split}
\end{align}
It is noteworthy that this map does not reproduce weakly equivalent variables, since only the geometric part of the constraints is included. This is not problematic for our applications, it just means that the clocks before and after the application of the map are two choices of sets for clocks which are not weakly equivalent.
In the following we will mostly only need the first order application of this map onto first order phase space functions $F[\delta\mathcal{A}^i_a,\delta\mathcal{E}_i^a]=\delta f[\delta\mathcal{A}^i_a,\delta\mathcal{E}_i^a]$, which is given by
\begin{align}\label{eq: vacuum dual map first order}
    \begin{split}
    \mathcal{O}^{\textrm{dual, vac} (1)}_{F,\{C_I^{\textrm{geo}}\}}\coloneqq & \delta f-\kappa\int\limits_{\mathbb{R}^3} d^3y \delta C_I^{\textrm{geo}}(\vec{y})\{\delta \mathcal{G}^I(\vec{y}),\delta f\}
    +O(\delta^2,\kappa^2).
    \end{split}
\end{align}
As above, the non weakly equivalence at the linearised level means that if we go on the hypersurface where $C_I=0\:\:\:\:\forall I$, thus for the geometric constraints $\delta C^{\rm geo}_I +\delta^2C^{\rm geo}_I=-\delta C_I^{\rm ph}-\delta^2 C^{\rm ph}_I$, the application of \eqref{eq: vacuum dual map first order} reads
\begin{align}\label{eq: weak inequavilenz for first order dual map}
    \mathcal{O}^{\textrm{dual,vac(1)}}_{F,\{C_I\}}\Big|_{C^{\rm geo}_I=0}=F+\kappa\int\limits_{\mathbb{R}^3} d^3y (\delta C_I^{\rm ph}+\delta^2 C^{\rm ph}_I)(\vec{y})\{\delta \mathcal{G}^I(\vec{y}),\delta f\}+O(\delta^2,\kappa^2)
\end{align}
which is in general unequal to $F$\footnote{Where $C_I^{\textrm{ph}} \coloneqq \left(\delta C^{\textrm{ph}}+\delta^2 C^{\textrm{ph}},\delta C_a^{\textrm{ph}}\right)$.}.
~\\
~\\
In our applications of the dual observable map we first consider the clock $\delta\widetilde{T}^a$ that does not commute with $\delta T$ and $\delta \Xi^j$. Therefore, we apply the dual observable map with respect to $\delta T$ and $\delta \Xi^j$ on $\delta\widetilde{T}^a$ yielding additional contributions linearly in $\delta G_i^{\textrm{geo}}$ and $\delta C$. We apply the vacuum dual observable map, since applying the full one \eqref{eq:linearised dual observable map} would produce further terms depending on electromagnetic degrees of freedom, which is what we do not want, to have the geometrical clocks consist of geometric degrees of freedom only. Also, we want to have first order reference fields, therefore we apply it only up to first order \eqref{eq: vacuum dual map first order}. The resulting quantity, by construction, Poisson commutes with all geometrical clocks and trivially commutes with the electromagnetic one, since it depends only on geometric degrees of freedom. Thus, we define the following diffeomorphism clock \cite{Fahn:2022zql}
\begin{align}\label{eq:Diffeo clock}
    \begin{split}
    \delta T^{a}(\vec{x},t) \coloneqq&\mathcal{O}^{\textrm{dual,vac}(1)}_{\delta\widetilde{T}^a(\vec{x},t),\{C_I\}}=\frac{1}{\kappa}\left(\delta_b^{a}\delta_c^{i}\partial^c-\frac{1}{2}\delta_b^{i}\partial^{a}+\delta^{ac}\delta_c^{i}\partial_b\right)(\delta \mathcal{E}_i^{b}\ast G^\Delta)(\vec{x},t)\\
    &+\frac{2\beta}{\kappa^2}\big[\frac{1}{2}\delta_b^{i}\partial^{a}\partial^{b}(\delta G^\textrm{geo}_i\ast G^{\Delta\Delta})(\vec{x},t)-\delta^{ab}\delta_b^{i}(\delta G^\textrm{geo}_i\ast G^\Delta)(\vec{x},t)\Big]\\
    &+\frac{1}{2\kappa^2}\partial^{a}\left((\delta C^{\textrm{geo}}\ast G^{\Delta\Delta}\right)(\vec{x},t),
    \end{split}
\end{align}
where $G^{\Delta \Delta}$ denotes the Green's function of the squared Laplacian, that is
\begin{align}\label{eq: greens fct Laplacian squared}
    G^{\Delta \Delta}(\vec{x}-\vec{y})=\int\limits_{\mathbb{R}^3} d^3 z\: G^{\Delta}(\vec{x}-\vec{z}) G^{\Delta}(\vec{y}-\vec{z}).
\end{align}
That the clocks $\delta T^a$ also commute with themselves is ensured by the fact that the term in the first line in \eqref{eq:Diffeo clock} commutes with $\delta C^{\textrm{geo}}$, is purely geometrical and thus obviously also commutes with $\delta T^{\textrm{U(1)}}$. Thus the set of clocks $\{\delta T, \delta T^a,\delta\Xi^i,T^{\textrm{U(1)}}\}$ is mutually commuting.
\subsubsection{Constructing an weakly equivalent set of constraints via the dual observable map}
The clocks are canonically conjugate to exactly one of the linearised constraints $(\delta C^{\textrm{geo}}, \delta C_a^{\textrm{geo}},\delta G_i^{\textrm{geo}}\\
,G^{\textrm{U(1)}})$. This means the next step is to ensure that the clocks commute with one of the corresponding constraints not only in lowest but also up to linear order in $\kappa,\delta$. Thus we need to make sure that the clocks also commute with the remaining linear and higher order contributions in the Hamiltonian, diffeomorphism and Gau\ss{} constraints. Explicitly, these are the following contributions 
$(\delta C^{\textrm{ph}}+\delta^2 C^{\textrm{ph}}+\delta^2 C^{\textrm{geo}}, \delta C_a^{\textrm{ph}}+\delta^2 C^{\textrm{geo}}_a,\delta^2 G^{\textrm{geo}})$, we do this by applying the dual observable map \eqref{eq:linearised dual observable map} onto them. The geometrical clocks do not commute with $(\delta^2C^{\textrm{ph}}, \delta^2C^{\textrm{geo}}, \delta^2 C_a^{\textrm{geo}},\delta^2 G_i^{\textrm{geo}})$ while the $U(1)$-Gau\ss{} clock does not commute with $(\delta C^{\textrm{ph}}+\delta^2 C^{\textrm{ph}}, \delta C_a^{\textrm{ph}})$. To achieve the desired Poisson algebra, we could either adjust the clocks in a suitable way or transform the constraints to a weakly equivalent form that has the desired Poisson algebra with our already constructed clocks. In this work, we will choose the latter approach. However, note that the other approach is also viable. Using the other approach would lead us to a different set of reference fields. To obtain the transformed constraints, we apply the dual observable map given in \eqref{eq:linearised dual observable map}. It is important that we do not consider only the vacuum version here, because in contrast to the case of the clocks, for the constraints we want weakly equivalent versions before and after the application of the map. This leads to the following set of constraints:
\begin{align}
\label{eq:primedConstraints}
    \begin{split}
    C^{\prime}=& \delta C^{\textrm{geo}}+\mathcal{O}^{\textrm{dual}}_{\delta C^{\textrm{ph}}+\delta^2 C,\{C_I\}},\\
    C^{\prime}_a=& \delta C^{\textrm{geo}}_a+\mathcal{O}^{\textrm{dual}}_{\delta C^{\textrm{ph}}_a+\delta^2 C_a,\{C_I\}},\\
    G^{\textrm{geo}\prime}_i =& \delta G^{\textrm{geo}}+\mathcal{O}^{\textrm{dual}}_{\delta^2 G^{\textrm{geo}},\{C_I\}},\\
    G^{\textrm{U(1)}\prime}=& G^{\textrm{U(1)}}.
    \end{split}
\end{align}
These primed constraints are, by construction, canonically conjugate to exactly one of the reference fields and have vanishing Poisson brackets with the others. Furthermore, they are weakly equivalent to the original set of constraints from which we started.
~\\
~\\
To calculate set of primed constraints in \eqref{eq:primedConstraints} explicitly, we first look at a property of the observable\footnote{The observbale map, maps a given phase space function to a corresponding phase space function which Poisson commutes with all constraints. For more detials see section \ref{Construction of Dirac observables with respect to the chosen clocks}.} that was proven in \cite{Thiemann:2004wk}, which states that the observable of a given phase space function is weakly equivalent to the same function evaluated at the corresponding observables. This also carries over to the dual observable map, which was not considered in \cite{Thiemann:2004wk}, as the main assumption that the two quantities used to construct the  observable map are canonically conjugate also holds in the dual case. In the model considered here this even holds as a strong equality, that is for a generic function on the linearised phase space $F(\delta\mathcal{A}_a^i,\delta\mathcal{E}^a_i,A_a,-E^a)$ we have
\begin{align}\label{eq:dual and observable on phase space function}
    \begin{split}
    \mathcal{O}_{F(\delta\mathcal{A}_a^i,\delta\mathcal{E}^a_i,A_a,-E^a),\{\mathcal{G}^I\}}=F(\mathcal{O}_{\delta\mathcal{A}_a^i,\{\mathcal{G}^I\}}, \mathcal{O}_{\delta\mathcal{E}^a_i,\{\mathcal{G}^I\}}, \mathcal{O}_{A_a,\{\mathcal{G}^I\}},\mathcal{O}_{-E^a,\{\mathcal{G}^I\}}),\\
    \mathcal{O}^{\textrm{dual}}_{F(\delta\mathcal{A}_a^i,\delta\mathcal{E}^a_i,A_a,-E^a),\{C_I\}}=F(\mathcal{O}^{\textrm{dual}}_{\delta\mathcal{A}_a^i,\{C_I\}}, \mathcal{O}^{\textrm{dual}}_{\delta\mathcal{E}^a_i ,\{C_I\}}, \mathcal{O}^{\textrm{dual}}_{A_a,\{C_I\}},\mathcal{O}^{\textrm{dual}}_{-E^a,\{C_I\}}).
    \end{split}
\end{align}
Hence, we only need to know the application of the dual observable map on the elementary variables ${\delta\mathcal{A}_a^i, \delta\mathcal{E}^a_i, A_a, -E^a}$. For the proof of these identities see appendix \ref{app: observable properties}.
~\\
~\\
We will start with the electromagnetic degrees of freedom. These commute with all geometrical clocks. Therefore, in the dual observable map, only the Poisson bracket with $T^{\textrm{U(1)}}$ is non-zero. It is also important to mention that the $U(1)$-Gau\ss{} clock depends linearly on the electromagnetic elementary variables. Thus, the Poisson bracket of the $U(1)$-Gau\ss{} clock with one of them is either zero or independent of the elementary electromagnetic variables. Therefore, only the first order of the iterative Poisson bracket in the power series can contribute. Overall, we get
\begin{align}
    \label{eq: dual map on electromagnetic connection}
    \begin{split}
    \mathcal{O}^{\textrm{dual}}_{A_a,\{C_I\}}(\vec{x},t)
    =& A_a(\vec{x},t)
    \end{split},\\
    \label{eq: dual map on electric field}
    \begin{split}
    \mathcal{O}^{\textrm{dual}}_{-E^a,\{C_I\}}(\vec{x},t)
    =& -E^a(\vec{x},t)+\partial^a G^{\textrm{U(1)}}(\vec{x},t)
    = -P_b^a E^b(\vec{x},t)
    \end{split}.
\end{align}
where
\begin{align}\label{eq: projector transverse subspace}
    P_b^aE^b=\delta_a^bE^b+\partial_b\partial^a\left(E^b\ast G^{\Delta}\right)
\end{align}
is the projector onto the transverse subspace \cite{Dittrich:2006ee}, for further details about the definition of these projectors see appendix \ref{app: projectors}.
~\\
~\\
Now we can use this to obtain the explicit form of the first order primed Hamiltonian and diffeomorphism constraints, that both commute with the $U(1)$-Gau\ss{} clock:
\begin{align}
    \label{eq: primed first order Hamilton constraint}
    \begin{split}
    \delta C^{\prime}\coloneqq & \delta C^{\textrm{geo}}+\mathcal{O}^{\textrm{dual}}_{\delta C^{\textrm{ph}},\{C_I\}}
    =\delta C^{\textrm{geo}}+\kappa T^{00}(-P^a_b E^b,A_a),
    \end{split}\\
    \label{eq: primed first order diffeo constraint}
    \begin{split}
    \delta C^{\prime}_a \coloneqq & \delta C^{\textrm{geo}}_a +\mathcal{O}^{\textrm{dual}}_{\delta C^{\textrm{ph}}_a,\{C_I\}}
    =\delta C^{\textrm{geo}}_a+\kappa T_{0a}(-P^a_b E^b,A_a).
    \end{split}
\end{align}
where we inserted $\delta C^{\textrm{ph}}=\kappa T^{00}$ and $\delta C_a^{\textrm{ph}}=\kappa T_{0a}$. These further contribution trivially commute with all geometric reference fields, since they only consist of electromagnetic degrees of freedom, while the geometric reference fields consist of geometric ones only.
~\\
~\\
To obtain the final form of the primed constraints up to second order, that we consider as our starting point for the construction of Dirac observables, we also need to apply the dual observable map to the second order Hamiltonian and diffeomorphism constraints. This will implement the abelianisation up to corrections that are higher in the order of $\kappa,\delta$ that we consider in this model. Here, we follow the same strategy as above and first construct the image of the geometrical elementary variables under the dual observable map and then we can apply \eqref{eq:dual and observable on phase space function}. Note that, since we want to calculate the second order contributions of the primed constraints, we only need to apply the dual observable map up to first order on the geometric elementary variables involved, otherwise higher-order terms would be generated. As above, we consider the linearised version of the dual observable map which is thus sufficient for our purposes. Applied to first-order functions $F[A_a,E^a,\delta\mathcal{A}_a^i,\delta\mathcal{E}^a_i] = f[A_a,E^a] + \delta f[A_a,E^a,\delta\mathcal{A}_a^i,\delta\mathcal{E}^a_i]$, it reads
\begin{align}\label{eq: first order dual map}
    \begin{split}
    \mathcal{O}^{\textrm{dual}(1)}_{F,\{C_I^\prime\}}=&\mathcal{O}^{\textrm{dual}}_{F,\{G^{\textrm{U(1)}}\}}-\kappa\int\limits_{\mathbb{R}^3} d^3y \delta C_I^\prime(\vec{y})\{\delta \mathcal{G}^I(\vec{y}),\delta f\}+O(\delta^2,\kappa^2).
    \end{split}
\end{align}
Moreover, we must already employ the primed first-order constraints \eqref{eq: primed first order Hamilton constraint} and \eqref{eq: primed first order diffeo constraint} here, as using the unprimed ones could produce terms that do not commute with the $U(1)$ Gau\ss{} clock. This yields
\begin{flalign}
    \label{eq: dual map on Ashtekar connection}
    \mathcal{O}^{\textrm{ dual}(1)}_{\delta\mathcal{A}_a^i,\{\mathcal{C}_I^\prime\}}(\vec{x},t)
    =&\frac{1}{2}\left(P_{la}P^{di}+\delta^d_l(\delta^i_a-P^i_a)+\delta_l^i(\delta_a^d-P^d_a)+\delta_a^dP_l^i-\delta_a^iP_l^d\right)\delta\mathcal{A}_d^l(\vec{x},t)\\ \nonumber
    &+\frac{1}{2}\epsilon_a^{\:\:cb}\delta_c^i\partial_b\left(T^{00}\left(-P_b^aE^b,A_a\right)\ast G^\Delta\right)(\vec{x},t)\\ \nonumber
    &-\beta\left(\delta_a ^g\delta_b^i\partial^b-\frac{1}{2}\delta_a^i\partial^g\right)\left(T_{0g}\left(-P_b^aE^b,A_a\right)\ast G^\Delta\right)(\vec{x},t)\\ \nonumber
    &-\frac{\beta}{2}\delta^i_b\partial^g\partial^b\partial_a\left(T_{0g}\left(-P_b^aE^b,A_a\right)\ast G^{\Delta\Delta}\right)(\vec{x},t),\\ 
    \label{eq: dual map on densifed triad}
    \mathcal{O}^{\textrm{dual} (1)}_{\delta\mathcal{E}^a_i,\{\mathcal{C}_I^\prime\}}(\vec{x},t)
    =& P_d^a \delta\mathcal{E}^d_i(\vec{x},t)\\ \nonumber
    &+\left(\delta_i^a\delta^c_j\delta_k^d\epsilon_l^{\:\:jk}\partial_c+\delta_k^d\epsilon_{il}^{\:\:\:\:k}\partial^a+\delta^{cj}\delta_k^a\delta_l^d\epsilon_{ji}^{\:\:\:\:k}\partial_c+\delta^{dj}\delta^a_k\delta^c_l\epsilon_{ij}^{\:\:\:\:k}\partial_c\right)\left(\delta\mathcal{A}_d^l\ast G^{\Delta}\right)(\vec{x},t)\\ \nonumber
    &+\delta^a_i\left(T^{00}\left(-P_b^aE^b,A_a\right)\ast G^\Delta\right)(\vec{x},t)+\beta\delta^{gj}\delta_k^a\epsilon_{ji}^{\:\:\:\:k}\left(T_{0g}\left(-P_b^aE^b,A_a\right)\ast G^\Delta\right)(\vec{x},t),
\end{flalign}
with the projector
\begin{align}
    \frac{1}{2}\left(P_{la}P^{di}+\delta^d_l(\delta^i_a-P^i_a)+\delta_l^i(\delta_a^d-P^d_a)+\delta_a^dP_l^i-\delta_a^iP_l^d\right),
\end{align}
which projects onto the 2 symmetric transverse traceless plus the 3 longitudinal degrees of freedom with respect to the spatial index a of the linearised Ashtekar-Barbero connection. For more details on the derivation of these terms see appendix \ref{app: action dual map}.
~\\
~\\
Using this and the properties \eqref{eq:dual and observable on phase space function}, we can calculate the application of the dual observable map on the second order constraints, where, up to now, we do not need their explicit form and write them as
\begin{align}\label{eq: transformed second order constraints}
    \begin{split}
    \delta^2 C^{\textrm{geo}\prime}\coloneqq &\delta^2 C^{\textrm{geo}}\left(\mathcal{O}^{\textrm{dual}(1)}_{\delta\mathcal{A}_a^i,\{ C_I^\prime\}}, \mathcal{O}^{\textrm{dual}(1)}_{\delta\mathcal{E}^a_i ,\{ C_I^\prime\}}\right),\\
    \delta^2 C^{\textrm{ph}\prime}\coloneqq & \delta^2 C^{\textrm{ph}}\left(\mathcal{O}^{\textrm{dual}(1)}_{\delta\mathcal{A}_a^i,\{C_I^\prime\}}, \mathcal{O}^{\textrm{dual}(1)}_{\delta\mathcal{E}^a_i ,\{ C_I^\prime\}}, \mathcal{O}^{\textrm{dual}}_{A_a,\{ C_I^\prime\}},\mathcal{O}^{\textrm{dual}}_{-E^a,\{C_I^\prime\}}\right),\\
    \delta^2 G_i^{\textrm{geo}\prime}\coloneqq &\delta^2 G_i^{\textrm{geo}}\left(\mathcal{O}^{\textrm{dual}(1)}_{\delta\mathcal{A}_a^i,\{ C_I^\prime\}}, \mathcal{O}^{\textrm{dual}(1)}_{\delta\mathcal{E}^a_i ,\{C_I^\prime\}}\right),\\
    \delta^2 C^{\textrm{geo}\prime}_a\coloneqq &\delta^2 C^{\textrm{geo}}_a\left(\mathcal{O}^{\textrm{dual}(1)}_{\delta\mathcal{A}_a^i,\{C_I^\prime\}}, \mathcal{O}^{\textrm{dual}(1)}_{\delta\mathcal{E}^a_i ,\{C_I^\prime\}}\right).
    \end{split}
\end{align}
Note that, even though we apply the dual observable map only up to first order on the geometric elementary variables, the primed constraints remain weakly equivalent to the unprimed ones, since the resulting correction is of third order.
\subsubsection{Final set of geometrical and Gau\ss{} clocks and constraints}
\label{sec:FinalSetCons+Clocks}
To summarise, the discussion and computations so far provide us with a set consisting of constraints and clocks that are suitable for the construction of Dirac observables. Furthermore, they will satisfy standard canonical Poisson brackets which provides the possibility to choose them later as elementary variables on the kinematical phase space in context of a Kucha\v{r} decomposition. As discussed in the two former subsections, this set was obtained by means of applying the dual vacuum observable map to the originally chosen clocks and the dual observable map to the constraints. This final set is given by
\begin{equation}
\label{eq:FinalSet}
 \{C^{'},C^{'}_a,G^{\textrm{geo}'}_i,G^{\textrm{U(1)}},\delta T,\delta T^a, \delta \Xi^i,T^{\textrm{U(1)}}\}.  
\end{equation}
Where the primed first order constraints are given in \eqref{eq: primed first order Hamilton constraint}, \eqref{eq: primed first order diffeo constraint} while the second order is given in \eqref{eq: transformed second order constraints}. The Hamiltonian clock is given in \eqref{Hamilton constraint clock}, the geometric Gauß clock in \eqref{geometrical Gauß constraints}, the U(1)-Gauß clock in \eqref{Electromagnetic clock} and the diffeomorphism clock in \eqref{eq:Diffeo clock}.
~\\
~\\
The set consists of eight canonically conjugate pairs, i.e. all other combinations strongly vanish. The set of constraints is weakly equivalent to the constraints we started with, and therefore they define the same constraint hypersurface. This is not true for the the set of clocks before and after the dual observable map has been applied, because when we applied the dual observable map(s) associated with certain clocks to some of the remaining clocks, we only considered the geometrical part of the Hamiltonian and the diffeomorphism constraint as a "dual" clock. Since $\delta C^{\rm geo}\approx -\delta C^{\rm ph}$ and $\delta C_a^{\rm geo}\approx -\delta C_a^{\rm ph}$ do not necessarily vanish individually, the clocks we started with and the final clocks in \eqref{eq:FinalSet} are not weakly equivalent. Our main motivation to apply the dual observable map to the clocks in this way is that otherwise the Poisson algebra of the physical degrees of freedom of the photon field becomes more complicated, which we want to avoid. However, this is not a problem since we can interpret two weakly non-equivalent choices for the clocks as two different choices for dynamical reference frames and we choose the one that is technically simpler in terms of the algebra of observables. In contrast to the set of constraints two choices of set of clocks do not necessarily have to be weakly equivalent.
~\\
~\\
As a last step we want to compare this set of constraints and clocks, as well as their construction with the one in \cite{Fahn:2022zql}. As already discussed, in \cite{Fahn:2022zql} they studied the coupling to a scalar field, rather than a vector field as in our case, but also formulated gravity in Ashtekar–Barbero variables. Similar to our construction, they apply the relational formalism, with the use of the dual observable map, to construct a set of canonically conjugate pairs of clocks and constraints. The first difference of the vector field to the scalar field case, is that the scalar field has no matter constraint, while for the vector field we have $G^{\textrm{U(1)}}$. Thus also no further matter clock is needed in the scalar field case. We constructed the matter clock $T^{\textrm{U(1)}}$ to have only electromagnetic degrees of freedom, while the geometrical clocks only have geometric ones and are the same as in \cite{Fahn:2022zql}. For the further construction we applied the dual observable map onto all parts of the constraints apart from the first order geometric ones. The same is done in \cite{Fahn:2022zql}, but since the scalar field considered there has no internal gauge degrees of freedom and its contribution in the linearised constraints by construction commutes with the geometrical clocks chosen in \cite{Fahn:2022zql}, the situation is simpler than here.
\subsection{Coordinate gauge fixing}\label{sec: Coordinate gauge fixing}
Before we continue, let us first discuss the coordinate gauge fixing that we will employ in this work. First, we choose w.l.o.g. $\xi^i\coloneqq 0,\:\gamma\coloneqq 0$. We will associate the remaining $(\tau,\sigma^a)$ with the temporal and spatial coordinates. For this purpose we choose $\tau=\overline{\tau}=t$ and $\sigma^a=\overline{\sigma}^a=x^a_\sigma$, where $x_\sigma^a$ is the unique solution for $T^a(\vec{x},t)=\sigma^a$. Thus, we choose all linear and higher order contributions to $\tau,\sigma^a$ to vanish, which corresponds to a specific coordinate gauge fixing at the level of the  linearised phase space\footnote{Although with slight different notation used in \cite{Fahn:2022zql}, the final form for $\tau,\sigma^a$ and $\xi^j$ in \cite{Fahn:2022zql} agrees with the choice made here.}. Summarising, we use the following choice for clock parameters:
\begin{align}
    \begin{split}
    \tau(\vec{x},t)=\overline{\tau}(\vec{x},t)\coloneqq t,\:\:\:\:\sigma^a(\vec{x},t)=\overline{\sigma}^a\coloneqq x^a_\sigma,\:\:\:\:\xi^i(\vec{x},t)\coloneqq0,\:\:\:\:\gamma(\vec{x},t)\coloneqq0.
    \end{split}
\end{align}
This choice is later reinserted in the Dirac observables, so we can understand them as functions of $(\vec{x}_\sigma,t)$. In the following, we will abuse the notation and drop $\sigma$ in $\vec{x}_\sigma$. Furthermore, in a more general way, we will refrain from inserting this exact choice of $\sigma^a$ and $\tau$ until we come to the quantisation.
\subsection{Construction of Dirac observables with respect to the chosen clocks}\label{Construction of Dirac observables with respect to the chosen clocks}
We can now construct the Dirac observables associated with the constraints \eqref{eq:FinalSet} with respect to the chosen set of reference fields (clocks). One option to obtain a set of Dirac observables is to apply the observable map to the elementary variables. The set of resulting Dirac observables is discussed in appendix \ref{app: action observable map}, which will not be the final set of Dirac observables that we will use in the later part of this work for the reason that we want to choose a set of Dirac observables which allows a Kucha\v{r} decomposition of the kinematical phase space, that is that the gauge and physical degrees of freedom can be clearly separated in the sense that the chosen clocks and the constraints can be chosen as elementary phase space variables. The latter requires that the Dirac observables do not only commute with the constraints but also with the set of clocks. This can be achieved by combining the observable map with the dual observable map. In general, the order in which the two maps are applied is relevant, but since for the system considered here the constraints and clocks form five conjugate pairs, as shown in appendix \ref{app: More details of the influence of the order in which we apply the observable and dual observable map}, the order does not matter. Furthermore, for later convenience, we will sometimes use just the vacuum part of the dual observable map \eqref{eq: vacuum dual map}, which as proven in \ref{app: More details of the influence of the order in which we apply the observable and dual observable map}, also commutes with the observable map, if applied on geometric variables. As the order is irrelevant in our case and we already applied the dual observable map to parts of the constraints, we proceed by first applying the dual and then the observable map. 
~\\
~\\
The observable map is defined similar to the dual one for which the role of the constraints and clocks is switched. The notion of partial observables was first introduced in \cite{Rovelli:2001bz} in the context of the relational formalism  and further developed in \cite{Dittrich:2004cb,Dittrich:2005kc} where the explicit form of the observable map was introduced. Moreover, in \cite{Vytheeswaran:1994np} in the context of gauge-unfixing a projector was introduced by means of which the observable map in \cite{Dittrich:2004cb,Dittrich:2005kc} can be expressed. In our case it is given, for a second order phase space function $F[A_a,E^a,\delta\mathcal{A}_a^i,\delta\mathcal{E}^a_i]=f[A_a,E^a]+\delta f[A_a,E^a,\delta\mathcal{A}_a^i,\delta\mathcal{E}^a_i]+\delta^2 f[A_a,E^a,\delta\mathcal{A}_a^i,\delta\mathcal{E}^a_i]$, up two second order, as follows\footnote{See equation \eqref{eq:definition kappa tilde} for the definition of $\tilde{\kappa}$.} \cite{Fahn:2022zql}
\begin{align}\label{eq:linearised observable map}
    \begin{split}
    \mathcal{O}_{F,\{\mathcal{G}^I\}}\coloneqq&[\exp{(\Tilde{\kappa}\int\limits_{\mathbb{R}^3} d^3y \beta^I(\vec{y})\{C_I^\prime(\vec{y}),.\})}F]\Big|_{\beta^I=\mathcal{G}^I}\\
    =&\mathcal{O}_{F,\{\mathcal{G}^{\textrm{U(1)}}\}}+\kappa\int\limits_{\mathbb{R}^3} d^3y \delta\mathcal{G}^I(\vec{y})\left(\{\delta C_I^\prime(\vec{y}),\delta f\}+\{\delta C_I^\prime(\vec{y}),\delta^2 f\}+\{\delta^2 C_I^\prime(\vec{y}),\delta f\}\right)\\
    &+\frac{\kappa^2}{2}\int\limits_{\mathbb{R}^3} d^3 y\int\limits_{\mathbb{R}^3} d^3 z \delta \mathcal{G}^I(\vec{y})\delta \mathcal{G}^J(\vec{z})\left(\{\delta C_J^\prime(\vec{z}),\{\delta^2 C_I^\prime(\vec{y}),\delta f\}\}+\{\delta^2 C_J^\prime(\vec{z}),\{\delta C_I^\prime(\vec{y}),\delta^2 f\}\}\right)\\
    &+O(\delta^3,\kappa^3).
    \end{split}
\end{align}
Likewise for the dual observable map \eqref{eq:linearised dual observable map}, the electromagnetic part, denoted by $\mathcal{O}_{F,\{\mathcal{G}^{\textrm{U(1)}}\}}$, cannot be expanded in $\kappa$, since in the post-Minkowskian approximation the photon sector remains unperturbed. Therefore, the complete infinite power series must be applied. Similar to the dual observable map, we will apply the observable map only to the electromagnetic elementary phase space variables, for which the series terminates after linear order.
~\\
~\\
The observable map \eqref{eq:linearised observable map} maps a function F to $\mathcal{O}_{F,\{\mathcal{G}^I\}}$ which, in general weakly, Poisson commutes with all constraints of \eqref{eq:FinalSet}. $\mathcal{O}_{F,\{\mathcal{G}^I\}}$ gives the value of $F$ at those values where the clocks $T^I$ take the values $\tau^I$. Since the constraints and clocks of \eqref{eq:FinalSet} consist of eight canonically conjugate pairs, with strongly vanishing Poisson brackets for all other combinations, in our case, $\mathcal{O}_{F,\{\mathcal{G}^I\}}$ strongly commutes with all constraints, in the order in perturbation theory we are interested in. For the same reason this is also the case for the above defined dual observable map \eqref{eq:linearised dual observable map}, as well as the vacuum dual observable map \eqref{eq: vacuum dual map}.
~\\
~\\
We will start the construction with the electromagnetic sector by applying first the dual observable map \eqref{eq:linearised dual observable map} and then the observable map \eqref{eq:linearised observable map} to $A_a,-E^a$. This leads to the following Dirac observables up to second order
\begin{flalign}
    (A_a)^{\textrm{phys}}(\vec{x},t)\coloneqq &\mathcal{O}_{\mathcal{O}^{\textrm{dual}}_{A_a,\{C_I^\prime\}},\{\mathcal{G}^I\}}(\vec{x},t)=\mathcal{O}_{A_a,\{\mathcal{G}^I\}}(\vec{x},t)\\
        \nonumber
        =&P^b_a A_b (\vec{x},t)+\kappa^2P_{ca}\left(\left(\delta T-\tau\right) E^c\right)(\vec{x},t)+\kappa^2P^c_a\left(\left(\delta T^g-\sigma^g\right) A_{[g,c]}\right)(\vec{x},t)\\ \nonumber
        &+\mathcal{O}^{(2)}_{A_a,\{\mathcal{G}^I\}}(\vec{x},t),\\
        (-E^a)^{\textrm{phys}}(\vec{x},t)\coloneqq &\mathcal{O}_{\mathcal{O}^{\textrm{dual}}_{-E^a,\{C_I^\prime\}},\{\mathcal{G}^I\}}(\vec{x},t)=P_b^a\mathcal{O}_{-E^b,\{\mathcal{G}^I\}}(\vec{x},t)\\\nonumber
        =& -P_b^aE^b-\frac{\kappa^2}{2}P^{a[c}\partial^{b]}\left(\left(\delta T-\tau\right) A_{[c,b]}\right)(\vec{x},t)-\kappa^2 P^a_{[g}\partial_{c]}\left(\left(\delta T^g-\sigma^g\right) E^c\right)(\vec{x},t)\\ \nonumber
        &+P_b^a\mathcal{O}^{(2)}_{-E^b,\{\mathcal{G}^I\}}(\vec{x},t),
\end{flalign}
where we do not explicitly wrote out the second order since we do not need it directly in the following\footnote{Note that the geometrical clocks have the order $O(\frac{1}{\kappa},\delta)$ and thus terms like $\kappa^2\delta T^I$ are of the order $O(\kappa,\delta)$.}. These variables  by construction Poisson commute with all constraints and clocks given in \eqref{eq:FinalSet}, in the perturbation theory order we are interested in. More details on the calculation can be found in appendix \ref{app: Dirac observables: electromagnetic sector}. We obtain the transverse electromagnetic degrees of freedom, along with additional terms linear in the geometrical clocks, stemming from the electromagnetic contributions to the Hamiltonian and diffeomorphism constraints. If we compute the spatial divergence of the photon field Dirac observables, we obtain the following result
\begin{align}\label{physical electromagnetic variables only transverse degrees of freedom}
    \begin{split}
    \partial^a(A_a)^{\textrm{phys}}(\vec{x},t)=&\partial^a\mathcal{O}^{(2)}_{A_a,\{\mathcal{G}^I\}}(\vec{x},t)\\
    \partial_a (-E^a)^{\textrm{phys}}(\vec{x},t)=&0
    \end{split}
\end{align}
since $\partial_aP^a_b=0=\partial^aP_a^b$ for all $b$. Thus, as expected in the linear phase space, the photon field Dirac observables do not involve longitudinal components. Note that if we would just have applied one of the dual observable map and observable map we would not get these transverse degrees of freedom. As for example the electric field is not altered by the observable map, only with also the application of the dual observable map it is projected on the transverse subspace.
~\\
~\\
Next we come to the construction of the geometric Dirac observables. Again we will first apply the dual map \eqref{eq:linearised dual observable map} and then the observable map \eqref{eq:linearised observable map}. Before we begin, let us once again consider the set of Dirac observables we are looking for: they should Poisson commute with all clocks and constraints specified in \eqref{eq:FinalSet}, where we have defined the geometrical clocks such that any contribution beyond linear order vanishes. We transformed the second order constraints $\delta^2 C_I$ to a weakly equivalent version $\delta^2 C_I^\prime$ which Poisson commute with all clocks. If we now take a closer look at the dual observable map \eqref{eq:linearised dual observable map}, constructed from \eqref{eq:FinalSet}, applied onto a first order function $F[\delta\mathcal{A}_a^i,\delta\mathcal{E}^a_i]=\delta f[\delta\mathcal{A}_a^i,\delta\mathcal{E}^a_i]$, it is given by
\begin{align}
    \begin{split}
    \mathcal{O}^{\textrm{dual}}_{F,\{C_I^\prime\}}\coloneqq&
    -\kappa\int\limits_{\mathbb{R}^3} d^3y \delta C_I^\prime(\vec{y})\{\delta \mathcal{G}^I(\vec{y}),\delta f\}-\kappa\int\limits_{\mathbb{R}^3} d^3y \delta^2 C_I^\prime(\vec{y})\{\delta \mathcal{G}^I(\vec{y}),\delta f\}\\
    &+O(\delta^3,\kappa^3)
    \end{split}
\end{align}
Note that since $\{\mathcal{G}^{\textrm{U(1)}}, F[\delta\mathcal{A}_a^i,\delta\mathcal{E}^a_i]\}=0\:\:\forall F$, the U(1)-Gauß constraint is not involved. On the other hand all $\{\delta\mathcal{G}^I,F[\delta\mathcal{A}_a^i,\delta\mathcal{E}^a_i]\}$ terms are always of order $O(\kappa^{-2}, \delta^0)$ and therefore have no geometric degrees of freedom. Thus the only phase space degrees of freedom are in $\delta^2 C_I^\prime$, which by construction Poisson commutes with all clocks. Therefore, we are adding additional contributions which are already invariant under the action of the clocks. Since our goal is to construct variables from $\delta\mathcal{A}_a^i,\:\delta\mathcal{E}_i^a$ which commute with the clocks in \eqref{eq:FinalSet} for purpose it is enough to only apply the first order of the dual observable map given by
\begin{align}\label{eq: linear order of dual observable map}
    \begin{split}
    \mathcal{O}^{\textrm{dual}(1)}_{F,\{C_I^\prime\}}\coloneqq&
    -\kappa\int\limits_{\mathbb{R}^3} d^3y \delta C_I^\prime(\vec{y})\{\delta \mathcal{G}^I(\vec{y}),\delta f\}+O(\delta^2,\kappa^2),
    \end{split}
\end{align}
onto $\delta\mathcal{A}_i^a,\:\delta\mathcal{E}^a_i$. The second order just adds further terms which Poisson commute with all clocks, but also adding more complexity to the constructed variables. Note that by that the constructed Dirac observables are not weakly equivalent to $\delta\mathcal{A}_i^a,\:\delta\mathcal{E}^a_i$. On the other hand if one includes this second order the Poisson algebra of the resulting Dirac observables gets way more complicated and by that also the quantisation will be way more complicated. More details can be found in appendix \ref{app: More details on the inclusion of the second order of the dual map into the construction}. Note that this discussion can be made analogously for the vacuum dual observable map \eqref{eq: vacuum dual map}.
~\\
~\\
Next we construct Dirac observables for the geometric sector. For later convenience (see section \ref{sec:AlgebraDiracObs}) we will apply only the vacuum dual observable map \eqref{eq: vacuum dual map first order} onto $\delta\mathcal{A}_a^i,\:\delta\mathcal{E}_i^a$. The resulting contribution will Poisson commute by construction with the geometrical clocks, since the geometrical clocks consist of geometric degrees of freedom only. On the other hand the resulting Dirac observables will have only geometric degrees of freedom and because of that will trivially Poisson commute with the U(1)-Gauß clock, since it has only electromagnetic ones. Thus the constructed Dirac observables will Poisson commute with all clocks and since we apply the observable map also with all constraints of \eqref{eq:FinalSet}. These Dirac observables read as follows
\begin{align}
    \label{reduced linearised physical Ashtekar connection}
    \left(\delta\mathcal{A}_a^i\right)^{\textrm{phys}}(\vec{x},t)= \mathcal{O}_{\mathcal{O}^{\textrm{dual, vac}(1)}_{\delta\mathcal{A}^i_a,\{C_I^\prime\}},\{\mathcal{G}^I\}}(\vec{x},t)=&P^{id}_{al}\delta\mathcal{A}_d^l(\vec{x},t)+\mathcal{O}^{(2)}_{\mathcal{O}^{\textrm{dual, vac}(1)}_{\delta\mathcal{A}^i_a,\{C_I^\prime\}},\{\mathcal{G}^I\}}(\vec{x},t) ,\\
    \label{reduced physical linearised densitised triad}
    \begin{split}(\delta\mathcal{E}_i^a)^{\textrm{phys}}(\vec{x},t)= \mathcal{O}^{\textrm{vac}}_{\mathcal{O}^{\textrm{dual, vac}(1)}_{\delta\mathcal{E}_i^a,\{C_I^\prime\}},\{\mathcal{G}^I\}}(\vec{x},t)=&P_{id}^{al}\delta\mathcal{E}^d_l(\vec{x},t)+\beta\epsilon_i^{\:\:lk}\delta_l^d\delta_k^a\partial_d\tau(\vec{x},t)\\
    &+\delta_i^b\delta^a_{[b}\partial_{c]}\sigma^c(\vec{x},t)+\mathcal{O}^{(2)}_{\mathcal{O}^{\textrm{dual, vac}(1)}_{\delta\mathcal{E}_i^a,\{C_I^\prime\}},\{\mathcal{G}^I\}}(\vec{x},t).
    \end{split}
\end{align}
Again we did not write down the second order explicitly and defined the projector onto the symmetric transverse traceless subspace\footnote{Here, $\ast$ denotes convolution, $G^\Delta$ the Green's function of $-\Delta$ (see equation \eqref{Greens function}) and $G^{\Delta\Delta}$ the Green's function of the squared Laplacian, see equation \eqref{eq: greens fct Laplacian squared}.}
\begin{align}\label{eq: STT projector position space}
    \begin{split}
    P^{ib}_{aj} X=&\frac{1}{2}\left(P^b_a P^i_j+P_{aj}P^{ib}-P_a^i P^b_j\right)X(\vec{x},t)\\
    =&\frac{1}{2}\Big(\left[\delta_{aj}\delta^{bi}+\delta_a^b\delta_j^i-\delta_a^i\delta_j^b\right]X(\vec{x},t)+\partial_a\partial^i\partial_j\partial^b\left(X\ast G^{\Delta\Delta}\right)(\vec{x},t)\\
    &+\left[\delta_{aj}\partial^i\partial^b+\delta^{ib}\partial_a\partial_j+\delta_a^b\partial^i\partial_j+\delta^i_j\partial_a\partial^b-\delta_a^i\partial_j\partial^b-\delta_j^b\partial_a\partial^i\right]\left(X\ast G^{\Delta}\right)(\vec{x},t)\Big).
    \end{split}
\end{align}
More details on the derivation can be found in \ref{app: Dirac observables: vacuum dual observable map and observable map}. As expected, both $(\delta\mathcal{A}_a^i)^{\textrm{phys}},\:(\delta\mathcal{E}_i^a)^{\textrm{phys}}$ consist of geometric degrees of freedom only. In the linear phase space $(\delta\mathcal{A}^i_a)^{\textrm{phys}}$ is given by the symmetric transverse traceless part of the Ashtekar-Barbero connection, while $(\delta\mathcal{E}^a_i)^{\textrm{phys}}$ is given by the symmetric transverse traceless part of the densitised triad. Note that, similar to the electromagnetic sector, we only get these symmetric transverse traceless degrees of freedom if we apply both, the dual observable map and the observable map.
~\\
~\\
Before continuing, we also analyse the Dirac observables that result from applying the full dual observable map, rather than its vacuum version. Thus we first apply the first order dual observable map \eqref{eq: linear order of dual observable map} and then the observable map \eqref{eq:linearised observable map} onto $\delta\mathcal{A}_a^i,\:\delta\mathcal{E}^a_i$, which leads to
\begin{align}
    \label{eq: primed physical Ashtekar connection}
    \begin{split}
        (\delta\mathcal{A}^i_a)^{\textrm{phys}\prime}(\vec{x},t)\coloneqq &\mathcal{O}_{\mathcal{O}^{\textrm{dual}(1)}_{\delta\mathcal{A}_a^i,\{C_I^\prime\}},\{\mathcal{G}^I\}}\\
        =& P_{al}^{id}\delta\mathcal{A}^l_a + \frac{1}{2}\epsilon_a^{\:\:cb}\delta_c^i\partial_b\left(T^{00}\left((-E^b)^{\textrm{phys}},(A_a)^{\textrm{phys}}\right)\ast G^\Delta\right)(\vec{x},t)\\
        &-\beta\left(\delta_a ^g\delta_b^i\partial^b-\frac{1}{2}\delta_a^i\partial^g\right)\left(T_{0g}\left((-E^b)^{\textrm{phys}},(A_a)^{\textrm{phys}}\right)\ast G^\Delta\right)(\vec{x},t)\\
        &-\frac{\beta}{2}\delta^i_b\partial^g\partial^b\partial_a\left(T_{0g}\left((-E^b)^{\textrm{phys}},(A_a)^{\textrm{phys}}\right)\ast G^{\Delta\Delta}\right)(\vec{x},t)\\
        &
        +\mathcal{O}^{(2)}_{\mathcal{O}^{\textrm{dual}(1)}_{\delta\mathcal{A}_a^i,\{C_I^\prime\}},\{\mathcal{G}^I\}}(\vec{x},t),
    \end{split}\\
    \label{eq: primed physical triad}
    \begin{split}
        (\delta\mathcal{E}_i^a)^{\textrm{phys}\prime}(\vec{x},t)\coloneqq &\mathcal{O}_{\mathcal{O}^{\textrm{dual}(1)}_{\delta\mathcal{E}_a^i,\{C_I^\prime\}},\{\mathcal{G}^I\}}\\
        =& P^{aj}_{ib}\delta\mathcal{E}_j^b+\delta^a_i\left(T^{00}\left((-E^b)^{\textrm{phys}},(A_a)^{\textrm{phys}}\right)\ast G^\Delta\right)(\vec{x},t)\\
        &+\beta\delta^{gj}\delta_k^a\epsilon_{ji}^{\:\:\:\:k}\left(T_{0g}\left((-E^b)^{\textrm{phys}},(A_a)^{\textrm{phys}}\right)\ast G^\Delta\right)(\vec{x},t)\\
        &+\beta\epsilon_i^{\:\:lk}\delta_l^d\delta_k^a\partial_d\tau(\vec{x},t)+\delta_i^b\delta^a_{[b}\partial_{c]}\sigma^c(\vec{x},t)\\
        &+\mathcal{O}^{(2)}_{\mathcal{O}^{\textrm{dual}(1)}_{\delta\mathcal{E}^a_i,\{C_I^\prime\}},\{\mathcal{G}^I\}}(\vec{x},t).
    \end{split}
\end{align}
Again, we did not write down the second order explicitly. Both $(\delta\mathcal{A}^i_a)^{\textrm{phys}\prime}$ and $(\delta\mathcal{E}_i^a)^{\textrm{phys}\prime}$ by construction Poisson commute with all clocks and constraints \eqref{eq:FinalSet} and are thus a set of Dirac observables that we are aiming at constructing. Again in the linear phase space, $(\delta\mathcal{A}^i_a)^{\textrm{phys}\prime}$ is given by the symmetric transverse traceless part of $\delta\mathcal{A}_a^i$, and $(\delta\mathcal{E}^a_i)^{\textrm{phys}\prime}$ is given by the symmetric transverse traceless part of $\delta\mathcal{E}_i^a$. $(\delta\mathcal{E}^a_i)^{\textrm{phys}\prime}$ also depends on the gauge parameters $\tau,\:\sigma^c$, if we insert the choice made in section \ref{sec: Coordinate gauge fixing}, $\tau=t,\:\sigma^c=x^c$, we get that $\partial_d\tau=0$ and $\delta_i^b\delta^a_{[b}\partial_{c]}x^c=2\delta_i^a$. Furthermore for both $(\delta\mathcal{A}_a^i)^{\textrm{phys}\prime}$ and $(\delta\mathcal{E}^a_i)^{\textrm{phys}\prime}$ contributions linear in the energy density and electromagnetic pointing vector depending on the Dirac observables $(A_a)^{\textrm{phys}}(-E^a)^{\textrm{phys}}$ occur. These contributions come from the application of the dual observable map and the electromagnetic part of the Hamilton and diffeomorphism constraint. For further details on the derivation see \ref{app: Dirac observables: (full) dual observable map and observable map}.
~\\
~\\
Overall we have constructed two sets of Dirac observables
\begin{align}\label{eq: sets of Dirac observables}
    \begin{split}
    S^\prime\coloneqq&\left((\delta\mathcal{A}^i_a)^{\textrm{phys}\prime},\:(\delta\mathcal{E}_i^a)^{\textrm{phys}\prime},\:(A_a)^{\textrm{phys}},\:(-E^a)^{\textrm{phys}}\right),\\
    S\coloneqq &\left((\delta\mathcal{A}^i_a)^{\textrm{phys}},\:(\delta\mathcal{E}_i^a)^{\textrm{phys}},\:(A_a)^{\textrm{phys}},\:(-E^a)^{\textrm{phys}}\right),
    \end{split}
\end{align}
which also commute with all clocks, for \eqref{eq:FinalSet}. Note that $(A_a)^{\textrm{phys}}$ and $\:(-E^a)^{\textrm{phys}}$ are the same for both sets.
~\\
~\\
Before we continue with the Poisson algebras of the individual sets, we want to make a connection between the sets. Using equations \eqref{eq: primed physical Ashtekar connection}, \eqref{eq: primed physical triad}, \eqref{reduced linearised physical Ashtekar connection} and \eqref{reduced physical linearised densitised triad} we get
\begin{align}\label{eq: transformation between sets of physical variables}
    \begin{split}
        (\delta\mathcal{A}^i_a)^{\textrm{phys}}(\vec{x},t)
        =& (\delta\mathcal{A}^i_a)^{\textrm{phys}\prime}(\vec{x},t)
        -\frac{1}{2}\epsilon_a^{\:\:cb}\delta_c^i\partial_b\big(T^{00}((-E^b)^{\textrm{phys}},(A_a)^{\textrm{phys}})\ast G^\Delta\big)(\vec{x},t)\\
        &+\beta\left(\delta_a ^g\delta_b^i\partial^b-\frac{1}{2}\delta_a^i\partial^g\right)\big(T_{0g}((-E^b)^{\textrm{phys}},(A_a)^{\textrm{phys}})\ast G^\Delta\big)(\vec{x},t)\\
        &+\frac{\beta}{2}\delta^i_b\partial^g\partial^b\partial_a\big(T_{0g}((-E)^{\textrm{phys}},(A_a)^{\textrm{phys}})\ast G^{\Delta\Delta}\big)(\vec{x},t),\\
    (\delta\mathcal{E}_i^a)^{\textrm{phys}}(\vec{x},t)=&(\delta\mathcal{E}_i^a)^{\textrm{phys}\prime}(\vec{x},t)-\delta^a_i\big(T^{00}((-E^b)^{\textrm{phys}},(A_a)^{\textrm{phys}})\ast G^\Delta\big)(\vec{x},t)
        \\
        &-\beta\delta^{gj}\delta_k^a\epsilon_{ji}^{\:\:\:\:k}\big(T_{0g}((-E^b)^{\textrm{phys}},(A_a)^{\textrm{phys}})\ast G^\Delta\big)(\vec{x},t).
    \end{split}
\end{align}
We constructed the two sets of Dirac observables \eqref{eq: sets of Dirac observables} with respect to the same constraints and clocks given in \eqref{eq:FinalSet}. One difference which can be used to characterise the geometric Dirac observables is that we used for the set $S$ the vacuum dual observable map \eqref{eq: vacuum dual map first order}, while for the set $S^\prime$ we used the 'full' dual observable map \eqref{eq:linearised dual observable map}, on $\delta\mathcal{A}_i^a,\:\delta\mathcal{E}^a_i$. In this work, the final set of Dirac observables was obtained by combining the observable map and its dual, both of which include the chosen set of clocks. One question that we leave for future work is whether it is possible to find a set of different clocks such that the same set of Dirac observables obtained here can be constructed using only the observable map with this other set of clocks and without applying the dual map.
\subsection{Poisson algebra of Dirac observables}
\label{sec:AlgebraDiracObs}
The next step is to calculate the Poisson algebra of the Dirac observables. For this, we will make use of the properties of the dual and observable map\footnote{In general the equal signs in the identities are only weak, but since \eqref{eq:FinalSet} are a set of 8 conjugate pairs, with strongly vanishing Poisson brackets for all other combinations, for our case it is a strong equal sign. See appendix \ref{app: observable properties} for more details.}
\begin{align}\label{pb equals dirac bracket under observable map}
    \begin{split}
        \left\{\mathcal{O}^{\textrm{dual}}_{f,\{C_I^\prime\}}, \mathcal{O}^{\textrm{dual}}_{f^\prime,\{C_I^\prime\}}\right\}=&\mathcal{O}^{\textrm{dual}}_{\{f,f^\prime\}^\ast,\{C_I^\prime\}},\\
        \left\{\mathcal{O}_{f,\{\mathcal{G}^I\}}, \mathcal{O}_{f^\prime,\{\mathcal{G}^I\}}\right\}=&\mathcal{O}_{\{f,f^\prime\}^\ast,\{\mathcal{G}^I\}},\\
    \end{split}
\end{align}
with the Dirac bracket $\{f,f^\prime\}^\ast$  associated with the constructed set of constraints and clocks \eqref{eq:FinalSet}. Note that this identity also applies for the vacuum dual observable map applied to geometric degrees of freedom only, for more details see appendix \ref{app: observable properties}. Since we constructed the constraints and clocks as (strongly) canonically conjugate pairs, the Dirac bracket takes the following form \cite{Fahn:2022zql}
\begin{align}\label{dirac Bracket}
    \begin{split}
    \{f(\vec{x},t),f^\prime(\vec{y},t)\}^\ast =& \{f,f^\prime\}+\Tilde{\kappa}\int\limits_{\mathbb{R}^3} d^3 z\{f(\vec{x},t),\mathcal{G}^I(\vec{z},t)\}\{C_I^\prime(\vec{z},t),f^\prime(\vec{y},t)\}\\
    &-\Tilde{\kappa}\int\limits_{\mathbb{R}^3} d^3 z\{f(\vec{x},t),C_I^\prime(\vec{z},t)\}\{\mathcal{G}^I(\vec{z},t),f^\prime(\vec{y},t)\}+O(\delta^3,\kappa^3).
    \end{split}
\end{align}
This especially means that if both $f$ and $f^\prime$ Poisson commute with the clocks or the constraints, the Dirac bracket reduces to the Poisson bracket.
~\\
~\\
Using this we now want to compute the algebra of the Dirac observables for the two sets of different Dirac observables $S^\prime$ and $S$ given in \eqref{eq: sets of Dirac observables}. Since the Dirac observables $(A_a)^{\textrm{phys}},\:(-E^a)^{\textrm{phys}}$ are the same in both sets, we start with their Poisson bracket, which is given by
\begin{align}
    \label{physical photon variables mixed PB}
    \begin{split}
    \{(A_a)^{\textrm{phys}}(\vec{x},t),\left(-E^b\right)^{\textrm{phys}}(\vec{y},t)\}=& \{\mathcal{O}_{\mathcal{O}^{\textrm{dual}}_{A_a,\{C_I^\prime\}},\{\mathcal{G}^I\}}(\vec{x},t),\mathcal{O}_{\mathcal{O}^{\textrm{dual}}_{-E^b,\{C_I^\prime\}},\{\mathcal{G}^I\}}(\vec{y},t)\}\\
    =&\mathcal{O}_{\{A_a(\vec{x},t),-P_c^bE^c(\vec{y},t)\},\{\mathcal{G}^I\}}\\
    =&P_c^b\mathcal{O}_{\delta_a^c\delta(\vec{x}-\vec{y}),\{\mathcal{G}^I\}}\\
    =&P_a^b\delta(\vec{x}-\vec{y}),
    \end{split}\\
    \label{physical photon variables self PB}
    \begin{split}
    \{(A_a)^{\textrm{phys}}(\vec{x},t),(A_b)^{\textrm{phys}}(\vec{y},t)\}=& 0 =\{(-E^a)^{\textrm{phys}}(\vec{x},t),(-E^b)^{\textrm{phys}}(\vec{y},t)\}.
    \end{split}
\end{align}
 In the first line we used the definition of $(A_a)^{\textrm{phys}},(-E^a)^{\textrm{phys}}$. In the second we utilised identity \eqref{pb equals dirac bracket under observable map}, that we already know the result of the application of the dual observable map to $A_a,-E^a$ (see \eqref{eq: dual map on electromagnetic connection}, \eqref{eq: dual map on electric field}) and the fact that, by construction, after the application of the dual observable map any quantity commutes with all clocks. As a consequence, the Dirac bracket reduces to the Poisson bracket. In the third line we pulled the projector out of the Poisson bracket and used that $\{A_a(\vec{x},t),-E^c(\vec{y},t)\}=\delta_a^c\delta(\vec{x}-\vec{y})$. In the fourth line we used that $\mathcal{O}_{\delta_a^c\delta(\vec{x}-\vec{y}),\{\mathcal{G}^I\}}=\delta_a^c\delta(\vec{x}-\vec{y})$, leading to the projector \eqref{eq: projector transverse subspace} acting on the delta distribution. The Poisson brackets of $(A_a)^{\textrm{phys}},(-E^a)^{\textrm{phys}}$ with itself can be calculated analogously and are both vanishing since $A_a$, as well as $-E^a$ Poisson commute with itself.
~\\
~\\
Next we want to calculate the remaining Poisson brackets of the sets $S$ and $S^\prime$. They are more involved and the exact derivation is given in the appendix \ref{app: More Details on the derivation of the Poisson algebra of the Dirac observables for the different sets}, but the basic steps and methods are the same as above. We will start with the set $S$ for which the Poisson algebra is given by
\begin{align}
    \label{physical ashtakar variables mixed PB}
    \begin{split}
    \{(\delta\mathcal{A}^i_a)^{\textrm{phys}}(\vec{x},t),(\delta\mathcal{E}_j^b)^{\textrm{phys}}(\vec{y},t)\}
    =&\frac{\beta}{\kappa}P_{aj}^{ib}\delta(\vec{x}-\vec{y}),
    \end{split}\\
    \label{physical ashtakar variables self PB}
    \begin{split}
    \{(\delta\mathcal{A}^i_a)^{\textrm{phys}}(\vec{x},t),(\delta\mathcal{A}^j_b)^{\textrm{phys}}(\vec{y},t)\}=&0=\{(\delta\mathcal{E}_i^a)^{\textrm{phys}}(\vec{x},t),(\delta\mathcal{E}_j^b)^{\textrm{phys}}(\vec{y},t)\},
    \end{split}\\
    \label{PB of physical triad and photon variables}
    \begin{split}
    \{(\delta\mathcal{E}_i^a)^{\textrm{phys}}(\vec{x},t),(-E^b)^{\textrm{phys}}(\vec{y},t)\}
    =&0 = \{(\delta\mathcal{E}_i^a)^{\textrm{phys}}(\vec{x},t),(A_b)^{\textrm{phys}}(\vec{y},t)\},
    \end{split}\\
    \label{PB of physical connection and photon variables}
    \begin{split}
    \{(\delta\mathcal{A}^i_a)^{\textrm{phys}}(\vec{x},t),(-E^b)^{\textrm{phys}}(\vec{y},t)\}
    =&0 = \{(\delta\mathcal{A}^i_a)^{\textrm{phys}}(\vec{x},t),(A_b)^{\textrm{phys}}(\vec{y},t)\}.
    \end{split}
\end{align}
More details on the calculation can be found in appendix \ref{app: PB of ashtekar and photon physical variables}. For the Poisson bracket of $(\delta\mathcal{A}^i_a)^{\textrm{phys}}$ with $(\delta\mathcal{E}_i^a)^{\textrm{phys}}$ we get the projector \eqref{eq: STT projector position space} acting on the delta distribution. All other Poisson brackets vanish.
~\\
~\\
Next we calculate the Poisson algebra for the Dirac observables of the set $S^\prime$, given in \eqref{eq: sets of Dirac observables}:
\begin{align}
    \label{Pb of full physical linearised ashtekar variables}
    \{(\delta\mathcal{A}_a^i)^{\textrm{phys}\prime}(\vec{x},t),(\delta\mathcal{E}^b_j)^{\textrm{phys}\prime}(\vec{y},t)\}=&\frac{\beta}{\kappa}P^{ib}_{aj}\delta(\vec{x}-\vec{y})+\Phi^1\left((A_a)^{\textrm{phys}},(-E^a)^{\textrm{phys}}\right)(\vec{x},\vec{y},t),\\
    \{(\delta\mathcal{A}_a^i)^{\textrm{phys}\prime}(\vec{x},t),(\delta\mathcal{A}_a^i)^{\textrm{phys}\prime}(\vec{y},t)\}=&\Phi^2\left((A_a)^{\textrm{phys}},(-E^a)^{\textrm{phys}}\right)(\vec{x},\vec{y},t),\\
    \{(\delta\mathcal{E}^a_i)^{\textrm{phys}\prime}(\vec{x},t),(\delta\mathcal{E}^b_j)^{\textrm{phys}\prime}(\vec{y},t)\}=&\Phi^3\left((A_a)^{\textrm{phys}},(-E^a)^{\textrm{phys}}\right)(\vec{x},\vec{y},t),\\
    \{(\delta\mathcal{A}_a^i)^{\textrm{phys}\prime}(\vec{x},t),(A_f)^{\textrm{phys}}(\vec{y},t)\}=&\Phi^4\left((A_a)^{\textrm{phys}},(-E^a)^{\textrm{phys}}\right)(\vec{x},\vec{y},t),\\
    \{(\delta\mathcal{A}_a^i)^{\textrm{phys}\prime}(\vec{x},t),(-E^f)^{\textrm{phys}}(\vec{y},t)\}=&\Phi^5\left((A_a)^{\textrm{phys}},(-E^a)^{\textrm{phys}}\right)(\vec{x},\vec{y},t),\\
    \{(\delta\mathcal{E}^a_i)^{\textrm{phys}\prime}(\vec{x},t),(A_f)^{\textrm{phys}}(\vec{y},t)\}=&\Phi^6\left((A_a)^{\textrm{phys}},(-E^a)^{\textrm{phys}}\right)(\vec{x},\vec{y},t),\\
    \label{last mixing PB of full physical variables}
    \{(\delta\mathcal{E}^a_i)^{\textrm{phys}\prime}(\vec{x},t),(-E^f)^{\textrm{phys}}(\vec{y},t)\}=&\Phi^7\left((A_a)^{\textrm{phys}},(-E^a)^{\textrm{phys}}\right)(\vec{x},\vec{y},t).
\end{align}
We always obtain a non-trivial function $\Phi^i$ on the right hand side. The exact expressions and calculations can be found in appendix \ref{app: derivation of Poisson algebra of full physical Variables}. The $\Phi^i$'s depend on the physical electromagnetic variables and none of them vanishes.
\subsection{Final choice of Dirac observables, clocks and constraints}\label{sec: Final choice of Dirac observables, clocks and constraints}
In the above sections we constructed a set of constraints and clocks \eqref{eq:FinalSet}, which consists of eight canonically conjugate pairs. Then we constructed, for these constraints and clocks, two sets of Dirac observables \eqref{eq: sets of Dirac observables} which Poisson commute with all clocks. In the last section we computed the Poisson algebra of both sets. It is obvious to see that the Poisson algebra for the set $S$ (see \eqref{physical ashtakar variables mixed PB}-\eqref{PB of physical connection and photon variables}) of Dirac observables has a much simpler structure than the one for $S^\prime$ (see \eqref{Pb of full physical linearised ashtekar variables}-\eqref{last mixing PB of full physical variables}). Hence, as far as the canonical quantisation is concerned the set $S$ is strongly preferred because for $S$ we do only need to quantise standard canonical Poisson brackets, whereas this is no longer the case for the second set $S^\prime$. 
Thus, we will choose the Dirac observable of the set $S$ for our next steps in this work. Overall, we have constructed a set of Dirac observables, clocks, and constraints given by
\begin{align}\label{New phase space variables up to second order}
\centering
    \begin{split}
    (\delta\mathcal{A}^i_a)^{\textrm{phys}},(\delta\mathcal{E}^i_a)^{\textrm{phys}},
    (A_a)^{\textrm{phys}},
    (-E^a)^{\textrm{phys}},\\
    C^\prime,C_a^\prime,G_i^{\textrm{geo}\prime},G^{\textrm{U(1)}},\delta T,\delta T^a, \delta\Xi^i,T^{\textrm{U(1)}}.
    \end{split}
\end{align}
The set consists of twelve canonically conjugate pairs, i.e. all other combinations strongly vanish in the linearised phase space. The first line represents the physical degrees of freedom, while the second one represents the gauge degrees of freedom. Therefore, we have constructed a Kucha\v{r} decomposition of the kinematical phase space, which allows a clear separation into sets of physical and gauge degrees of freedom. As we study the linearised dynamics and each application of a Poisson bracket reduces the order in  $\kappa$ by one and the order in $\delta$ by two, we provide the quantities up to second order in perturbation theory.
~\\
~\\
Before we construct the physical Hamiltonian, we first discuss the differences of the Dirac observables $(\delta\mathcal{A}_a^i)^{\textrm{phys}},\:(\delta\mathcal{E}^a_i)^{\textrm{phys}}$ to the ones chosen in \cite{Fahn:2022zql}. In \cite{Fahn:2022zql} the geometric Dirac observables are chosen as
\begin{align}\label{eq: scalar field dirac observables}
    \begin{split}
        \left(\delta\mathcal{A}^i_a\right)^{\textrm{phys,scalar}}&\coloneqq P^{ic}_{ak}\mathcal{O}_{\delta\mathcal{A}^k_c, \{\delta\mathcal{G}^I\}}
        ,\\
    \left(\delta\mathcal{E}^b_j\right)^{\textrm{phys,scalar}}&\coloneqq P^{bl}_{jd}\mathcal{O}_{\delta\mathcal{E}_l^d, \{\delta\mathcal{G}^I\}}.
    \end{split}
\end{align}
Hence, in \cite{Fahn:2022zql} the projector is manually inserted to project the Dirac observables obtained from the observable map onto the symmetric transverse traceless degrees of freedom. Next,  we the application of the observable map \eqref{eq:linearised observable map} yields
\begin{align}
    \begin{split}
        \left(\delta\mathcal{A}^i_a\right)^{\textrm{phys,scalar}} = &P^{ic}_{ak}\delta\mathcal{A}_c^k+\kappa P^{ic}_{ak}\int\limits_{\mathbb{R}^3} d^3y\delta\mathcal{G}^I(\vec{y})\{\delta C_I^{\textrm{geo}}(\vec{y}),\delta\mathcal{A}_c^k\}+P^{ic}_{ak}\mathcal{O}^{(2)}_{\delta\mathcal{A}^k_c, \{\delta\mathcal{G}^I\}}\\
        =& P^{ic}_{ak}\delta\mathcal{A}_c^k+\kappa\int\limits_{\mathbb{R}^3} d^3y\delta\mathcal{G}^I(\vec{y})\{\delta C_I^{\textrm{geo}}(\vec{y}),P^{ic}_{ak}\delta\mathcal{A}_c^k\}+P^{ic}_{ak}\mathcal{O}^{(2)}_{\delta\mathcal{A}^k_c, \{\delta\mathcal{G}^I\}},
    \end{split}
\end{align}
and similar for $\left(\delta\mathcal{E}^b_j\right)^{\textrm{phys,scalar}}$. In the first line we used the definition of the observable map and that $\delta\mathcal{A}_a^i$ Poisson commutes with $\delta C^{\textrm{ph}},\delta C_a^{\textrm{ph}}$. In the second line we used that we can pull the projector $P^{ic}_{ak}$ into the Poisson bracket. In the sections above we showed that $P^{ic}_{ak}\delta \mathcal{A}_c^k,\:P^{bl}_{jd}\delta\mathcal{E}^d_l$ can be obtained in linear order from the combination of the observable map and its dual. Thus $P^{ic}_{ak}\delta \mathcal{A}_c^k,\:P^{bl}_{jd}\delta\mathcal{E}^d_l$ by construction Poisson commutes with all linearised geometric constraints and geometrical clocks, the latter are identical to the geometrical clocks used in \cite{Fahn:2022zql}. This especially means that $\{\delta C_I(\vec{y}),P_{ak}^{ic}\delta\mathcal{A}^k_c\}=0$ (and similar for $P^{bl}_{jd}\delta\mathcal{E}^d_l$). 
\begin{align}
    \begin{split}
        \left(\delta\mathcal{A}^i_a\right)^{\textrm{phys,scalar}}&= P^{ic}_{ak}\delta\mathcal{A}^k_c+P^{ic}_{ak}\mathcal{O}^{(2)}_{\delta\mathcal{A}^k_c, \{\delta\mathcal{G}^I\}},\\
    \left(\delta\mathcal{E}^b_j\right)^{\textrm{phys,scalar}}&\coloneqq P^{bl}_{jd}\delta\mathcal{E}_l^d+P^{bl}_{jd}\mathcal{O}^{(2)}_{\delta\mathcal{E}_l^d, \{\delta\mathcal{G}^I\}},
    \end{split}
\end{align}
where we used that the zeroth order observable map is just the identity and we do need the explicit form of the second order. If we compare with \eqref{reduced linearised physical Ashtekar connection} and \eqref{reduced physical linearised densitised triad} we see that the first order of the Dirac observables agree. The second order differs in general but is, as discussed, not explicitly needed in our work, as well as in \cite{Fahn:2022zql}. Furthermore, in \cite{Fahn:2022zql} they derive for their Dirac observables \eqref{eq: scalar field dirac observables} the same Poisson algebra as we derived in section \ref{sec:AlgebraDiracObs} for $(\delta\mathcal{A}_a^i)^{\textrm{phys}},\:(\delta\mathcal{E}^a_i)^{\textrm{phys}}$. Therefore, for our purpose in the order in perturbation theory we are interested in, we can say that our geometric Dirac observables are equivalent to the ones in \cite{Fahn:2022zql}.
\subsection{Physical Hamiltonian}
With all of this work done, we can now construct the physical Hamiltonian. Once the dynamical reference fields, that is clocks have been chosen and the temporal reference, here $\delta T$ has been identified, by construction the physical Hamiltonian generates the evolution of the all Dirac observables with respect the physical temporal coordinate $\tau$. For the model under consideration this can be made explicit by looking at the primed Hamiltonian constraint (see \eqref{eq: transformed second order constraints} and \eqref{eq: primed first order Hamilton constraint}):
\begin{align}\label{eq: physical Hamiltonian construction deparametrised system argument}
    \begin{split}
    C^\prime=& \delta C^{\textrm{geo}}+\delta C^{\textrm{ph}\prime}+\delta^2 C^{\textrm{ph}\prime}+\delta^2C^{\textrm{geo}\prime}+O(\delta^3,\kappa^3)\\
    =& P_{\delta T}+\kappa T^{00}\left(-P_b^a E^b,A_a\right)+\delta^2 C^{\textrm{ph}}\left(\mathcal{O}^{\textrm{dual}(1)}_{\delta\mathcal{A}_a^i,\{C_I^\prime\}}, \mathcal{O}^{\textrm{dual}(1)}_{\delta\mathcal{E}^a_i ,\{ C_I^\prime\}}, \mathcal{O}^{\textrm{dual}}_{A_a,\{ C_I^\prime\}},\mathcal{O}^{\textrm{dual}}_{-E^a,\{C_I^\prime\}}\right)\\
    &+\delta^2 C^{\textrm{geo}}\left(\mathcal{O}^{\textrm{dual}(1)}_{\delta\mathcal{A}_a^i,\{ C_I^\prime\}}, \mathcal{O}^{\textrm{dual}(1)}_{\delta\mathcal{E}^a_i ,\{ C_I^\prime\}}\right)+O(\delta^3,\kappa^3)\\
    =&P_{\delta T}+\kappa T^{00}\left(-P_b^a E^b,A_a\right)-\frac{\kappa^2}{2}W_{abc}^i\mathcal{O}^{\textrm{dual}(1)}_{\delta\mathcal{E}_i^c,\{C_I^\prime\}} T^{ab}\left(-P_b^a E^b,A_a\right)\\
    &+\delta^2 C^{\textrm{geo}}\left(\mathcal{O}^{\textrm{dual}(1)}_{\delta\mathcal{A}_a^i,\{ C_I^\prime\}}, \mathcal{O}^{\textrm{dual}(1)}_{\delta\mathcal{E}^a_i ,\{ C_I^\prime\}}\right)+O(\delta^3,\kappa^3)\\
    \coloneqq& P_{\delta T}+h+O(\delta^3,\kappa^3),
    \end{split}
\end{align}
here, $P_{\delta T}=\delta C^{\textrm{geo}}$ represents the canonical momentum for the Hamiltonian clock. As already discussed, $C^\prime$ is weakly equivalent to the original Hamiltonian constraint up to second order because we applied the dual observable map onto it up to second order. We apply it onto second order functions $\delta^2 C^{\textrm{geo}}$ and $\delta^2 C^{\textrm{ph}}$, consisting of combinations of $\delta\mathcal{A}_a^i,\:\delta\mathcal{E}_i^a$. Thus it is sufficient for our purpose, using \eqref{eq:dual and observable on phase space function}, to only apply the linear order of the dual observable map \eqref{eq: linear order of dual observable map} onto $\delta\mathcal{A}_a^i,\:\delta\mathcal{E}_i^a$. The second order would lead to third order terms. As we can see the Hamiltonian constraint deparametrises since the function $h$ does not depend on the temporal clock $\delta T$. As a consequence, the physical Hamiltonian will be independent of the physical time $\tau$. The final physical Hamiltonian is then obtained by applying the observable map onto $\frac{1}{\kappa}h$ and integrating over all spatial coordinates\footnote{Since this factor $\frac{1}{\kappa}$ is also in front of $C$, this does not alter our arguments regarding the order in perturbation theory, $C^\prime$ is still weakly canonically conjugate to $C$ up to second order.}
\begin{align}\label{eq: physical Hamiltonian for S prime}
    \begin{split}
    H^{\textrm{phys}\prime}\coloneqq& \int\limits_{\mathbb{R}^3} d^3 x\mathcal{O}_{\frac{1}{\kappa}h,\{\mathcal{G}^I\}}\\
    =&\int\limits_{\mathbb{R}^3} d^3 x\Big[T^{00}((-E^a)^{\textrm{phys}},(A_a)^{\textrm{phys}}\big))\\
    &-\frac{\kappa}{2}W_{abc}^i\big(\delta\mathcal{E}_i^c)^{\textrm{phys}\prime} T^{ab}((-E^a)^{\textrm{phys}},(A_a)^{\textrm{phys}}\big)\\
    &+\frac{\kappa^2}{2}\Big[\left(\delta_l^a\delta_m^b-\delta_l^b\delta^a_m\right)(\delta\mathcal{A}_a^l)^{\textrm{phys}\prime}(\delta\mathcal{A}_b^m)^{\textrm{phys}\prime}\\
    &-\frac{\beta^2+1}{\beta^2}\left(\delta_l^a\delta_m^b-\delta_l^b\delta^a_m\right)\left((\delta\mathcal{A}_a^l)^{\textrm{phys}\prime}-(\delta\Gamma_a^l)^{\textrm{phys}\prime}\right)\left((\delta\mathcal{A}_b^m)^{\textrm{phys}\prime}-(\delta\Gamma_b^m)^{\textrm{phys}\prime}\right)\\
    &+(\delta\mathcal{A}^i_{[b,a]})^{\textrm{phys}\prime}\epsilon_i^{\:\:jk}\left(\delta_j^a(\delta\mathcal{E}_k^b)^{\textrm{phys}\prime}+\delta_k^b(\delta\mathcal{E}_j^a)^{\textrm{phys}\prime}-\frac{1}{2}\delta_j^a\delta_k^b\delta_d^n(\delta\mathcal{E}_n^d)^{\textrm{phys}\prime}\right)\Big]\Big],
    \end{split}
\end{align}
where we defined
\begin{align}\label{LinearisedSpinconnectionintermsoftriads1 primed Dirac observables}
    \begin{split}
        (\delta\Gamma_a^i)^{\textrm{phys}\prime}\coloneqq-\frac{1}{2}\epsilon^{ijk}\delta_k^b\left(-\delta_a^l\delta_c^j\partial_b+\delta_c^j\delta_b^l\partial_a-\delta_{ac}\delta^l_j\partial_b+\delta_a^j\delta_c^l\partial_b\right)(\delta\mathcal{E}_l^c)^{\textrm{phys}\prime}.
    \end{split}
\end{align}
More details on the derivation can be found in appendix \ref{app: more details on the derivation of the physical Hamiltonian}. This form of the physical Hamiltonian generates the physical time evolution for the Dirac observables of $S^\prime$ with respect to the physical time $\tau$, the parameter the Hamiltonian clock $\delta T$ takes. The first term in \eqref{eq: physical Hamiltonian for S prime} is analogous to the Maxwell Hamiltonian on Minkowski spacetime in the uncoupled case, but here it depends on the Dirac observables $(A_a)^{\textrm{phys}},\:(-E^a)^{\textrm{phys}}$ of the coupled system. The second term describes a interaction Hamiltonian between the Dirac observables $(\delta\mathcal{E}^a_i)^{\textrm{phys}\prime}$ and $(A_a)^{\textrm{phys}},\:(-E^a)^{\textrm{phys}}$. The last three lines give the free evolution of the Dirac observables $(\delta\mathcal{A}_a^i)^{\textrm{phys}\prime},\:(\delta\mathcal{E}^a_i)^{\textrm{phys}\prime}$.
~\\
~\\
The reason why in this construction we obtain the physical Hamiltonian for the Dirac observables of $S^\prime$ lies in the fact that in the construction of $C^\prime$ the dual map, rather than its vacuum version, is applied. If then the observable map is applied on this expression, using identity \eqref{eq:dual and observable on phase space function}, one gets the Dirac observables of $S^\prime$. If we had instead applied the vacuum dual map in the construction of $C^\prime$, we would have ended up with the Dirac observables of $S$ in the resulting Hamiltonian. However, in that case $C^\prime$ would not be weakly equivalent to $C$ in the non-vacuum case, and therefore the Hamiltonian constructed in that way can only be interpreted as a physical Hamiltonian in the vacuum case.
~\\
~\\
As already discussed, the Poisson algebra of the Dirac observables of set $S^\prime$ (see \eqref{Pb of full physical linearised ashtekar variables}-\eqref{last mixing PB of full physical variables}) is quite complicated therefore we want to choose the other set $S$ of Dirac observables. Hence, we also need to 
construct the physical Hamiltonian generating the time evolution for the set $S$. With the result above, this can be easily done by using the relation between the different Dirac observables, already presented in \eqref{eq: transformation between sets of physical variables} and \eqref{eq: physical Hamiltonian for S prime}. Thus, the next step is to rewrite the physical Hamiltonian in terms of the Dirac observables $(\delta\mathcal{E}^a_i)^{\textrm{phys}}$ and $(\delta\mathcal{A}_a^i)^{\textrm{phys}}$, inserting \eqref{eq: transformation between sets of physical variables} into \eqref{eq: physical Hamiltonian for S prime}. 
As already discussed $\delta^2 C^{\textrm{ph}\prime}$ and $\delta^2 C^{\textrm{geo}}$ are second order functions, therefore, for our purpose it is enough to only insert the first order of 
$(\delta\mathcal{E}^a_i)^{\textrm{phys}}$ and $(\delta\mathcal{A}_a^i)^{\textrm{phys}}$, which read
\begin{align}
    (\delta\mathcal{E}^a_i)^{\textrm{phys}}=P^{al}_{id}\delta\mathcal{E}_l^d,\hspace{0.2in}(\delta\mathcal{A}_a^i)^{\textrm{phys}}=P_{al}^{id}\delta\mathcal{A}^l_d.
\end{align}
Using this the physical Hamiltonian which generates the physical time evolution for the Dirac observables of set $S$ with respect to the physical time $\tau$, the parameter the Hamiltonian clock $\delta T$ takes, can be written as
\begin{align}\label{eq: physical Hamiltonian for set S}
    H^{\textrm{phys}} \coloneqq & H^{\textrm{phys}}_{\textrm{Maxwell}}+H^{\textrm{phys}}_{\textrm{Grav}}+H^{\textrm{phys}}_I+\kappa U,
\end{align}
more details on the derivation can be found in appendix \ref{app: more details on the derivation of the physical Hamiltonian}. Here, we defined the free evolution of the Dirac observables $(A_a)^{\textrm{phys}},\:(-E^a)^{\textrm{phys}}$, which is given by
\begin{align}\label{eq: free evolution for electromagnetic Dirac observable}
    \begin{split}
        H^{\textrm{phys}}_{\textrm{Maxwell}} \coloneqq \int\limits_{\mathbb{R}^3} d^3 xT^{00}((-E^b)^{\textrm{phys}},(A_a)^{\textrm{phys}})(\vec{x},t)
    \end{split},
\end{align}
and the free evolution of the Dirac observables $(\delta\mathcal{A}^i_a)^{\textrm{phys}},\:(\delta\mathcal{E}_i^a)^{\textrm{phys}}$
\begin{align}\label{eq: geometric free evolution position space}
    \begin{split}
        H^{\textrm{phys}}_{\textrm{Grav}}\coloneqq \kappa\int\limits_{\mathbb{R}^3} d^3x\frac{1}{2\beta^2}\Big(P^{bd}_{nl}\delta\mathcal{A}_d^l P^{nf}_{bk}\delta\mathcal{A}_f^k-(\beta^2+1)P^{jl}_{bh}\delta\mathcal{E}_l^h\Delta P^{bk}_{jd}\delta\mathcal{E}_k^d+2\epsilon_c^{\:\:bg}P^{nf}_{bl}\delta\mathcal{A}^l_f\partial_g P_{nd}^{ck}\delta\mathcal{E}_k^d\Big),
    \end{split}
\end{align}
where, as discussed, their first order is sufficient. Next we have the true interaction Hamiltonian
\begin{align}\label{eq: true interaction Hamiltonian position space}
    H^{\textrm{phys}}_I\coloneqq &\kappa\int\limits_{\mathbb{R}^3} d^3x \delta_b^i\delta_{ac}T^{ab}\left((-E^a)^{\textrm{phys}},(A_a)^{\textrm{phys}}\right)P^{cj}_{id}\delta\mathcal{E}_j^d.
\end{align}
And lastly we have the gravitationally induced self-interaction
\begin{align}\label{eq: self interaction position space}
    \begin{split}
    U \coloneqq  \int\limits_{\mathbb{R}^3} d^3x\Big[&(T_{0a}\ast G^\Delta)\frac{\partial^a\partial^b}{4}(T_{0b}\ast G^\Delta)+T_{0a}\delta^{ab}(T_{0b}\ast G^\Delta)-\frac{3}{4}T^{00}(T^{00}\ast G^\Delta)-T^{00}\Big].
    \end{split}
\end{align}
All energy-momentum tensor components are to be understood as functions of the Dirac observables $(A_a)^{\textrm{phys}},\:(-E^a)^{\textrm{phys}}$. The entire expression $U$ scales with $\kappa$ and thus, as expected, the self-interaction disappears when there is no gravity ($\kappa=0$). From this, it is easy to see that the term is introduced exclusively by the coupling to linearised gravity. The same applies to the modifications in the algebra of Dirac observables\footnote{Because in the vacuum case the dual observable map is always its vacuum version and thus the geometric Dirac observables for the set $S^\prime$ are the same as the one for $S$.}.
~\\
~\\
Comparing the physical Hamiltonian for the Dirac observables of $S^\prime$ \eqref{eq: physical Hamiltonian for S prime} and the one for $S$ \eqref{eq: physical Hamiltonian for set S}, we observe that the self-interaction term is absent in \eqref{eq: physical Hamiltonian for S prime}. One might therefore argue that this physical Hamiltonian is considerably simpler and that the set $S^\prime$ should be preferred for further investigation. However, if we look at the Poisson algebra of $S^\prime$ (see section \ref{sec:AlgebraDiracObs}), we see that the self-interaction is instead encoded in the Poisson algebra rather than in the physical Hamiltonian itself. Thus, for $S^\prime$ one obtains a simple physical Hamiltonian but a complicated Poisson algebra, whereas for $S$ the situation is reversed. Since the quantisation of both the algebra and the physical Hamiltonian associated with $S$ is well understood, as will be discussed in the next section, we will use this set in our approach. The set $S^\prime$ would in principle also be viable. However, this would require to canonically quantise non-standard canonical Poisson brackets, which is in general an non-trivial task.
~\\
~\\
Before we continue we want to compare the classical physical Hamiltonian \eqref{eq: physical Hamiltonian for set S} with the results of \cite{Fahn:2022zql} and \cite{Lagouvardos:2020laf}. Even though in \cite{Fahn:2022zql} a scalar field is studied we get structurally similar terms in the physical Hamiltonian. They also get a true interaction \eqref{eq: true interaction Hamiltonian position space} and self interaction \eqref{eq: self interaction position space} contribution, depending on the energy density and the pointing vector of the matter system. These differ because of the different underlying matter system, but expressed in terms of the energy-momentum tensor, they are structurally the same. For the free evolution of the geometric Dirac observables \eqref{eq: geometric free evolution position space} we get the same expression, which is not surprising since, as discussed in section \ref{sec: Final choice of Dirac observables, clocks and constraints}, our geometric variables are equivalent to theirs in the perturbation order we are interested in. In \cite{Lagouvardos:2020laf} also linearised gravity coupled to Maxwell theory is studied. In comparison to our approach, ADM variables are used, as well as a gauge fixing. Even though they use a different formulation than in our approach, the resulting physical Hamiltonian is similar to ours. They get a free evolution for the gauge fixed ADM and electromagnetic variables, as well as a true and self interaction term, depending on the electromagnetic energy density and pointing vector in the chosen gauge fixing. The advantage of formulating the model in terms of Dirac observables is that it is valid in any chosen gauge, but the interpretation of the Dirac observables in relation to the original kinematical variables varies depending on the choice of gauge.
\section{Master equation for photons}
\label{sec:Meq}
In this section we will derive the master equation for photons with linearised gravity as the environment. As a first step we will reformulate the model in Fourier space. After that, we will apply a Fock quantisation of the model, which then allows to apply the projection operator formalism in the context of open quantum systems to obtain the final master equation. In the latter step we need to compute environmental correlation functions for which we  consider an initial thermal state for the gravitational environment.
~\\
~\\
Our first goal of this section is to calculate the dynamics of of the Dirac observables of the set $S$, given by the physical Hamiltonian \eqref{eq: physical Hamiltonian for set S}, up to second order. Thus we have to calculate the Poisson bracket of $\left(\delta\mathcal{A}^i_a\right)^{\textrm{phys}},\: \left(\delta\mathcal{E}_i^a\right)^{\textrm{phys}},\:\left(A_a\right)^{\textrm{phys}},\:\left(-E^a\right)^{\textrm{phys}}$ with the physical Hamiltonian \eqref{eq: physical Hamiltonian for set S}. As discussed, the physical Hamiltonian \eqref{eq: physical Hamiltonian for set S} is a second order function and the Poisson bracket reduces the order of $\kappa$ by one and the order of $\delta$ by two, which can be seen from the definition of the Poisson bracket in equation \eqref{physical ashtakar variables mixed PB}. Therefore it is sufficient for our purpose to work only with the first order of $\left(\delta\mathcal{A}^i_a\right)^{\textrm{phys}},\: \left(\delta\mathcal{E}_i^a\right)^{\textrm{phys}},\:\left(A_a\right)^{\textrm{phys}},\:\left(-E^a\right)^{\textrm{phys}}$. The Poisson bracket of the second order would lead, with the second order Hamiltonian, again to a second order contribution.
~\\
~\\
We start by transforming the first order of $\left(\delta\mathcal{A}^i_a\right)^{\textrm{phys}},\: \left(\delta\mathcal{E}_i^a\right)^{\textrm{phys}},\:\left(A_a\right)^{\textrm{phys}},\:\left(-E^a\right)^{\textrm{phys}}$ and also the physical Hamiltonian \eqref{eq: physical Hamiltonian for set S} into Fourier space. Given an orthonormal basis $\Big\{\frac{1}{|\vec{k}|}{k}_a, m_a(\vec{k}),\overline{m}_a(\vec{k})\Big\}$ in Fourier space, we have the following decomposition of the Dirac observables (see appendix \ref{app:Fourier space} for further details)
\begin{align}\label{Fourier expansion of physical Dirac observables}
    \begin{split}
    (A_a)^{\textrm{phys}(1)}(\vec{k},t)&\coloneqq \left( A^+\right)^{\textrm{phys}}(\vec{k},t)m_a(\vec{k})+\left(A^-\right)^{\textrm{phys}}(\vec{k},t)\overline{m}_a(\vec{k}),\\
    \left(- E^a\right)^{\textrm{phys}(1)}(\vec{k},t)&\coloneqq \left(-E^+\right)^{\textrm{phys}}(\vec{k},t)m^a(\vec{k})+\left(- E^-\right)^{\textrm{phys}}(\vec{k},t)\overline{m}^a(\vec{k}),\\
    P^{ib}_{aj}\delta\mathcal{A}_b^j(\vec{k},t)&\coloneqq \delta\mathcal{A}^+(\vec{k},t)m_a(\vec{k})m^i(\vec{k})+\delta \mathcal{A}^-(\vec{k},t)\overline{m}_a(\vec{k})\overline{m}^i(\vec{k}),\\
    P^{aj}_{ib}\delta\mathcal{E}_j^b(\vec{k},t)&\coloneqq \delta\mathcal{ E}^+(\vec{k},t)m^a(\vec{k})m_i(\vec{k})+\delta\mathcal{E}^-(\vec{k},t)\overline{m}^a(\vec{k})\overline{m}_i(\vec{k}).
    \end{split}
\end{align}
Here, $A^\pm$ and $\delta\mathcal{A}^\pm$ denote the different polarisations for the photon field and gravitational Dirac observables and likewise $E^\pm$ and $\delta\mathcal{E}^\pm$ for their conjugate momenta. Note that for the electromagnetic Dirac observables this decomposition is only possible since they only have transverse degrees of freedom in the linear phase space (see \eqref{physical electromagnetic variables only transverse degrees of freedom}). Given the set of independent physical observables in Fourier space, we can use these variables to express the physical Hamiltonian in Fourier space as well. We will do this for the different parts of the physical Hamiltonian \eqref{eq: physical Hamiltonian for set S} separately.
~\\
~\\
We will start with $H_{\textrm{Maxwell}}^{\textrm{phys}}$ (see \eqref{eq: free evolution for electromagnetic Dirac observable}), which is just the energy density of the photon field depending on the physical electromagnetic Dirac observables of the coupled system. It reads in Fourier space
\begin{align}\label{Maxwell Hamiltonian derived as physical Hamiltonian}
    \begin{split}
    H_{\textrm{Maxwell}}^{\textrm{phys}}\coloneqq\int\limits_{\mathbb{R}^3} d^3 k \frac{1}{2}\sum_{u\in\{\pm\}}\Big[&(E^u)^{\textrm{phys}}(\vec{k},t)(E^u)^{\textrm{phys}}(-\vec{k},t)\\
    &+||\vec{k}||^2\left(A^u\right)^{\textrm{phys}}(\vec{k},t)\left(A^u\right)^{\textrm{phys}}(-\vec{k},t)\Big].
    \end{split}
\end{align}
Note that only this contribution represents the free evolution of the photon field, as the self-interacting term \eqref{eq: self interaction position space}, as well as the true interaction term \eqref{eq: true interaction Hamiltonian position space} also non-trivially contributes to their equations of motion.
~\\
~\\
Next we come to the free evolution of the Dirac observables $(\delta\mathcal{A}_a^i)^{\textrm{phys}},\:(\delta\mathcal{E}^a_i)^{\textrm{phys}}$ given by $H_{\textrm{Grav}}^{\textrm{phys}}$. Using \eqref{Fourier expansion of physical Dirac observables} it reads
\begin{align}\label{eq: geometric free evolution fourier}
    \begin{split}
        H_{\textrm{Grav}}^{\textrm{phys}} = \kappa\int\limits_{\mathbb{R}^3} d^3k\Big[&\sum_{u\in \{\pm\}}\frac{1}{2\beta^2}\Big(\delta\mathcal{A}^u(\vec{k},t)\delta\mathcal{A}^u(-\vec{k},t)\\
    &+||\vec{k}||^2(\beta^2+1) \delta\mathcal{E}^u(\vec{k},t)\delta\mathcal{E}^u(-\vec{k},t)\\
    &+2 u ||\vec{k}|| \delta\mathcal{E}^u(\vec{k},t)\delta\mathcal{A}^u(-\vec{k},t)\Big)\Big].
    \end{split}
\end{align}
Again, in general the true interaction term \eqref{eq: true interaction Hamiltonian position space} has a non trivial effect on $(\delta\mathcal{A}_a^i)^{\textrm{phys}}$. Lastly we have the true interaction $H^{\textrm{phys}}_I$ \eqref{eq: true interaction Hamiltonian position space} and the self interaction $U$ \eqref{eq: self interaction position space}, which read using \eqref{Fourier expansion of physical Dirac observables}
\begin{align}
    \label{eq: true interaction Hamitlonian fourier space}
    \begin{split}
    H^{\textrm{phys}}_I =& \kappa\int\limits_{\mathbb{R}^3} d^3k T^{ab}(\vec{k},t)\delta_b^i\delta_{ac}P^{cj}_{id}\delta\mathcal{E}_j^d(-\vec{k},t),
    \end{split}\\
    \label{self interaction classical level}
    \begin{split}
    U =&\int\limits_{\mathbb{R}^3} d^3 k \Big[ T^{00}(\vec{k},t)-\frac{1}{\Omega_k^2}\left(\frac{3}{4}T^{00}(\vec{k},t)T^{00}(-\vec{k},t)-T_{0a}(\vec{k},t)\left(\delta^{ab}-\frac{\hat{k}^a\hat{k}^b}{4}\right)T_{0b}(-\vec{k},t)\right)\Big],
    \end{split}
\end{align}
where all energy-momentum tensor's components are to be considered as functions of
$(A_a)^{\textrm{phys}},\:(-E^a)^{\textrm{phys}}$.
~\\
~\\
Since the free evolution $H^{\textrm{phys}}_{\textrm{Maxwell}}$ is a harmonic oscillator for each polarisation and mode of $(A_a)^{\textrm{phys}},\:(-E^a)^{\textrm{phys}}$ we obtain, for each, the equation of motion of a harmonic oscillator with frequency $\Omega_{\vec{k}} = \sqrt{k^a k_a}$ (where $||\vec{k}||^2 \hat{=} -\Delta$). Therefore, the time evolution for each polarisation is given by a mode expansion
\begin{align}\label{mode expansion of the elctromagnetic polarisations}
    \begin{split}
    \left(A^\pm\right)^{\textrm{phys}}(\vec{k},t)&=a_k^\pm e^{-i\Omega_k t}+c_k^\pm e^{i\Omega_k t},\\
    \left(- E^\pm\right)^{\textrm{phys}}(\vec{k},t)&= \left(\dot{ A}^\pm\right)^{\textrm{phys}}(\vec{k},t)=(-i)\Omega_k\left(a_k^\pm e^{-i\Omega_k t}-c_k^\pm e^{i\Omega_k t}\right)
    \end{split}
\end{align}
with arbitrary complex functions $a_k^\pm$ and $c_k^\pm$ of the wave vector. For the benefit of the reader and to keep our notation simple, we will employ an abuse of notation by not explicitly writing the wave vector as a vector if it occurs as an index, such as for instance $\Omega_{{k}} \coloneqq \Omega_{\vec k} $.
~\\
By imposing that $\left(A^\pm\right)^{\textrm{phys}}$ and $\left(-E^\pm\right)^{\textrm{phys}}$ are real functions, we obtain the condition $c_k^\pm = \overline{a}_{-k}^\pm$, where $\overline{a}$ denotes complex conjugation. Utilising this, along with a rescaling $a_k^\pm \rightarrow \sqrt{2\Omega_k} a_k^\pm$, the polarisation functions now read
\begin{align}\label{electromagnetic polarisation functions 2}
    \begin{split}
    \left(A^\pm\right)^{\textrm{phys}}(\vec{k},t) &=\sqrt{\frac{1}{2\Omega_k}}\left(a_k^\pm e^{-i\Omega_k t}+\overline{a}_{-k}^\pm e^{i\Omega_k t}\right),\\
    \left(- E^\pm\right)^{\textrm{phys}}(\vec{k},t) &=(-i)\sqrt{\frac{\Omega_k}{2}} \left(a_k^\pm e^{-i\Omega_k t}-\overline{a}_{-k}^\pm e^{i\Omega_k t}\right).
    \end{split}
\end{align}
To ensure that the right hand side yields the correct Poisson algebra of $(A_a)^{\textrm{phys}}$ and $(-E^a)^{\textrm{phys}}$, we impose the following Poisson brackets for the functions $a_k^\pm$ and $(\overline{a}_k^\pm)$:
\begin{align}\label{electromagnetic Poisson algebra of ladder classical}
    \{a_k^u,\overline{a}_p^r\}=i\delta^{ur}\delta(\vec{k}-\vec{p}),\:\:\:\:\:\:\{a_k^u,a_p^r\}=0=\{\overline{a}_k^u,\overline{a}_p^r\}.
\end{align}
Using the relations $m_a(\vec{k})m^a(\vec{k})=0=\overline{m}_a(\vec{k})\overline{m}^a(\vec{k}),\:m_a(\vec{k})\overline{m}^a(\vec{k})=1,\: m_a(\vec{k})=\overline{m}^a(-\vec{k})$ (see appendix \ref{app:Fourier space}), it is now easy to prove that the choice \eqref{electromagnetic Poisson algebra of ladder classical} does indeed satisfy the standard canonical Poisson brackets of the Dirac observables of the photon field in \eqref{eq:elementrayPBelectro}.
~\\
~\\
The formulation for the geometrical degrees of freedom works similarly and can be found in~\cite{Fahn:2022zql}. Here, we can also apply a mode expansion to the geometrical Dirac observables, for each polarisation 
\begin{align}\label{mode expansion of the geometrical variables 2}
    \begin{split}
    \delta\mathcal{E}^u(\vec{k},t)&=\sqrt{\frac{1}{2\kappa\Omega_k}}\left(b_k^u e^{-i\Omega_k t}+\overline{b}_{-k}^u e^{i\Omega_k t}\right),\\
    \delta\mathcal{A}^u(\vec{k},t)&=\sqrt{\frac{\Omega_k}{2\kappa}}\left((i\beta\mp1) b_k^u e^{-i\Omega_k t}+(-i\beta\mp1) \overline{b}_{-k}^u e^{i\Omega_k t}\right),
    \end{split}
\end{align}
with the complex functions $b_k^\pm,\:\overline{b}_k^\pm$. Similar to the Maxwell case, the condition for their Poisson algebra has the following form: 
\begin{align}\label{geometric Poisson algebra of ladder classical}
    \{b_k^u,\overline{b}_p^r\}=i\delta^{ur}\delta(\vec{k}-\vec{p}),\:\:\:\:\:\:\{b_k^u,(b_p^r)\}=0=\{\overline{b}_k^u,\overline{b}_p^r\}.
\end{align}
It turns out that the geometrical free Hamiltonian is simpler and appears more similar to the electromagnetic case when formulated in terms of a new variable\footnote{In terms of $\{\delta\mathcal{E}^{\rm phys},\delta\mathcal{C}\}$ the last line in the Hamiltonian \eqref{eq: geometric free evolution fourier} vanishes, casting it into a form similar to the Maxwell Hamiltonian where $\delta\mathcal{E}^{\rm phys}$ takes over the role of $A^{\rm phys}$ and $\delta\mathcal{C}$ takes over the role of $E^{\rm phys}$.} $\delta \mathcal{C}_a^i$. Therefore, we define a new quantity depending on $(\delta\mathcal{A}_a^i)^{\textrm{phys}},\:(\delta\mathcal{E}^a_i)^{\textrm{phys}}$ \cite{Fahn:2022zql}
\begin{align}\label{new variable for ashtekar connection}
    \delta \mathcal{C}_a^i(k,t)=\left(\delta\mathcal{C}^+m_a m^i+\delta\mathcal{C}^-\overline{m}_a\overline{m}^i\right)(k,t),\:\:\:\:\:\delta\mathcal{C}^{\pm}(k,t)=-\frac{1}{\beta}\left(\delta\mathcal{A}^{\pm}\pm \Omega_k\delta\mathcal{E}^{\pm}\right)(k,t).
\end{align}
We now treat, without loss of generality, $\delta\mathcal{C}_a^i,\:(\delta\mathcal{E}^a_i)^{\textrm{phys}}$ as fundamental physical geometrical observables.
~\\
~\\
The final step is to express the Hamiltonian using the mode-expanded variables:
\begin{align}\label{physical Hamiltonian in terms of modes}
    \begin{split}
    H^{\textrm{phys}} &=\int\limits_{\mathbb{R}^3} d^3 k\Big( \frac{\Omega_k}{2}\sum_{u\in\{\pm\}}\left[\overline{b}_k^u b_k^u + b_k^u \overline{b}_k^u +\overline{a}_k^u a_k^u + a_k^u \overline{a}_k^u\right]\\
   &+ \sqrt{\frac{\kappa}{2\Omega_k}}\sum_{u\in\{\pm\}} (P^u(-\vec{k}))_{ab} T^{ab}(-\vec{k},t)\left(b_k^u e^{-i\Omega_k t}+ \overline{b}_{-k}^u e^{i\Omega_k t}\right)+\kappa U(\vec{k},t)\Big).
   \end{split}
\end{align}
Here, we defined the projections $(P^u(\vec{k}))^a_b=\overline{m}_b(u\vec{k})\overline{m}^a(u\vec{k})$, which projects onto the transverse traceless subspace (for more details see appendix \ref{app: projectors}). The first line gives the free Hamiltonian for \eqref{Fourier expansion of physical Dirac observables} without the coupling of gravity and electromagnetism. The second part is the real interaction term. The last part
is the gravitationally induced self-interaction of the photon field.
\subsection{Fock quantisation of the model}\label{sec:Fockquant}
Next we will apply a Fock quantisation to the reduced phase space so that the physical Hilbert space of the model will be the tensor product of the bosonic Fock spaces of the physical photon and gravitational wave degrees of freedom. As usual, we will quantise the Hamiltonian operator in perturbation theory and first define the free Hamiltonians of the photon and the gravitational wave sector and then quantise the interactions in terms of polynomials of the corresponding annihilation and creation operators. We apply a Fock quantisation to the Dirac observables we constructed with respect to the constraints and clocks in \eqref{eq:FinalSet}
\begin{align}
    \left((A_a)^{\textrm{phys}},(-E^a)^{\textrm{phys}}\right),\:\left(\delta\mathcal{C}_a^i,(\delta\mathcal{E}_i^a)^{\textrm{phys}}\right).
\end{align}
We denote operators and operator-valued distributions by a hat. The corresponding commutator algebra which is consistent with the classical symplectic structure in \eqref{eq:elementrayPBelectro},\eqref{eq:elementrayPBgeo} reads
\begin{align}\label{eq: commutator algebra}
    \begin{split}
        [\left(\widehat{ A}^\pm\right)^{\textrm{phys}}(\vec{x},t),\left(-\widehat{E}^\pm\right)^{\textrm{phys}}(\vec{y},t)]&=i\delta^3(\vec{x}-\vec{y})\mathbbm{1},\\
        [\widehat{\delta \mathcal{C}}^\pm(\vec{x},t),\widehat{\delta \mathcal{E}}^\pm(\vec{y},t)]&=i\frac{\beta}{\kappa}\delta^3(\vec{x}-\vec{y})\mathbbm{1},
    \end{split}
\end{align}
where all remaining commutators vanish. This means that for each polarisation field we get a bosonic algebra. Furthermore, we constructed, at the classical level, the Dirac observables such that Dirac observables in the Maxwell and gravitational sector mutually Poisson commute, which carries over to the corresponding commutators of the operator-valued distributions.
~\\
~\\
In terms of the mode expansion, the quantised Dirac observables read
\begin{align}\label{quantsied field in mode expansion}
    \begin{split}
     &\left(\widehat{A}_a\right)^{\textrm{phys}}(\vec{x},t) =\int\limits_{\mathbb{R}^3}\frac{d^3 k}{(2\pi)^\frac{3}{2}}\sqrt{\frac{1}{2\Omega_k}}\sum_{u\in\{\pm\}}\left(\overline{m}_a(-u\vec{k})\hat{a}_k^u e^{-i\Omega_k t+i\vec{k}\vec{x}}+\overline{m}_a(u\vec{k})(\hat{a}_{k}^u)^\dagger e^{i\Omega_k t-i\vec{k}\vec{x}}\right),\\
    &\left(-\widehat{ E}^a\right)^{\textrm{phys}}(\vec{x},t) =(-i)\int\limits_{\mathbb{R}^3} \frac{d^3 k}{(2\pi)^\frac{3}{2}}\sqrt{\frac{\Omega_k}{2}}\sum_{u\in\{\pm\}}\left( \overline{m}^a(-u\vec{k})\hat{a}_k^u e^{-i\Omega_k t+i\vec{k}\vec{x}}-\overline{m}^a(u\vec{k})(\hat{a}_{k}^u)^\dagger e^{i\Omega_k t-i\vec{k}\vec{x}}\right),\\
    &(\widehat{\delta\mathcal{E}}^a_i)^\textrm{phys}(\vec{x},t)=\int\limits_{\mathbb{R}^3} \frac{d^3 k}{(2\pi)^\frac{3}{2}} \sqrt{\frac{1}{2\kappa\Omega_k}}\sum_{u\in\{\pm\}}\left((P^u(-k))^a_i \hat{b}_k^u e^{-i\Omega_k t+i\vec{k}\vec{x}}+(P^r(k))^a_i(\hat{b}_{k}^u)^\dagger e^{i\Omega_k t-i\vec{k}\vec{x}}\right),\\
    &\widehat{\delta\mathcal{C}}^i_a(\vec{x},t)=(-i)\int\limits_{\mathbb{R}^3} \frac{d^3 k}{(2\pi)^\frac{3}{2}}  \sqrt{\frac{\Omega_k}{2\kappa}}\sum_{u\in\{\pm\}}\left((P^u(-k))^i_a \hat{b}_k^u e^{-i\Omega_k t+i\vec{k}\vec{x}}-(P^r(k))^i_a(\hat{b}_{k}^u)^\dagger e^{i\Omega_k t-i\vec{k}\vec{x}}\right).
    \end{split}
\end{align}
Here we introduced the set of annihilation and creation operators $\hat{a}_k^\pm,\: \left(\hat{a}_k^\pm\right)^\dagger,\:\hat{b}_k^\pm,\: \left(\hat{b}_k^\pm\right)^\dagger$ that satisfies the following algebra:
\begin{align}\label{commutation relation of smeared annihilation and creation operator}
    \begin{split}
        \left[\hat{a}^u_k,\left(\hat{a}^r_p\right)^\dagger\right]=\delta^{ur}\delta^3 (\vec{k}-\vec{p})\mathbbm{1},\:\left[\hat{b}^u_k,\left(\hat{b}^r_p\right)^\dagger\right]&=\delta^{ur}\delta^3 (\vec{k}-\vec{p})\mathbbm{1},
    \end{split}
\end{align}
where all other commutators between $\hat{a}_k^\pm,\: \left(\hat{a}_k^\pm\right)^\dagger,\:\hat{b}_k^\pm,\: \left(\hat{b}_k^\pm\right)^\dagger$ vanish. With this choice, \eqref{eq: commutator algebra} is also fulfilled. This means that we also have the distributional algebra character for the quantised fields. Thus the fields defined in \eqref{quantsied field in mode expansion} are also operator valued distributions.
~\\
~\\
With the use of this, we can construct the quantised free Hamiltonian. This is done by just replacing the functions $ b_k^u,\: \overline{b}_k^u,\: a_k^u,\:\overline{a}_k^u$ in the first line of \eqref{physical Hamiltonian in terms of modes} by their corresponding operator valued distributions $ \hat{b}_k^u,\: (\hat{b}_k^u)^\dagger,\: \hat{a}_k^u,\:(\hat{a}_k^u)^\dagger$. As it is usually done in quantum field theory we quantise using a normal ordering for the operator-valued distributions $ \hat{b}_k^u,\: (\hat{b}_k^u)^\dagger,\: \hat{a}_k^u,\:(\hat{a}_k^u)^\dagger$. Overall, the normal ordered free Hamiltonian operator on the bosonic Fock space reads
\begin{align}\label{normal ordered free Hamiltonian}
    \::\widehat{H}^{\textrm{free}}:
    \coloneqq\int\limits_{\mathbb{R}^3} d^3 k\: \Omega_k \left[(\hat{b}_k^+)^\dagger \hat{b}_k^+ + (\hat{b}_k^-)^\dagger \hat{b}_k^- +(\hat{a}_k^+)^\dagger \hat{a}_k^+ + (\hat{a}_k^-)^\dagger \hat{a}_k^-\right].
\end{align}
The next step is to Fock quantise the interaction and self-interaction parts of the classical physical Hamiltonian \eqref{physical Hamiltonian in terms of modes}. We do this in the same way as for the free Hamiltonian, by applying a normal ordering and inserting the operator valued distributions
\begin{align}\label{quantised Hamiltonian}
    \begin{split}
    :\widehat{H}:\:&\coloneqq\int\limits_{\mathbb{R}^3} d^3 k\Big( \Omega_k\left[(\hat{b}_k^+)^\dagger \hat{b}_k^+ + (\hat{b}_k^-)^\dagger \hat{b}_k^- +(\hat{a}_k^+)^\dagger \hat{a}_k^+ + (\hat{a}_k^-)^\dagger \hat{a}_k^-\right]\\
   &+ \sqrt{\frac{\kappa}{2\Omega_k}}\sum_{u\in\{\pm\}} (P^u(-\vec{k}))_{ab} :\hat{T}^{ab}:(-\vec{k},t)\left(\hat{b}_k^u e^{-i\Omega_k t}+ (\hat{b}_{-k}^u)^\dagger e^{i\Omega_k t }\right)+\kappa :\hat{U}:(\vec{k},t)\Big).
   \end{split}
\end{align}
In this work, we choose the normal ordering in such a way that if we have a product of two or more operator-valued distributions, we first evaluate the product and then apply the normal ordering. The same approach is taken in \cite{Fahn:2022zql}, while \cite{Lagouvardos:2020laf} takes the opposite approach. In \cite{Fahn:2024fgc}, the effects of different normal orderings for a scalar field coupled to linearised gravity are discussed and investigated. The components of the energy-momentum tensor are understood as polynomials of the annihilation and creation operators $\hat{a}_k,\:\hat{a}_k^\dagger$ on the physical Hilbert space, specifically they consist of combinations of two creation or annihilation operators for different modes.
~\\
~\\
To compare the results better with already existing models, we rewrite the Hamiltonian in the following form:
\begin{align}\label{quantised Hamiltonian 2}
    \begin{split}
    :\widehat{H}:\:&\coloneqq\int\limits_{\mathbb{R}^3} d^3 k\Big( \Omega_k\left[(\hat{b}_k^+)^\dagger \hat{b}_k^+ + (\hat{b}_k^-)^\dagger \hat{b}_k^- +(\hat{a}_k^+)^\dagger \hat{a}_k^+ + (\hat{a}_k^-)^\dagger \hat{a}_k^-\right]\\
   &+ \sqrt{\frac{\kappa}{2\Omega_k}}\sum_{u\in\{\pm\}} \left(\hat{b}_k^u \widehat{J}_u^\dagger(\vec{k},t) e^{-i\Omega_k t}+ (\hat{b}_{k}^u)^\dagger \widehat{J}_u(\vec{k},t) e^{i\Omega_k t}\right)+\kappa :\widehat{U}:(\vec{k},t)\Big)
   \end{split}
\end{align}
with the operators
\begin{align}\label{new operator for simpliefied Hamiltonian}
    \widehat{J}_r(\vec{k},t)=(P^r(\vec{k}))_{ab} :\widehat{T}^{ab}:(\vec{k},t).
\end{align}
In this way, we formulate the physical Hamiltonian operator in a similar way to \cite{Fahn:2022zql, Lagouvardos:2020laf}. In \cite{Fahn:2022zql} a scalar field is coupled to linearised gravity. The difference to the scalar field is encoded in the operators $\widehat{J}_r$ and the self-interaction $\widehat{U}$. Furthermore, the free Hamiltonian operator consists of a set of two ladder operators for the photon part, since the photon degrees of freedom have two polarisations. In \cite{Lagouvardos:2020laf} they study gravity in ADM variables coupled to Maxwell theory in a gauge fixing. Even though the basic formulations are different, their Hamiltonian is structurally the same and they define a similar operator $\widehat{J}_r$. This will be discussed in more detail when we study the exact expression of $\widehat{J}_r$.
~\\
~\\
As we employed canonical quantisation, which is based on the mode expansions coming from dynamics arising from the uncoupled evolution of gravity and photon field, the operators defined so far are effectively formulated in the interaction picture. This means that their time evolution is precisely given by the free Hamiltonian \eqref{normal ordered free Hamiltonian}, while the evolution of the states is given by the full Hamiltonian \eqref{quantised Hamiltonian 2}. The Hamiltonian in the Schrödinger picture is given by evaluating \eqref{quantised Hamiltonian 2} at $t=0$, yielding
\begin{align}\label{quantised Hamiltonian in Schrödinger picture}
    \begin{split}
    :\widehat{H}_{\textrm{Schrödinger}}:\:\coloneqq\int\limits_{\mathbb{R}^3} d^3 k\Big(& \Omega_k\left[(\hat{b}_k^+)^\dagger \hat{b}_k^+ + (\hat{b}_k^-)^\dagger \hat{b}_k^- +(\hat{a}_k^+)^\dagger \hat{a}_k^+ + (\hat{a}_k^-)^\dagger \hat{a}_k^-\right]\\
   &+ \sqrt{\frac{\kappa}{2\Omega_k}}\sum_{u\in\{\pm\}} \left(\hat{b}_k^u \hat{J}_u^\dagger(\vec{k},0)+ (\hat{b}_{k}^u)^\dagger \widehat{J}_u(\vec{k},0)\right)+:\widehat{U}:(\vec{k},0)\Big).
   \end{split}
\end{align}
Based on this, we discuss in the next subsection how a master equation that encodes the effective dynamics of the photon field in an environment of linearised gravity can be obtained.
\subsection{Derivation of the master equation for photons in the projection operator formalism}
\label{sec:DerivMeq}
In this section, we will derive the final master equation for Maxwell theory coupled to linearised gravity using the time convolutionless (TCL) master equation (for a derivation see e.g. \cite{Breuer:2007juk, Zwanzig:1960gvu, Nakajima:1958pnl, Giesel:2022pzh,shibata1977generalized,chaturvedi1979time,shibata1980expansion}) truncated at second order in the framework of the projection operator formalism. To interpret the coupled system as an open quantum system, we consider the physical sector of Maxwell theory as the system of interest with Hilbert space $\mathcal{H}_{\textrm{S}}$ and the physical sector of linearised gravity as the environment with Hilbert space $\mathcal{H}_{\textrm{E}}$. Since for different polarisations the set of annihilation and creation operators mutually commutes, the physical Hilbert space is a tensor product of the bosonic Fock Hilbert spaces for each polarisation of the electromagnetic and gravitational waves. Thus, $\mathcal{H}=\mathcal{H}_{\textrm{E}}\otimes\mathcal{H}_{\textrm{S}}$, which is one of the assumptions necessary for the application of the projection operator formalism.
With the open quantum system interpretation, we decompose the Hamiltonian operator \eqref{quantised Hamiltonian in Schrödinger picture} into three parts
\begin{align}\label{definition of Hamiltonians for open quantum systems}
    \begin{split}
    \widehat{H}_{\textrm{S}}&\coloneqq\int\limits_{\mathbb{R}^3} d^3 k\Omega_k\left((\hat{a}_k^+)^\dagger \hat{a}_k^+ + (\hat{a}_k^-)^\dagger \hat{a}_k^- \right),\\
    \widehat{H}_{\textrm{E}}&\coloneqq\int\limits_{\mathbb{R}^3} d^3 k\Omega_k\left((\hat{b}_k^+)^\dagger \hat{b}_k^+ + (\hat{b}_k^-)^\dagger \hat{b}_k^- \right),\\
    \widehat{H}_I &\coloneqq \sqrt{\kappa} \widehat{H}^{\textrm{TI}}+\kappa\widehat{U},
    \end{split}
\end{align}
where the coupling parameter $\sqrt{\kappa}$ of the interaction Hamiltonian \eqref{quantised Hamiltonian in Schrödinger picture} has been factored out. From now on, we will drop the notation :.: for the normal ordering by an abuse of notation. The true interaction, in contrast to to the self-interactions of the photon field, is encoded in the contribution $\widehat{H}^{\textrm{TI}}$ given by the second term in \eqref{quantised Hamiltonian in Schrödinger picture}.
~\\
~\\
For the further investigation, we switch to the interaction picture, which we denote by a tilde, and we assume that the environmental density matrix $\widehat{\rho}_{\textrm{E}}$ is given by a thermal state, which is by definition stationary in time:
\begin{align}\label{Gibbs state 2}
    \widehat{\tilde{\rho}}_{\textrm{E}}=\widehat{\rho}_{\textrm{E}} = \frac{e^{-\beta \widehat{H}_{\textrm{E}}}}{\textrm{tr}_{\textrm{E}} \left(e^{-\beta \widehat{H}_{\textrm{E}}}\right)}\coloneqq \frac{1}{Z_{\textrm{E}}}\exp{\left(-\beta\int\limits_{\mathbb{R}^3} d^3 k \Omega_k\sum_{u\in\{\pm\}}\hat{n}_k^u\right)},
\end{align}
where we defined the partition sum $Z_{\textrm{E}}=\textrm{tr}_{\textrm{E}}\left(\exp{\left(-\beta\int\limits_{\mathbb{R}^3} d^3 k \Omega_k\sum_{u\in\{\pm\}}\hat{n}_k^u\right)}\right)$, the occupation number operator valued distribution $\hat{n}_k^u=\left(\hat{b}_k^u\right)^\dagger\hat{b}_k^u$ and $\beta=\frac{1}{k_B T}$ with Boltzmann constant $k_B$ and temperature parameter $T$ that characterises the thermal state. Such a thermal state of gravitational waves can, for example, be a description of a gravitational environment consisting of a primordial thermal gravitational wave background, for which a temperature of $T=0.9$ K is assumed in cosmological models \cite{Giovannini:2019oii} and used for instance in the application to neutrino oscillations in \cite{Domi:2024ypm}. For the choice $T=0$ the thermal state describes a vacuum state. 
~\\
~\\
The next step is now to interpret the system as an open quantum system. The interaction Hamiltonian is in lowest order of the order $\sqrt{\kappa}$. The reason for this is that the quantised geometric Dirac observables scale with a factor $\frac{1}{\sqrt{\kappa}}$ (see \eqref{quantsied field in mode expansion}). This means in the sense of the open quantum system formulation our coupling parameter is given by $\sqrt{\kappa}$, which implies that if we now truncate at second order, we keep terms up to order $(\sqrt{\kappa})^2=\kappa$.
~\\
~\\
Before we can use the projection operator technique with the time convolutionless approach, we need to assume that the initial state is separable $\widehat{\Tilde{\rho}}(0)=\widehat{\rho}_{\textrm{S}}(0)\otimes\widehat{\rho}_{\textrm{E}}$. 
In addition, the coupling parameter $\sqrt{\kappa}$ has to be sufficiently small. For a more detailed discussion on the assumptions and derivation see \cite{Fahn:2022zql,Giesel:2022pzh,Breuer:2007juk}.
~\\
~\\
The TCL master equation truncated at second order for the physical sector of the photon field coupled to linearised gravity has the form
\begin{align}\label{TCL Master equation without environmental trace calculation}
    \frac{\partial}{\partial t} \widehat{\Tilde{\rho}}_\textrm{S}(t)=-i\kappa[\widehat{\Tilde{U}}(t),\widehat{\Tilde{\rho}}_\textrm{S}(t)]-\kappa\int_{t_0}^t ds \:\textrm{tr}_{\textrm{E}}\left\{[\widehat{\Tilde{H}}_\textrm{TI}(t),[\widehat{\Tilde{H}}_\textrm{TI}(s),\widehat{\Tilde{\rho}}_\textrm{S}(t)\otimes  \widehat{\rho}_{\textrm{E}}]]\right\}.
\end{align}
where $\widehat{\rho}_{\textrm{S}}=\textrm{tr}_{E}\left({\widehat{\rho}}\right)$, which denotes the partial trace over the environmental degrees of freedom of the total density matrix.
~\\
The final step is to explicitly calculate the environmental correlation functions. To do this, we need to consider the true interaction Hamiltonian in position space:
\begin{align}\label{true interaction for environmental trace}
    \widehat{\Tilde{H}}_\textrm{TI}(t)=\int\limits_{\mathbb{R}^3} d^3 x \widehat{\Tilde{T}}^{ab}(\vec{x},t)(\widehat{\Tilde{\delta\mathcal{E}}_{a}^i})^\textrm{phys}(\vec{x},t)\delta_{ib},
\end{align}
where we absorbed a factor $\frac{1}{\sqrt{\kappa}}$ by working with the quantised Dirac observables $\left(\widehat{\Tilde{\delta\mathcal{E}}_{a}^i}\right)^{\textrm{phys}}$ rather than $\hat{b}_k,\: \hat{b}_k^\dagger$ (see equation \eqref{quantsied field in mode expansion}). Using this, we get for the second term in the TCL master equation
\begin{align}
    \begin{split}
    \textrm{tr}_{\textrm{E}}\left\{[\widehat{\Tilde{H}}_\textrm{TI}(t),[\widehat{\Tilde{H}}_\textrm{TI}(s),\widehat{\Tilde{\rho}}_\textrm{S}(t)\otimes  \widehat{\rho}_{\textrm{E}}]]\right\}=& \int\limits_{\mathbb{R}^3} d^3x\int\limits_{\mathbb{R}^3} d^3 y \:\:\textrm{tr}_{\textrm{E}}\Big\{\Big[\widehat{\Tilde{T}}^{ab}(\vec{x},t)(\widehat{\Tilde{\delta\mathcal{E}}}_{ab})^\textrm{phys}(\vec{x},t),\\
    &[\widehat{\Tilde{T}}^{ab}(\vec{y},s)(\widehat{\Tilde{\delta\mathcal{E}}}_{ab})^\textrm{phys}(\vec{y},s),\widehat{\Tilde{\rho}}_\textrm{S}(t)\otimes  \widehat{\rho}_{\textrm{E}}]\Big]\Big\}.
    \end{split}
\end{align}
The trace only affects the gravitational degrees of freedom $\left(\widehat{\Tilde{\delta\mathcal{E}}}_{ab}\right)^{\textrm{phys}}$. This means we can pull out the spatial energy-momentum tensor $\widehat{\Tilde{T}}^{ab}$ and $\widehat{\Tilde{\rho}}_{\textrm{S}}$ from the environmental trace. By evaluating the commutator and by using the invariance of the trace under cyclic permutations, we get for the right hand side
\begin{align}\label{term to calculate for environmental trace}
    \begin{split}
    \int\limits_{\mathbb{R}^3} d^3x\int\limits_{\mathbb{R}^3} d^3 y &\Big(\left(\widehat{\Tilde{T}}^{ab}(\vec{x},t)\widehat{\Tilde{T}}^{cd}(\vec{y},s)\widehat{\Tilde{\rho}}_\textrm{S}(t)-\widehat{\Tilde{T}}^{cd}(\vec{y},s)\widehat{\Tilde{\rho}}_\textrm{S}(t)\widehat{\Tilde{T}}^{ab}(\vec{x},t)\right)\\
    &\hspace{0.5in}\cdot\textrm{tr}_{\textrm{E}}\left\{(\widehat{\Tilde{\delta\mathcal{E}}}_{ab})^\textrm{phys}(\vec{x},t)(\widehat{\Tilde{\delta\mathcal{E}}}_{cd})^\textrm{phys}(\vec{y},s)\widehat{\rho}_{\textrm{E}}\right\}\\
    &+\left(\widehat{\Tilde{\rho}}_\textrm{S}(t)\widehat{\Tilde{T}}^{cd}(\vec{y},s)\widehat{\Tilde{T}}^{ab}(\vec{x},t)-\widehat{\Tilde{T}}^{ab}(\vec{x},t)\widehat{\Tilde{\rho}}_\textrm{S}(t)\widehat{\Tilde{T}}^{cd}(\vec{y},s)\right)\\
    &\hspace{0.5in}\cdot\textrm{tr}_{\textrm{E}}\left\{(\widehat{\Tilde{\delta\mathcal{E}}}_{cd})^\textrm{phys}(\vec{y},s)(\widehat{\Tilde{\delta\mathcal{E}}}_{ab})^\textrm{phys}(\vec{y},t)\widehat{\rho}_{\textrm{E}}\right\}\Big).
    \end{split}
\end{align}
The terms in the second and fourth line of equation \eqref{term to calculate for environmental trace} can be rewritten in terms of thermal Wightman functions:
\begin{align}\label{thermal wightmann functions}
    G^>_{abcd}(\vec{x}-\vec{y},t-s)&\coloneqq <(\widehat{\Tilde{\delta\mathcal{E}}}_{ab})^\textrm{phys}(\vec{x},t)(\widehat{\Tilde{\delta\mathcal{E}}}_{cd})^\textrm{phys}(\vec{y},s)>_{\textrm{E}},\\
    G^<_{abcd}(\vec{x}-\vec{y},t-s)&\coloneqq <(\widehat{\Tilde{\delta\mathcal{E}}}_{cd})^\textrm{phys}(\vec{y},s)(\widehat{\Tilde{\delta\mathcal{E}}}_{ab})^\textrm{phys}(\vec{x},t)>_{\textrm{E}},
\end{align}
where $<...>_{\textrm{E}}\coloneqq\textrm{tr}_{\textrm{E}} (...\widehat{\rho}_{\textrm{E}})$. 
A detailed derivation of these functions can be found in \cite{Fahn:2022zql}. The results can be directly transferred to the model considered here, as both models include the same choice of linearised gravity as the environment. The use of a thermal state for the environment and the linearity of $(\widehat{\Tilde{\delta\mathcal{E}}}_{ab})^\textrm{phys}$ in the gravitational creation and annihilation operators implies that the Wightman functions only depend on the difference of the coordinates of the $(\widehat{\Tilde{\delta\mathcal{E}}}_{ab})^\textrm{phys}$.
~\\
~\\
Using this, we can finally formulate the master equation for Maxwell's theory coupled to linearised gravity in the interaction picture:
\begin{align}
    \begin{split}
    \frac{\partial}{\partial t} \widehat{\Tilde{\rho}}_\textrm{S}(t)=-i\kappa[\widehat{\Tilde{U}}(t),\widehat{\Tilde{\rho}}_\textrm{S}(t)]-\kappa \int\limits_{\mathbb{R}^3} d^3x\int\limits_{\mathbb{R}^3} d^3 y \sum_{r\in\{\pm\}}
    &\Big[\left(\widehat{\Tilde{J}}_r(\vec{x},t)\widehat{\Tilde{J}}_r(\vec{y},s)\widehat{\Tilde{\rho}}_\textrm{S}(t)-\widehat{\Tilde{J}}_r(\vec{y},s)\widehat{\Tilde{\rho}}_\textrm{S}(t)\widehat{\Tilde{J}}_r(x,t)\right) \\
    &\cdot G^>(\vec{x}-\vec{y},t-s)\\
    &+\left(\widehat{\Tilde{\rho}}_\textrm{S}(t)\widehat{\Tilde{J}}_r(\vec{y},s)\widehat{\Tilde{J}}_r(\vec{x},t)-\widehat{\Tilde{J}}_r(\vec{x},t)\widehat{\Tilde{\rho}}_\textrm{S}(t)\widehat{\Tilde{J}}_r(\vec{y},s)\right)\\
    &\cdot G^>(\vec{x}-\vec{y},t-s)\Big]  .  
    \end{split}
\end{align}
\subsection{The final master equation}
By transforming back to the Schrödinger picture and with some further rewriting, we end up with the following form of the master equation \cite{Fahn:2022zql}
\begin{align}\label{final master equation 2}
    \begin{split}
    \frac{\partial}{\partial t} \widehat{\rho}_\textrm{S}(t)=&-i[\widehat{H}_S
    +\kappa\widehat{U},\widehat{\rho}_\textrm{S}(t)]\\
    &-\kappa \int\limits_{\mathbb{R}^3} \frac{d^3 k}{2\Omega_k} \sum_{r\in\{\pm\}}\Big([\widehat{J}_r^\dagger(\vec{k}),\widehat{\Tilde{J}}_r(\vec{k},t)\widehat{\rho}_s(t)]+N(\Omega_k)[\widehat{J}_r^\dagger(\vec{k}),[\widehat{\Tilde{J}}_r(\vec{k},t),\widehat{\rho}_S(t)]]+\textrm{h.c.}\Big),
    \end{split}
\end{align}
where $N(\Omega_k)=\textrm{tr}_{\textrm{E}}\left(\hat{n}_k\widehat{\rho}_{\textrm{E}}\right) = \frac{1}{e^{\beta \Omega_k}-1}$ is the Bose-Einstein distribution depending on the frequency of the gravitational waves $\Omega_k$ and the environment's temperature parameter $T$. Equation \eqref{final master equation 2} permits to furthermore directly identify the influence of the vacuum fluctuations and the thermal constituent of the gravitational environment: while the first term in the second line is present independently of the choice of the temperature parameter $T$, the second term vanishes when considering the vacuum case $k_B T = \frac{1}{\beta}\to 0$.
~\\
~\\
The structure of the Master equation is very similar to the one for a scalar field in \cite{Fahn:2022zql} even though the underlying matter fields, which are coupled, differ. The differences become apparent if we look at the explicit form of the involved system operators. The system Hamiltonian $H_{\textrm{S}}$ and the self-interaction are easy to compare since they consists of the contribution of the corresponding energy-momentum tensors, which obviously differ for the different underlying matter systems. More interesting is the true interaction which is encoded in
\begin{align}
    \widehat{J}_r(\vec{k})=&-\int\limits_{\mathbb{R}^3} \frac{d^3 p}{(2\pi)^\frac{3}{2}}\sum_{a=1}^{4}\hat{j}_r^a(\vec{k},\vec{p}),\\
    \widehat{\Tilde{J}}_r(\vec{k},t)=&-\int\limits_{\mathbb{R}^3} \frac{d^3 p}{(2\pi)^\frac{3}{2}}\sum_{a=1}^{4}\hat{j}_r^a(\vec{k},\vec{p})f(\Omega_k+\omega_k(\vec{k},\vec{p});t),\\
    f(\omega,t)=&\int_0^t ds e^{-i\omega (t-s)}=\frac{i}{\omega}(e^{-i\omega t}-1)
\end{align}
with the operator valued distributions
\begin{align}\label{small j definition}
    \begin{split}
    \hat{j}_r^1(\vec{k},\vec{p})=&\sum_{u,s\in\{\pm\}}(\hat{a}_{p-k}^u)^\dagger \hat{a}_{p}^s M_1^{ab}(\vec{k},\vec{p},u,s)(P^r(\vec{k}))_{ab}\:\:\:\:\:\:\:\:\:\:\:\:\:\:\omega_1 (\vec{k},\vec{p})=\Omega_{p}-\Omega_{k-p},\\
    \hat{j}_r^2(\vec{k},\vec{p})=&\sum_{u,s\in\{\pm\}}(\hat{a}_{-p}^s)^\dagger \hat{a}_{k-p}^u M_1^{ab}(\vec{k},\vec{p},u,s)(P^r(\vec{k}))_{ab}\:\:\:\:\:\:\:\:\:\:\:\omega_2 (\vec{k},\vec{p})=\Omega_{k-p}-\Omega_{p}, \\
    \hat{j}_r^3(\vec{k},\vec{p})=&\sum_{u,s\in\{\pm\}}\hat{a}_{p}^s \hat{a}_{k-p}^u M_2^{ab}(\vec{k},\vec{p},u,s)(P^r(\vec{k}))_{ab}\:\:\:\:\:\:\:\:\:\:\:\:\:\:\:\:\:\:\:\omega_3 (\vec{k},\vec{p})=\Omega_{p}+\Omega_{k-p}, \\
    \hat{j}_r^4(\vec{k},\vec{p})=&\sum_{u,s\in\{\pm\}}(\hat{a}_{-p}^s)^\dagger (\hat{a}_{p-k}^u)^\dagger M_2^{ab}(\vec{k},\vec{p},u,s)(P^r(\vec{k}))_{ab}\:\:\:\:\:\:\omega_4 (\vec{k},\vec{p})=-\Omega_{p}-\Omega_{k-p},
    \end{split}
\end{align}
where we defined the tensors $M_1^{ab},\:M^{ab}_2$ which encode the vector index structure associated with the photons
\begin{align}\label{eq: matrices that encode the index structure}
    \begin{split}
    M_1^{ab}(\vec{k},\vec{p},u,r)\coloneqq&\Big(-\frac{\sqrt{\Omega_p\Omega_{k-p}}}{2}\overline{m}^a(-u\vec{p})\overline{m}^b(r(\vec{p}-\vec{k}))\\
    &-\frac{1}{2\sqrt{\Omega_p\Omega_{k-p}}}p^{[a}(k-p)^{[b}\overline{m}^{c]}(-u\vec{p})\overline{m}^{d]}(r(\vec{p}-\vec{k}))\delta_{cd}\Big),\\
    M_2^{ab}(\vec{k},\vec{p},u,r)\coloneqq&\Big(\frac{\sqrt{\Omega_p\Omega_{k-p}}}{2}\overline{m}^a(-u\vec{p})\overline{m}^b(r(\vec{p}-\vec{k}))\\
    &-\frac{1}{2\sqrt{\Omega_p\Omega_{k-p}}}p^{[a}(k-p)^{[b}\overline{m}^{c]}(-u\vec{p})\overline{m}^{d]}(r(\vec{p}-\vec{k}))\delta_{cd}\Big).
    \end{split}
\end{align}
Compared to \cite{Fahn:2022zql}, the effect of the different underlying systems become apparent in \eqref{small j definition} and \eqref{eq: matrices that encode the index structure}. First, the photon field consists of two polarisations. This leads to a sum over these polarisations in \eqref{small j definition}. The fact that we have coupled a vector field instead of a scalar field is encoded in the tensors $M_1^{ab},\:M^{ab}_2$. In addition to the results in \cite{Fahn:2022zql}, here the result also includes contributions in which no contraction with the momenta is involved. The reason for this is simply that the scalar field has no spatial indices by definition and therefore has a trivial expansion in this Fourier basis. Next we compare with \cite{Lagouvardos:2020laf}, although they use a different notation, their index structure is similar to our results. This is not surprising since they also couple a vector field, but we will discuss the comparison to their results in more detail later in this article.
~\\
~\\
Despite these differences in the operator valued distributions $ j^l_r $ for $ l\in{1,2,3,4} $, the behaviour of them is similar to that in \cite{Fahn:2022zql}, since the contraction of the tensors $ M_1^{ab},\:M^{ab}_2 $ with the projector leaves no index open. This especially means that the Master equation can be brought into a form similar to \cite{Fahn:2022zql}
\begin{align}
    \mathcal{D}[\widehat{\rho}_S]=\frac{\kappa}{2}\sum_{r;a,b}\int\limits_{\mathbb{R}^3}\int\limits_{\mathbb{R}^3}\int\limits_{\mathbb{R}^3} \frac{d^3k d^3p d^3l}{(2\pi)^\frac{6}{2}}\left\{\Delta_{ab}(\vec{p},\vec{l};\vec{k},t)\left[\hat{j}_r^b(\vec{k},\vec{l})\widehat{\rho}_S(t),\hat{j}_r^a(\vec{k},\vec{p})^\dagger\right]+h.c.\right\},
\end{align}
with
\begin{align}\label{delta operator}
    \begin{split}
    \Delta_{ab}(\vec{p},\vec{l};\vec{k},t)\coloneqq &\frac{1}{\Omega_k}\left[(N(\Omega_k)+1)f(\Omega_k+\omega_b(\vec{k},\vec{l});t)+N(\Omega_k)f(-\Omega_k+\omega_b(\vec{k},\vec{l});t)\right]\\
    =&2\int_0^t ds G^>(\vec{k},t-s)e^{-i\omega_b(\vec{k},\vec{l})(t-s)}.
    \end{split}
\end{align}
We can combine $\Delta_{ab}$ into two functions
\begin{align}\label{S operator}
    \begin{split}
    S_{ab}(\vec{p},\vec{l};\vec{k},t):=&\frac{1}{2i}\left(\Delta_{ab}(\vec{p},\vec{l};\vec{k},t)-\Delta_{ba}^*(\vec{l},\vec{p};\vec{k},t)\right)\\
    =&\frac{1}{2\Omega_k}\Big[(N(\Omega_k)+1)\left\{\frac{e^{-i(\Omega_k+\omega_b(\vec{k},\vec{l}))t}-1}{\Omega_k+\omega_b(\vec{k},\vec{l})}+\frac{e^{i(\Omega_k+\omega_a(\vec{k},\vec{p}))t}-1}{\Omega_k+\omega_a(\vec{k},\vec{p})}\right\}\\
    &-N(\Omega_K)\left\{\frac{e^{i(\Omega_k-\omega_b(\vec{k},\vec{l}))t}-1}{\Omega_k-\omega_b(\vec{k},\vec{l})}+\frac{e^{-i(\Omega_k-\omega_a(\vec{k},\vec{p}))t}-1}{\Omega_k-\omega_a(\vec{k},\vec{p})}\right\}\Big],
    \end{split}
    \\
    \begin{split}\label{R operator}
    R_{ab}(\vec{p},\vec{l};\vec{k},t)\coloneqq &\left(\Delta_{ab}(\vec{p},\vec{l};\vec{k},t)+\Delta_{ba}^*(\vec{l},\vec{p};\vec{k},t)\right)\\
    =&\frac{i}{\Omega_k}\Big[(N(\Omega_k)+1)\left\{\frac{e^{-i(\Omega_k+\omega_b(\vec{k},\vec{l}))t}-1}{\Omega_k+\omega_b(\vec{k},\vec{l})}-\frac{e^{i(\Omega_k+\omega_a(\vec{k},\vec{p}))t}-1}{\Omega_k+\omega_a(\vec{k},\vec{p})}\right\}\\
    &-N(\Omega_K)\left\{\frac{e^{i(\Omega_k-\omega_b(\vec{k},\vec{l}))t}-1}{\Omega_k-\omega_b(\vec{k},\vec{l})}-\frac{e^{-i(\Omega_k-\omega_a(\vec{k},\vec{p}))t}-1}{\Omega_k-\omega_a(\vec{k},\vec{p})}\right\}\Big].
    \end{split}
\end{align}
Now the dissipator can be split into two parts
\begin{align}
    \mathcal{D}[\widehat{\rho}_S(t)]=-i\kappa[\widehat{H}_\textrm{LS},\widehat{\rho}_S(t)]+\mathcal{D}_\textrm{first}[\widehat{\rho}_S(t)],
\end{align}
with the Lamb shift Hamiltonian
\begin{align}
    \widehat{H}_\textrm{LS}= \frac{1}{2}\int\limits_{\mathbb{R}^9}\frac{d^3 k d^3p d^3 l}{(2\pi)^3}\sum_{r;a,b} S_{ab}(\vec{p},\vec{l};\vec{k},t) \hat{j}_r^a(\vec{k},\vec{p})^\dagger \hat{j}_r^b(\vec{k},\vec{l}),
\end{align}
and the effective dissipator (similar to the first standard form \cite{Fahn:2022zql})
\begin{align}\label{effective Dissipator}
    \begin{split}
    \mathcal{D}_\textrm{first}[\widehat{\rho}_S(t)]:=\frac{\kappa}{2}\int\limits_{\mathbb{R}^9}\frac{d^3 k d^3p d^3 l}{(2\pi)^3}\sum_{r;a,b} R_{ab}(\vec{p},\vec{l};\vec{k},t)\Big(&\hat{j}_r^b(\vec{k},\vec{l})\widehat{\rho}_S(t) \hat{j}_r^a(\vec{k},\vec{p})^\dagger\\
    &-\frac{1}{2}\left\{\hat{j}_r^a(\vec{k},\vec{p})^\dagger \hat{j}_r^b(\vec{k},\vec{l}),\widehat{\rho}_S(t)\right\}\Big).
    \end{split}
\end{align}
Therefore the Master equation can be rewritten in the following form
\begin{align}\label{final not Lindblad Master equation}
    \begin{split}
    \frac{\partial}{\partial t} \widehat{\rho}_\textrm{S}(t)=&-i[\widehat{H}_S+\kappa\widehat{U},\widehat{\rho}_\textrm{S}(t)]-i\kappa[\widehat{H}_\textrm{LS},\widehat{\rho}_S(t)]\\
    &+\mathcal{D}_\textrm{first}[\widehat{\rho}_S(t)],
    \end{split}
\end{align}
where the unitary evolution is given by the first line and the non-unitary evolution by the second.
~\\
~\\
As mentioned before, the structure of the Master equation is very similar to the one in \cite{Fahn:2022zql} and therefore it is not surprising that the master equation obtained here is also not of strict Lindblad form because $R_{ab}$ in \eqref{effective Dissipator} still depends on time. A more detailed discussion on whether a Markov approximation, which presents together with a rotating wave approximation the possibility to achieve a Lindblad form \cite{Breuer:2007juk}, can be applied in the relativistic field theory case, can be found in \cite{Fahn:2022zql}. First results in this direction are discussed in \cite{Fahn:2024fgc}, where at least for the ultra-relativistic limit an application of the Markov approximation can be justified under certain conditions. More general, the question whether the master equation for relativistic systems can be formulated in Lindblad form is a topic currently discussed in the literature for instance in \cite{Matsumura:2023cni, Diosi:2022lpq, Milburn:2005vz, Menicucci:2011nc}. 
~\\
~\\
Next, we aim at comparing our result with the existing literature \cite{Fahn:2022zql}, \cite{Lagouvardos:2020laf} and \cite{Anastopoulos:2013zya}. We have already discussed the similarities and differences to \cite{Fahn:2022zql} and showed that the final master equation we derived is structurally consistent with the one in \cite{Fahn:2022zql}. Therefore, the comparison with the other cited models can be made by analogy. The main points are that our model is defined in Ashtekar-Barbero variables, whereas the literature uses ADM variables. Instead of a gauge fixing, we used the relational formalism to construct physical observables. The normal ordering was chosen by taking the products first and then applying the normal ordering. In \cite{Lagouvardos:2020laf} and \cite{Anastopoulos:2013zya} the opposite approach is taken. Unlike \cite{Lagouvardos:2020laf}, where they also study a vector field, we have not derived a Lindblad type master equation, since the discussion of the applicability of Markov and rotating wave approximation for a linearised gravitational environment is still not completely answered, see \cite{Fahn:2024fgc} for more discussion. Apart from that, the structure of operators and commutator-anticommutator brackets of the derived master equation \eqref{final master equation 2} is consistent with that in \cite{Lagouvardos:2020laf}. Furthermore, since a different normal ordering is chosen, the self-interaction and $\widehat{J}_r$ operator valued distributions are different. Due to the structural similarity to the master equation in \cite{Fahn:2022zql}, we expect that also in the photonic case divergent terms are present in the master equation. Hence a renormalisation is needed to tame them. Such a renormalisation is expected to be feasible similarly to the one in \cite{Fahn:2024fgc}, where the model from \cite{Fahn:2022zql} is treated.
\section{Conclusions}\label{sec:Conclusions}
In this work, we have discussed the field-theoretical derivation of a quantum master equation that describes the effective dynamics of photons under the influence of a thermal gravitational wave environment. To this end, a photon field coupled to linearised gravity was considered at the classical level in a post-Minkowski approximation.
~\\
Compared to existing open QFT models, which used quantised linearised gravity and a photon field \cite{Lagouvardos:2020laf}, a scalar field \cite{Anastopoulos:2013zya, Blencowe:2012mp}, or a general boson field \cite{Oniga:2015lro}, formulated in ADM variables, the models here are expressed in Ashtekar-Barbero variables, similar to the work in \cite{Fahn:2022zql}. The reason for this is to provide the model in a form in which, in future work, an LQG-inspired quantisation can also be applied to the environment described by linearised gravity.
~\\
Another difference from the results of \cite{Anastopoulos:2013zya,Blencowe:2012mp} is that in this work we have derived the physical phase space by constructing Dirac observables for the final set of physical degrees of freedom, rather than applying a gauge fixing. An advantage of this approach is that the model is valid in any chosen gauge, but the interpretation of the Dirac observables may vary for different gauge choices.
~\\
For this purpose, the framework of the relational formalism was applied, in which dynamical reference fields are introduced, commonly referred to as (relational) clocks, and in relation to which the dynamics of the remaining degrees of freedom are formulated. Here, it was applied within the framework of perturbation theory, taking into account that we are working in a linearised phase space. Similar to \cite{Fahn:2022zql}, geometrical clocks were chosen for the Hamiltonian and the spatial diffeomorphism constraint in order to provide a concept of physical time and physical spatial coordinates. Due to the U(1) gauge constraint in Maxwell's theory, an additional clock needed to be  introduced, which resembles a Lorenz-like gauge fixing when restricted to a gauge fixing surface. Compared to \cite{Fahn:2022zql}, we have gained new insights into the chosen set of Dirac observables and their properties. In order to obtain a set of mutually commuting clocks in which additionally each clock is canonically conjugate to one of the constraints, the clock associated with the Maxwell U(1) gauge constraint was constructed using purely electromagnetic degrees of freedom.
~\\
~\\
 A more detailed analysis of the construction of Dirac observables and their corresponding algebra in comparison to \cite{Fahn:2022zql} showed that a combination of the observable map and its dual directly generates Dirac observables as the two symmetric transverse traceless degrees of freedom in the gravitational sector. Note that this is not achieved by applying the observable map alone. For this reason, the symmetric transverse traceless projector in \cite{Fahn:2022zql} was introduced manually and not constructed entirely relationally. Similarly, in \cite{Anastopoulos:2013zya,Lagouvardos:2020laf}, the projector must be introduced manually by constructing because they use a  gauge fixing. The same result and comparison also apply to the Dirac observables for the photon field.
~\\
~\\
Given the final set of chosen Dirac observables, a physical Hamiltonian was obtained from the observable map that can be expressed solely in terms of Dirac observables, where the electromagnetic part was entirely formulated in terms of the energy-momentum tensor, depending on the electromagnetic Dirac observables. The Hamiltonian can be split into four parts, where two parts describe the individual, isolated evolution of the photon field and gravitational Dirac observables respectively, the third part encodes their interaction and the last part, which is denoted as self-interaction, only depends on electromagnetic Dirac observables but would be absent without the coupling to linearised gravity.
~\\
~\\
In addition, we discussed a second set of Dirac observables and compared the corresponding algebra and the resulting physical Hamiltonian with the other choice. It turns out that the two choices differ in how the above-mentioned self-interaction term is incorporated into the model. In the first choice, which we used in this work, the self-interaction is part of the physical Hamiltonian, ensuring a simple form of the Poisson algebra of Dirac observables. In the second choice, the physical Hamiltonian  is simplified and the self-interaction term is absent, but at the cost of entering the Dirac observables algebra and making it more complicated. The reason we have opted for the first choice here is that in this case, reduced phase space quantisation is simpler, as we only need to quantise standard canonical Poisson brackets. Note that in the context of formulating Feynman diagrams for the master equation in \cite{Fahn:2024fgc} it was shown that the presence of the last part can be interpreted as a relic of the use of the non-covariant Hamiltonian theory: when deriving Feynman rules for this non-covariant formulation of linearised gravity whose propagator and vertices only have spatial indices, then this part leads to a non-local interaction term for the matter. When restoring covariance in a next step by defining a suitable propagator and vertices with spacetime indices, then this self-interaction term is precisely the contribution needed to be able to form a covariant propagator for linearised gravity and one basic covariant interaction vertex with matter, similarly to for instance QED in a non-covariant gauge. For the relation of the covariant harmonic gauge and the geometrical clocks chosen here in the case of vacuum linearised gravity at the classical level see \cite{Giesel:2024xtb}. 
~\\
~\\
To obtain the final master equation in the quantum theory, a Fock quantisation of the chosen Dirac observables was carried out. Following the projection operator technique, we then derived a field-theoretical quantum master equation for photons, based on the quantisation of the electromagnetic Dirac observables, treating the quantisation of the physical gravitational degrees of freedom as environment. The environmental degrees of freedom were traced out using a thermal state characterised by a temperature parameter. This master equation was discussed in different equivalent formulations that feature the identification of vacuum and thermal contributions as well as a form similar to the first standard form of a master equation with, however, still time-dependent coefficients. This time dependency could be removed by the application of a Markov approximation whose applicability is, however, not yet fully proven for a linearised gravitational environment, see also the discussion in \cite{Fahn:2024fgc}, where a validity condition for single ultra-relativistic scalar particles was given. Also, the application of the rotating wave approximation, required to diagonalise the first standard form, poses similar challenges. In contrast to \cite{Anastopoulos:2013zya,Lagouvardos:2020laf}, where these two approximations are employed, the master equation here therefore does not have a Lindblad form.
~\\
~\\
In order to link of these results to specific applications, the dynamics of a single photon predicted by the final field-theoretical master equation of this work is one of the next steps that we want to consider in future work. While a similar procedure including a renormalisation and a discussion of the physical effects of the Markov and rotating wave approximation was carried out in \cite{Fahn:2024fgc} for a scalar field, the electromagnetic field now provides a more realistic physical application. While we expect that several steps and arguments from \cite{Fahn:2024fgc} can be taken over, as they mainly concern the gravitational environment which is identical in the present work, there are still some points worth of further investigation, such as for instance the renormalisation of the linearised system now containing the massless electromagnetic field. As suggested by \cite{Fahn:2024fgc}, such a renormalisation might remove contributions from the master equation which are responsible for a decoherence effect on the diagonal elements of the photon density matrix in momentum basis (populations) derived in \cite{Oniga:2015lro} arising from vacuum fluctuations of the linearised gravitational environment.
~\\
~\\
A further work we want to address in future work is a formulation of an analogous model for fermions.
To the knowledge of the authors, a derivation of a master equation for this scenario has so far not been discussed starting from the field-theoretical, relativistic action and treating gravity quantised but using a non-relativistic model with classical gravity for a spin-1/2-particle like in \cite{Asprea:2020fuy} or a quantum mechanical toy model for a neutrino coupled to a bath of quantised harmonic oscillator that model the gravitational field as \cite{Domi:2024ypm}. In this context, the fact that the contribution of matter, in this case the photon field, was written in the form of the energy-momentum tensor, which allows direct generalisation to fermions, as well as the use of Ashtekar-Barbero variables, could be useful. Furthermore, such a master equation for fermions provides the possibility to bridge better to  phenomenological models as for instance \cite{Benatti:2000ph,Lisi:2000zt,Guzzo:2014jbp,BalieiroGomes:2018gtd,Lessing:2023uxb} and the microscopic model in  \cite{Domi:2024ypm}. The latter suggest the manifestation of the gravitational decoherence effect as a damping in the neutrino oscillations which allows to set bounds on decoherence parameters by current neutrino oscillations experiments.
~\\
~\\
The current model based on Fock quantisation will also provide the necessary insight that we need to compare models with different quantisation procedures such as Fock and for instance loop quantum gravity (LQG) inspired ones, where the latter requires models formulated in terms of  Ashtekar-Barbero variables.  While decoherence effects in symmetry reduced models of LQG have been studied in the last years for example in the context of black holes (see e.g. \cite{Feller:2016zuk}, or more phenomenologically motivated where a minimal area gap is implemented at an effective level \cite{Fahn:2025fxl,Fahn:2025jxh}), the model in the present work and the model in \cite{Fahn:2022zql} allow in principle a direct application of LQG inspired techniques for linearised gravity following \cite{Ashtekar:1991mz,Varadarajan:2002ht}. The main technical challenges that we expect for such a model will be the computation of environmental correlation functions. A first step to face this is to consider quantum mechanical toy models in the framework of polymerised quantum mechanics, see for instance \cite{Giesel:2022pzh} for an open scattering model. In future work we plan to consider quantum mechanical toy models that are related to the QFT model considered in \cite{Fahn:2022zql} and in this work. One model that we will consider in future work is still quantised in a Schrödinger quantisation with a bath of harmonic oscillators but with a non-standard interaction of the system and environment that involves Weyl operators instead of position operators \cite{Weylpaper}. In a second step it is planned that the model is quantised in the framework of polymer quantum mechanics \cite{Polypaper}, an LQG inspired quantisation in quantum mechanics, and to compare the characteristic features of both models might give first insights on how the corresponding open QFT models might differ and what physical implications they provide for gravitationally induced decoherence in specific applications.

\begin{acknowledgments}
M.J.F. is partially supported by the INFN grant FLAG. R.K. thanks the Villigst foundation for financial support. K.G. is grateful for the hospitality of Perimeter Institute where part of this work was carried out. Research at Perimeter Institute is supported in part by the Government of Canada through the Department of Innovation, Science and Economic Development and by the Province of Ontario through the Ministry of Colleges and Universities. This work was supported by a grant from the Simons Foundation (1034867, Dittrich). The authors would like to acknowledge the contribution of the COST Action CA23130 ``Bridging high and low energies in search of quantum gravity (BridgeQG)". The authors have benefited from the activities of COST Action CA23115: Relativistic Quantum Information, funded by COST (European Cooperation in Science and Technology).
\end{acknowledgments}

\begin{appendices}
\renewcommand{\thesection}{A.\Roman{section}}
\renewcommand{\thesubsection}{\thesection.\arabic{subsection}}
\renewcommand{\thesubsubsection}{\thesubsection.\arabic{subsubsection}}
\renewcommand{\theequation}{\thesection.\arabic{equation}}

\makeatletter
\renewcommand{\p@subsection}{}
\renewcommand{\p@subsubsection}{}
\makeatother
\section{Properties of the dual and observable map}\label{app: observable properties}
In this section, we take a closer look at the following properties of the observable and dual observable map. For the observable map, the following identities hold (for two arbitrary phase space functions $f$ and $f^\prime$):
\begin{align}\label{eq app: Observable map identities}
    \begin{split}
    \left\{\mathcal{O}_{f,\{\mathcal{G}^I\}}, \mathcal{O}_{f^\prime,\{\mathcal{G}^I\}}\right\}\approx&\mathcal{O}_{\{f,f^\prime\}^{\ast},\{\mathcal{G}^I\}},\\
    \mathcal{O}_{f,\{\mathcal{G}^I\}}\mathcal{O}_{f^\prime,\{\mathcal{G}^I\}}\approx&\mathcal{O}_{f f^\prime,\{\mathcal{G}^I\}},
    \end{split}
\end{align}
where $\{f,f^\prime\}^\ast$ is the Dirac bracket generated by the constraints and reference fields. The proof for these characteristics can be found in \cite{Thiemann:2004wk} and uses a generic set of non Abelian first-order constraints. A weak abelianisation method is employed to construct a set of weakly commuting constraints. The work in \cite{Thiemann:2004wk} considers the non field theory case, which can however be easily generalised to field theory by simple integration. In this section, we will proof that these identities hold for our system with strong equal signs and also for the dual observable map, up to the order we are interested in.
~\\
~\\
We constructed a set of constraints $C_I^\prime$ and reference fields $\mathcal{G}^J$ (see \eqref{eq:FinalSet}), which strongly commute with each other except for their canonically conjugate up to $O(\delta^2,\kappa^2)$ contributions:
\begin{align}\label{eq: poisson algebra of clocks and constraints for appendix proof}
    \begin{split}
    \{C_I^\prime(\vec{x},t),C_J^\prime (\vec{y},t)\}=&0 +O(\delta^2,\kappa^2)\:\:\forall I,J,\vec{x},\vec{y},\\
    \{\mathcal{G}^I(\vec{x},t),\mathcal{G}^J(\vec{y},t)\}=&0+O(\delta^2,\kappa^2)\:\:\forall I,J,\vec{x},\vec{y},\\
    \{\mathcal{G}^I(\vec{x},t),C^\prime_J(\vec{y},t)\}=&\frac{1}{\kappa}\delta_J^I\delta(\vec{x}-\vec{y})+O(\delta^2,\kappa^2)\:\:\forall I,J,\vec{x},\vec{y}.
    \end{split}
\end{align}
In the main text, we use the identities \eqref{eq app: Observable map identities} to construct the Dirac observables and physical Hamiltonian up to second order in $\delta,\kappa$. Thus, we also need to derive the identities \eqref{eq app: Observable map identities} up to second order here but can neglect terms that are higher in the perturbations.
~\\
~\\
For further investigation, we define the Hamiltonian vector fields corresponding to the constraints and reference fields
\begin{align}\label{eq: Hamitlonian vector fields for dual and observable map}
    \chi_{C_I^\prime(\beta_I)}[f]\coloneqq &\int\limits_{\mathbb{R}^3} d^3x\beta_I(\vec{x})\{C_I^\prime(\vec{x},t),f\},\\
    \chi_{\mathcal{G}^J(\gamma^J)}[f]\coloneqq &\int\limits_{\mathbb{R}^3} d^3x\gamma^J(\vec{x})\{\mathcal{G}^J(\vec{x},t),f\}.
\end{align}
With suitable phase space functions $\beta$ and $\gamma$, it is straightforward to prove the following properties using the Poisson algebra of the constraints and reference fields \eqref{eq: poisson algebra of clocks and constraints for appendix proof}
\begin{align}\label{eq: Properties of hamiltonian vector fields}
    \begin{split}
    \chi_{C_I^\prime(\beta_I)}[C_J^\prime]=&0+O(\delta^3,\kappa^2)\:\:\forall I,J,\\
    \chi_{C_I^\prime(\beta_I)}[\mathcal{G}^J]=&\beta^J\:\:\forall I,J,\\
    \chi_{\mathcal{G}^J(\gamma^J)}[\mathcal{G}^I]=&0+O(\delta^3,\kappa^2)\:\:\forall I,J,\\
    \chi_{\mathcal{G}^J(\gamma^J)}[C_I^\prime]=&\gamma_I\:\:\forall I,J.
    \end{split}
\end{align}
To define the observable and dual observable map using the Hamiltonian vector fields, we define the n-times application of the Hamiltonian vector fields as follows
\begin{align}\label{eq: n-times Hamitlonian vector fields for dual and observable map}
    \chi_{C_I^\prime(\beta_I)}^n[f]\coloneqq &\int\limits_{\mathbb{R}^3}...\int\limits_{\mathbb{R}^3} (d^3x)^n(\beta_I(\vec{x}))^n\{C_I^\prime(\vec{x},t),f\}_{(n)},\\
    \chi_{\mathcal{G}^J(\gamma^J)}^n[f]\coloneqq &\int\limits_{\mathbb{R}^3}...\int\limits_{\mathbb{R}^3} (d^3x)^n(\gamma^J(\vec{x}))^n\{\mathcal{G}^J(\vec{x},t),f\}_{(n)},
\end{align}
with the nested Poisson bracket $\{f,g\}_{(n)}$ defined by $\{g,f\}_0 = f,\:\: \{g,f\}_{(n+1)}=\{g,\{g,f\}_{(n)}\}$. Using this, we can write the observable map \eqref{eq:linearised observable map} and its dual \eqref{eq:linearised dual observable map} in the following way
\begin{align}\label{eq: Observable map defined with hamiltonian vectorfield}
    \mathcal{O}_{f,\{\mathcal{G}^K\}}=\sum_{I}\sum_{n=0}^\infty\frac{\kappa^n}{n!}\left(\chi^n_{C_I^\prime(\beta_I)}[f]\right)\Big|_{\beta_I=\mathcal{G}_I},\\
    \label{eq: Dual map defined with hamiltonian vectorfield}
    \mathcal{O}_{f,\{C_K^\prime\}}^{\textrm{dual}}=\sum_{I}\sum_{n=0}^\infty(-1)^n\frac{\kappa^n}{n!}\left(\chi^n_{\mathcal{G}^I(\gamma^I)}[f]\right)\Big|_{\gamma_I=C_I^\prime}.
\end{align}
These definitions are similar to those in \cite{Thiemann:2004wk}. If we compare with the proof in \cite{Thiemann:2004wk} for the identity \eqref{eq app: Observable map identities}, we see that we need to check how the order in which we apply Hamiltonian vector fields affects the result. To do this, it is sufficient to consider the action of only two Hamiltonian vector fields on each other, as we can repeatedly apply the identities we derive. Due to the sum over the index $I$ in both expressions, we have contributions in both maps where the Hamiltonian vector fields corresponding to different clocks or constraints act. In our case, in contrast to \cite{Thiemann:2004wk}, these form an abelian Poisson algebra up to $O(\delta^2,\kappa^2)$ contributions \eqref{eq: poisson algebra of clocks and constraints for appendix proof}. Hence, we investigate the following expression for $I \neq J$
\begin{align}\label{eq: proof for vanishing commutator of two Hamiltonian vector fields}
    \begin{split}
    \left(\chi_{C_I^\prime(\beta_I)}\left(\chi_{C_J^\prime(\beta_J)}[f]\right)\Big|_{\beta^J=\mathcal{G}^J}\right)\Big|_{\beta^I=\mathcal{G}^I}=& \int\limits_{\mathbb{R}^3}d^3y\left(\chi_{C_I^\prime(\beta_I)}\left(\mathcal{G}_J(\vec{y},t)\{C_J^\prime(\vec{y},t),f\}\right)\right)\Big|_{\beta^I=\mathcal{G}^I}\\
    =&\int\limits_{\mathbb{R}^3}\int\limits_{\mathbb{R}^3}d^3 xd^3y\mathcal{G}_I(\vec{x},t)\mathcal{G}_J(\vec{y},t)\{C_I^\prime(\vec{y},t),\{C_J^\prime(\vec{x},t),f\}\}+O(\delta^3,\kappa^2)\\
    =& \int\limits_{\mathbb{R}^3}\int\limits_{\mathbb{R}^3}d^3 xd^3y\mathcal{G}_I(\vec{x},t)\mathcal{G}_J(\vec{y},t)\Big(\{C_J^\prime(\vec{x},t),\{C_I^\prime(\vec{y},t),f\}\}\\
    &+\{f,\{C_J^\prime(\vec{x},t),C_I^\prime(\vec{y},t)\}\}\Big)+O(\delta^3,\kappa^2)\\
    =&\int\limits_{\mathbb{R}^3}\int\limits_{\mathbb{R}^3}d^3 xd^3y\mathcal{G}_J(\vec{y},t)\{C_J(\vec{y},t),\mathcal{G}_I(\vec{x},t)\{C_I^\prime(\vec{x},t),f\}\}\\
    &+O(\delta^3,\kappa^2)\\
    =&\int\limits_{\mathbb{R}^3}d^3 y \mathcal{G}_J(\vec{y},t)\{C_J(\vec{y},t),\left(\chi_{C_I(\beta_I)}[f]\right)\Big|_{\beta_I=\mathcal{G}_I}\}+O(\delta^3,\kappa^2)\\
    =&\left(\chi_{C_J^\prime(\beta_J)}\left(\chi_{C_I^\prime(\beta_I)}[f]\right)\Big|_{\beta^I=\mathcal{G}^I}\right)\Big|_{\beta^J=\mathcal{G}^J}+O(\delta^3,\kappa^2).
    \end{split}
\end{align}
In the first and second lines, we used the definition of the Hamiltonian vector fields \eqref{eq: Hamitlonian vector fields for dual and observable map}. In the third line, we used the Jacobi identity for the Poisson bracket. In the second, fourth and fifth line, we used the Poisson algebra \eqref{eq: poisson algebra of clocks and constraints for appendix proof} and that $I\neq J$. In the fifth and sixth lines, we again used the definition of the Hamiltonian vector fields. The end result proves that, in the perturbation order we are interested in, the order in which we apply the Hamiltonian vector fields for different constraints does not matter. The same statement follows similarly for the case $I=J$ using \eqref{eq: poisson algebra of clocks and constraints for appendix proof}. Furthermore, the proof that this is also the case for the Hamiltonian vector fields associated with the reference fields can be made analogously. Upon comparing this result with the proof in \cite{Thiemann:2004wk}, it becomes evident that the identities \eqref{eq app: Observable map identities} hold with a strong equality sign in our case, since we also have one in \eqref{eq: proof for vanishing commutator of two Hamiltonian vector fields}. Furthermore, the proof can be analogously extended to the dual observable map. Overall, we get the following identities:
\begin{align}\label{dual map characteristic}
    \begin{split}
        \left\{\mathcal{O}^{\textrm{dual}}_{f,\{C_I\}}, \mathcal{O}^{\textrm{dual}}_{f^\prime,\{C_I\}}\right\}=&\mathcal{O}^{\textrm{dual}}_{\{f,f^\prime\}^\ast,\{C_I\}}+O(\delta^3,\kappa^3),\\
        \mathcal{O}^{\textrm{dual}}_{f,\{C_I\}}\mathcal{O}^{\textrm{dual}}_{f^\prime,\{C_I\}}=&\mathcal{O}^{\textrm{dual}}_{f f^\prime,\{C_I\}}+O(\delta^3,\kappa^3), 
    \end{split}
\end{align}
as well as \eqref{eq app: Observable map identities} with a strong equality sign.
~\\
~\\
From the last line and the last line of \eqref{eq app: Observable map identities}, we can also follow for a phase space function $F(\delta\mathcal{A}_a^i,\delta\mathcal{E}^a_i,A_a,-E^a)$ under the action of the observable map or its dual
\begin{align}
    \begin{split}
    \mathcal{O}_{F(\delta\mathcal{A}_a^i,\delta\mathcal{E}^a_i,A_a,-E^a),\{\mathcal{G}^I\}}=F(\mathcal{O}_{\delta\mathcal{A}_a^i,\{\mathcal{G}^I\}}, \mathcal{O}_{\delta\mathcal{E}^a_i,\{\mathcal{G}^I\}}, \mathcal{O}_{A_a,\{\mathcal{G}^I\}},\mathcal{O}_{-E^a,\{\mathcal{G}^I\}}) +O(\delta^3,\kappa^3),\\
    \mathcal{O}^{\textrm{dual}}_{F(\delta\mathcal{A}_a^i,\delta\mathcal{E}^a_i,A_a,-E^a),\{C_I\}}=F(\mathcal{O}^{\textrm{dual}}_{\delta\mathcal{A}_a^i,\{C_I\}}, \mathcal{O}^{\textrm{dual}}_{\delta\mathcal{E}^a_i ,\{C_I\}}, \mathcal{O}^{\textrm{dual}}_{A_a,\{C_I\}},\mathcal{O}^{\textrm{dual}}_{-E^a,\{C_I\}})+O(\delta^3,\kappa^3).
    \end{split}
\end{align}
The last question of this section concerns the vacuum dual observable map \eqref{eq: vacuum dual map}, which we used for constructing the geometric Dirac observables (see section \ref{Construction of Dirac observables with respect to the chosen clocks}) of set $S$. Here, only the geometric part of the constraints is used. These vacuum constraints trivially commute with all electromagnetic degrees of freedom and therefore also with the electromagnetic reference field and constraint. Additionally, they Poisson commute with themselves in the linear phase space \cite{Ashtekar:1991mz}. This means that the set of constraints and clocks from which the vacuum dual observable map is built, again has the structure \eqref{eq: poisson algebra of clocks and constraints for appendix proof}. Therefore, in this vacuum case as well, the order in which we apply Hamiltonian vector fields corresponding to different reference fields does not matter. From this, it follows that the identities \eqref{dual map characteristic} also hold for the vacuum dual observable map \eqref{eq: vacuum dual map}.
\section{Order of application of the observable and dual observable map}\label{app: More details of the influence of the order in which we apply the observable and dual observable map}
In this section, we investigate whether the order of application of the dual observable map and the observable map is relevant. In general, this is the case, but in the following, we will prove that if \eqref{eq: poisson algebra of clocks and constraints for appendix proof} holds, the order of application does not matter, up to the perturbation order we are interested in. For the proof we will use the definitions and notations from appendix \ref{app: observable properties}. We will begin by writing down the expression for acting first with the observable map and then with its dual on a general phase space function $f$, using \eqref{eq: Observable map defined with hamiltonian vectorfield} and \eqref{eq: Dual map defined with hamiltonian vectorfield}
\begin{align}\label{eq: first dual then observable map on general function}
    \mathcal{O}^{\textrm{dual}}_{\mathcal{O}_{f,\{\mathcal{G}^K\}},\{C_K^\prime\}}=\sum_{I,J}\sum_{n,m=0}^{\infty}(-1)^n\frac{\kappa^{n+m}}{n!m!}\left(\chi^n_{\mathcal{G}^J(\gamma^J)}\left(\chi^m_{C_I^\prime(\beta_I)}[f]\right)\Big|_{\beta^I=\mathcal{G}^I}\right)\Big|_{\gamma_J=C_J^\prime}.
\end{align}
With identity \eqref{eq: proof for vanishing commutator of two Hamiltonian vector fields} and the similar one for the Hamiltonian vector fields corresponding to the reference fields, the m-th power of the Hamiltonian vector fields can be rewritten as follows
\begin{align}\label{eq: rewriting of n-times Hamiltonian vector fields for proof}
    \sum_I \left(\chi^m_{C_I^\prime(\beta_I)}[f]\right)\Big|_{\beta^I=\mathcal{G}^I} =& \sum_I \left(\chi_{C_I^\prime(\beta_I)}\left(\chi^{m-1}_{C_I^\prime(\beta_I)}[f]\right)\Big|_{\beta^I=\mathcal{G}^I}\right)\Big|_{\beta^I=\mathcal{G}^I}+O(\delta^3,\kappa^2),\\
    \sum_J \left(\chi^n_{\mathcal{G}^J(\gamma^J)}[f]\right)\Big|_{\gamma_J=C_J^\prime} =& \sum_J \left(\chi^{n-1}_{\mathcal{G}^J(\gamma^J)}\left(\chi_{\mathcal{G}^J(\gamma^J)}[f]\right)\Big|_{\gamma_J=C_J^\prime}\right)\Big|_{\gamma_J=C_J^\prime}+O(\delta^3,\kappa^2).
\end{align}
If we insert these two equations into \eqref{eq: first dual then observable map on general function}, it is easy to see that if we could commute $\left(\chi_{C_I^\prime(\beta_I)}[f]\right)\Big|_{\beta^I=\mathcal{G}^I}$ and $\left(\chi_{\mathcal{G}^J(\gamma^J)}[f]\right)\Big|_{\gamma_J=C_J^\prime}$, we could repeat this process iteratively until we get to the expression for first applying the dual observable map and then the observable map. Therefore, the order in which we apply them would not matter. Thus we take a closer look at the following
\begin{align}\label{eq: proof that order of observable and dual map does not matter}
    \begin{split}
    \sum_{I,J}\left(\chi_{C_I^\prime(\beta_I)}\left(\chi_{\mathcal{G}^J(\gamma^J)}[f]\right)\Big|_{\gamma_J=C_J^\prime}\right)\Big|_{\beta^I=\mathcal{G}^I}=&\sum_I\int\limits_{\mathbb{R}^3}d^3y \left(\chi_{C_I^\prime(\beta_I)}\left(C_J^\prime(\vec{y},t)\{\mathcal{G}^J(\vec{y},t),f\}\right)\right)\Big|_{\beta^I=\mathcal{G}^I}\\
    &+O(\delta^3,\kappa^2)\\
    =&\int\limits_{\mathbb{R}^3}\int\limits_{\mathbb{R}^3}d^3 xd^3y\mathcal{G}^I(\vec{x},t)C_J^\prime(\vec{y},t)\{C_I^\prime(\vec{x},t),\{\mathcal{G}^J(\vec{y},t),f\}\}\\
    &+O(\delta^3,\kappa^2)\\
    =& \int\limits_{\mathbb{R}^3}\int\limits_{\mathbb{R}^3}d^3 xd^3y \mathcal{G}^I(\vec{x},t)C_J^\prime(\vec{y},t)\Big(\{\mathcal{G}^J(\vec{y},t),\{C_I^\prime(\vec{x},t),f\}\}\\
    &+\{f,\{\mathcal{G}^J(\vec{y},t),C_I^\prime(\vec{x},t)\}\}\Big)+O(\delta^3,\kappa^2)\\
    =& \int\limits_{\mathbb{R}^3}\int\limits_{\mathbb{R}^3}d^3 xd^3y \mathcal{G}^I(\vec{x},t)C_J^\prime(\vec{y},t)\Big(\{\mathcal{G}^J(\vec{y},t),\{C_I^\prime(\vec{x},t),f\}\}\\
    &+\frac{1}{\kappa}\{f,\delta_I^J\delta(\vec{y}-\vec{x})\}\}\Big)+O(\delta^3,\kappa^2)\\
    =&\int\limits_{\mathbb{R}^3}\int\limits_{\mathbb{R}^3}d^3 xd^3yC_J^\prime(\vec{y},t)\{\mathcal{G}^J(\vec{y},t),\mathcal{G}^I(\vec{x},t)\{C_I^\prime(\vec{x},t),f\}\}\\
    &+O(\delta^3,\kappa^2)\\
    =&\int\limits_{\mathbb{R}^3}d^3y C_J^\prime(\vec{y},t)\{\mathcal{G}^J(\vec{y},t),\left(\chi_{C_I(\beta_I)}[f]\right)\Big|_{\beta_I=\mathcal{G}_I}\}+O(\delta^3,\kappa^2)\\
    =&\sum_{I,J}\left(\chi_{\mathcal{G}^J(\gamma^J)}\left(\chi_{C_I^\prime(\beta_I)}[f]\right)\Big|_{\beta^I=\mathcal{G}^I}\right)\Big|_{\gamma_J=C_J^\prime}+O(\delta^3,\kappa^2).
    \end{split}
\end{align}
In the first and second lines, we used the definition of the Hamiltonian vector fields \eqref{eq: Hamitlonian vector fields for dual and observable map}. In the third line, we used the Jacobi identity for the Poisson bracket. In the second and fourth line, we used \eqref{eq: poisson algebra of clocks and constraints for appendix proof}. In the fifth and sixth lines, we used again \eqref{eq: poisson algebra of clocks and constraints for appendix proof}, as well as the fact that the Poisson bracket of every phase space function with a constant vanishes. In the last two lines, we used again the definition of the Hamiltonian vector fields. From the end result, one can see that the order in which we apply $\left(\chi_{C_I^\prime(\beta_I)}[f]\right)\Big|_{\beta^I=\mathcal{G}^I}$ and $\left(\chi_{\mathcal{G}^J(\gamma^J)}[f]\right)\Big|_{\gamma_J=C_J^\prime}$ does not matter. As discussed above, it directly follows from this that the observable map and its dual commute, in the perturbation order we are interested in.
~\\
~\\
Again, the last step is to discuss this for the vacuum dual observable map \eqref{eq: vacuum dual map}. If we apply the observable map corresponding to the full set of constraints and clocks to a function with only geometric degrees of freedom, only the geometric part of the constraints can contribute to the Poisson bracket. Since the corresponding reference fields also consist solely of geometric degrees of freedom, the result will also consist only of geometric degrees of freedom. This means that in this setup, the electromagnetic part of the constraints $\delta C^{\textrm{ph}},\:\delta^2 C^{\textrm{ph}},\:\delta C_a^{\textrm{ph}},\:G^{\textrm{U(1)}}$ never contributes. From this, when acting on geometric degrees of freedom only, we could also build the observable map without loss of generality from the vacuum constraints. Therefore both maps act, as working in the vacuum. With this, it is easy to see that the order in which we apply the vacuum dual observable map and full observable map does not matter, if we apply them to a function consisting only of geometric degrees of freedom.
\section{Fourier transformation}\label{app:Fourier space}
In this section we will briefly introduce the Fourier transformation and the chosen Fourier basis used in the main text. This section follows the presentation in \cite{Ashtekar:1991mz}. First, we define the Fourier transformation for a generic phase space function $f$
\begin{align}\label{Fourier Trafo}
    f(\vec{x},t)=\int_\mathbb{R}\frac{d^3k}{(2\pi)^\frac{3}{2}}f(\vec{k},t) e^{i\vec{k}\vec{x}},\:\:\:\:\:\:f(\vec{k},t)=\int_\mathbb{R}\frac{d^3x}{(2\pi)^\frac{3}{2}}f(\vec{x},t) e^{-i\vec{k}\vec{x}},
\end{align}
where $\vec{k}$ is the wave vector.
~\\
~\\
The Fourier basis is constructed as an orthonormal basis from the following three vectors
\begin{align}\label{Fourier basis}
    \hat{k}^a\coloneqq\frac{k^a}{||\vec{k}||},\:\:\:\:m^a(\vec{k}),\:\:\:\overline{m}^a(\vec{k}),
\end{align}
where we interpret the plane perpendicular to $\vec{k}$ as the complex plane spanned by the vectors $m^a(\vec{k})$ and $\overline{m}^a(\vec{k})$. The overbar denotes the complex conjugation. The norm and metric of the real vector space are preserved in the complex one. Thus, $m_a$ is the ordinary transpose of $m^a$ without complex conjugation. The orthonormality of this basis is given by
\begin{align}\label{orthonormality Fourier basis}
    \begin{split}
    \hat{k}^a\hat{k}_a &=m^a(\vec{k})\overline{m}_a(\vec{k})=\overline{m}^a (\vec{k}) m_a(\vec{k})=1,\\
    \hat{k}^a m_a(\vec{k}) &=\hat{k}^a\overline{m}_a(\vec{k})=m^a (\vec{k}) m_a(\vec{k})=\overline{m}^a (\vec{k}) \overline{m}_a(\vec{k})=0.
    \end{split}
\end{align}
We also fix the orientation of the basis vectors such that
\begin{align}\label{orientation Fourir Basis}
    \begin{split}
    \epsilon_{abc}\hat{k}^a m^b(\vec{k})\overline{m}^c(\vec{k})&=-i,\\
    m_a(\vec{k})&=\overline{m}_a(-\vec{k}).
    \end{split}
\end{align}
Using this, we can expand any Fourier transformed tensor field in this orthonormal basis in momentum space. $\delta \mathcal{A}^i_a,\:\delta \mathcal{E}^a_i,\:A_a,\:-E^a$ read in this basis as follows \cite{Ashtekar:1991mz,Fahn:2022zql}
\begin{align}\label{Fourier expanded phase space variables}
    \begin{split}
    \delta\mathcal{E}_i^a(\vec{k})\coloneqq&\delta\mathcal{E}^+(\vec{k})m^a(\vec{k})m_i(\vec{k})+\delta\mathcal{E}^-(\vec{k}) \overline{m}^a(\vec{k})\overline{m}_i(\vec{k})+\delta\mathcal{E}^1(\vec{k}) \hat{k}^a(\vec{k})m_i(\vec{k})+\delta\mathcal{E}^{\overline{1}}(\vec{k}) \hat{k}^a\overline{m}_i(\vec{k})\\
    &+\delta\mathcal{E}^2(\vec{k}) m^a(\vec{k})\hat{k}_i+\delta\mathcal{E}^{\overline{2}}(\vec{k}) \overline{m}^a(\vec{k})\hat{k}_i+\delta\mathcal{E}^3(\vec{k}) \hat{k}^a \hat{k}_i+\delta\mathcal{E}^4(\vec{k})m^a(\vec{k})\overline{m}_i(\vec{k})\\
    &+\delta\mathcal{E}^5(\vec{k})\overline{m}^a(\vec{k})m_i(\vec{k}),\\
    \delta\mathcal{A}^i_a(\vec{k})\coloneqq&\delta\mathcal{A}^+(\vec{k})m_a(\vec{k})m^i(\vec{k})+\delta\mathcal{A}^-(\vec{k}) \overline{m}_a(\vec{k})\overline{m}^i(\vec{k})+\delta\mathcal{A}^1(\vec{k}) \hat{k}_a(\vec{k})m^i(\vec{k})+\delta\mathcal{A}^{\overline{1}}(\vec{k}) \hat{k}_a\overline{m}^i(\vec{k})\\
    &+\delta\mathcal{A}^2(\vec{k}) m_a(\vec{k})\hat{k}^i+\delta\mathcal{A}^{\overline{2}}(\vec{k}) \overline{m}_a(\vec{k})\hat{k}^i+\delta\mathcal{A}^3(\vec{k}) \hat{k}_a \hat{k}^i+\delta\mathcal{A}^4(\vec{k})m_a(\vec{k})\overline{m}^i(\vec{k})\\
    &+\delta\mathcal{A}^5(\vec{k})\overline{m}_a(\vec{k})m^i(\vec{k}),\\
    A_a(\vec{k},t)\coloneqq& A^+(\vec{k},t)m_a(\vec{k})+A^-(\vec{k},t)\overline{m}_a(\vec{k})+A^1(\vec{k},t)\hat{k}_a,\\
    -E^a(\vec{k},t)\coloneqq&-E^+(\vec{k},t)m^a(\vec{k})-E^-(\vec{k},t)\overline{m}^a(\vec{k})-E^1(\vec{k},t)\hat{k}^a.\\
    \end{split}
\end{align}
Before we continue, we investigate the Poisson bracket of $\delta\mathcal{A}^i_a(\vec{k}),\:\delta\mathcal{E}^a_i(\vec{k})$. For this we just insert the back transformation \eqref{Fourier Trafo} and use the Poisson algebra of the position space
\begin{align}\label{Geometrical poisson algebra in Fourier space}
    \begin{split}
    \{\delta \mathcal{A}^i_a(\vec{k},t),\delta\mathcal{E}_j^b(\vec{p},t)\}=&\int \frac{d^3 x}{(2\pi)^{\frac{3}{2}}}\int \frac{d^3 y}{(2\pi)^{\frac{3}{2}}}e^{-i(\vec{k}\cdot\vec{x}+\vec{p}\cdot\vec{y})}\{\delta\mathcal{A}^i_a(\vec{x},t),\delta\mathcal{E}^b_j(\vec{y},t)\}\\
    =&\int \frac{d^3 x}{(2\pi)^{\frac{3}{2}}}\int \frac{d^3 y}{(2\pi)^{\frac{3}{2}}}e^{-i(\vec{k}\cdot\vec{x}+\vec{p}\cdot\vec{y})}\frac{\beta}{\kappa}\delta^i_j\delta_a^b\delta(\vec{x}-\vec{y})\\
    =&\int \frac{d^3 x}{(2\pi)^{\frac{3}{2}}}e^{-i\vec{x}\cdot(\vec{k}+\vec{p})}\frac{\beta}{\kappa}\delta^i_j\delta_a^b\\
    =&\frac{\beta}{\kappa}\delta^i_j\delta^b_a\delta(\vec{k}+\vec{p}),\\
    \{A_a(\vec{k},t),-E^b(\vec{p},t)\}=&\delta_b^a\delta(\vec{k}+\vec{p}),
    \end{split}
\end{align}
where the Poisson bracket for the electromagnetic variables can be derived analogously. For the vanishing Poisson brackets in position space the results also vanish in Fourier space.
~\\
~\\
With this result, the Poisson algebra for the individual functions in \eqref{Fourier expanded phase space variables} can be calculated 
\begin{align}\label{Algebra polarisation functions}
    \begin{split}
    \{\delta\mathcal{A}^r(\vec{k},t),\delta\mathcal{E}^u(\vec{p},t)\}=&\frac{\beta}{\kappa}\delta^{ru}\delta(\vec{k}+\vec{p})\:\:\:\:\:\:r,u\in\{+,-,3,4,5\},\\
    \{\delta\mathcal{A}^m(\vec{k},t),\delta\mathcal{E}^n(\vec{p},t)\}=&-\frac{\beta}{\kappa}\delta^{mn}\delta(\vec{k}+\vec{p})\:\:\:\:\:\:m,n\in\{1,\overline{1},2,\overline{2}\},\\
    \{A^r(\vec{k},t),-E^u(\vec{p},t)\}=&\delta^{ru}\delta(\vec{k}+\vec{p})\:\:\:\:\:\:r,u\in\{+,-\},\\
    \{A^1(\vec{k},t),-E^1(\vec{p},t)\}=&-\delta(\vec{k}+\vec{p}),\\
    \end{split}
\end{align}
all other combinations vanish.
\section{Projectors in position and Fourier space}\label{app: projectors}
In this section we will give more details on the projectors, in position and Fourier space, used in the main text. For further details on these projectors see e.g. \cite{Dittrich:2006ee}. We will start with the projector onto the transverse subspace, which is defined for an arbitrary vector field $X^b$ as
\begin{align}
    P^a_b X^b \coloneqq \delta^a_bX^b+\partial^a\partial_b\left(X^b\ast G^\Delta\right).
\end{align}
It acts by projecting a variable onto the transverse subspace with respect to the open index. This can be calculated using the Helmholtz decomposition $X^b = X_\perp^b + X_L^{\:\:,b}$, where $X_L^{\:\:,b}$ represents the longitudinal part and $X_\perp^b$ denotes the transverse part, with $\partial_a X^a_\perp=0$. Applying the projector to $X^b$ yields
\begin{align}
    \begin{split}
    P_b^aX^b=&\delta_b^aX^b+\partial_b\partial^a\left(\left(X_\perp^b+X_L^{\:\:,b}\right)\ast G^\Delta\right)\\
    =&X^b+\partial_b\partial^a\partial^b\left(X_L\ast G^\Delta\right)\\
    =&X^a-X_L^{\:\:,a}=X_\perp^a.
    \end{split}
\end{align}
From this, along with the fact that the projector is symmetric, the following identities can be derived
\begin{align}\label{identities transverse projector}
    \partial_a P^a_b=&0=\partial^bP_b^a,\\
    \epsilon_a^{\:\:bc}P^a_b=&0,\\
    P^a_a X^b=&2X^b.
\end{align}
In this work, we will utilise an abuse of notation by also writing the projectors with internal indices. These projectors should be interpreted as follows
\begin{align}
    P_j^i X^b\delta_b^j \coloneqq \delta_j^iX^b\delta_b^j+\delta^a_j\delta_b^i\partial_a\partial^b\left(X^b\delta_b^j\ast G^\Delta\right).
\end{align}
Utilising the transverse projector, we can construct the projector onto the symmetric transverse traceless subspace
\begin{align}
    \begin{split}
    P^{ib}_{aj} X=&\frac{1}{2}\left(P^b_a P^i_j+P_{aj}P^{ib}-P_a^i P^b_j\right)X(\vec{x},t)\\
    =&\frac{1}{2}\Big(\left[\delta_{aj}\delta^{bi}+\delta_a^b\delta_j^i-\delta_a^i\delta_j^b\right]X(\vec{x},t)+\partial_a\partial^i\partial_j\partial^b\left(X\ast G^{\Delta\Delta}\right)(\vec{x},t)\\
    &+\left[\delta_{aj}\partial^i\partial^b+\delta^{ib}\partial_a\partial_j+\delta_a^b\partial^i\partial_j+\delta^i_j\partial_a\partial^b-\delta_a^i\partial_j\partial^b-\delta_j^b\partial_a\partial^i\right]\left(X\ast G^{\Delta}\right)(\vec{x},t)\Big).
    \end{split}
\end{align}
For simplicity, in this work, we mostly use the formulation with the transverse projector. Given the identities of the transverse projector, the following identities are straightforward to prove
\begin{align}\label{eq: properties sst projector}
    \begin{split}
    \partial_a P_{ib}^{aj}=&0 \textrm{\:\:\:\:and equally for all other contractions with partial derivatives},\\
    P^{aj}_{ab}=&0=P^{aj}_{ij},\\
    P^{aj}_{jb}=& P^a_b,\\
    P^{aj}_{ia}=& P^i_j,\\
    \epsilon_{a}^{\:\:ic}P^{aj}_{ib}=&0,\\
    \epsilon_{\:\:j}^{b\:\:c}P^{aj}_{ib}=&0.
    \end{split}
\end{align}
The final step is to formulate the projectors in Fourier space. For this purpose, we will utilise the basis introduced in appendix \ref{app:Fourier space}, as well as the fact that $G^\Delta(\vec{k})=\frac{1}{||\vec{k}||^2}$. This yields the following expressions
\begin{align}
    P^a_b(\vec{k})&=m^a(\vec{k})\overline{m}_b(\vec{k})+\overline{m}^a(\vec{k}) m_b(\vec{k})=\delta^a_b-\hat{k}^a\hat{k}_b,\\
    P_{ib}^{aj}(\vec{k})&\coloneqq\overline{m}^a(\vec{k})\overline{m}_i(\vec{k})m^j(\vec{k})m_b(\vec{k})+m^a(\vec{k})m_i(\vec{k})\overline{m}^j(\vec{k})\overline{m}_b(\vec{k}).
\end{align}
\section{Application of the observable map and its dual}\label{app:action dual and observable map}
In the following, we will briefly discuss the key points regarding the application of the observable map \eqref{eq:linearised observable map}, the dual observable map \eqref{eq:linearised dual observable map} and the vacuum dual observable map \eqref{eq: vacuum dual map}. For our purposes in this work, we solely require the detailed expression for the first order. Therefore, we will not calculate the explicit form of the second one. Also, to be more general, we will not insert the choice for the gauge fixing parameters discussed in section \ref{sec: Coordinate gauge fixing}, but leave them general.
\subsection{Application of the vacuum dual observable map}\label{app: action vacuum dual map}
We will start with the application of the vacuum dual observable map \eqref{eq: vacuum dual map} onto $\delta\mathcal{A}_a^i,\:\delta\mathcal{E}_i^a,\:A_a,\:E^a$. Since the vacuum dual observable map is purely geometric, its action on $A_a,-E^a$ is trivial. For the geometric variables, we get the following
\begin{align}
    \begin{split}
    \mathcal{O}^{\textrm{dual, vac}(1)}_{\delta\mathcal{A}_a^i,\{\mathcal{C}_I^\prime\}}(\vec{x},t)
    =&\delta\mathcal{A}_a^i(\vec{x},t)+\frac{1}{2\kappa}\epsilon_a^{\:\:cb}\delta_c^i\partial_b\left(\delta C^{\textrm{geo}}\ast G^\Delta\right)(\vec{x},t)\\
    &-\frac{\beta}{\kappa}\left(\delta_a ^g\delta_b^i\partial^b-\frac{1}{2}\delta_a^i\partial^g\right)\left(\delta C_g^{\textrm{geo}}\ast G^\Delta\right)(\vec{x},t)\\
    &-\frac{\beta}{2\kappa}\delta^i_b\partial^g\partial^b\partial_a\left(\delta C_g^{\textrm{geo}}\ast G^{\Delta\Delta}\right)(\vec{x},t)\\
    =&\frac{1}{2}\left(P_{la}P^{di}+\delta^d_l(\delta^i_a-P^i_a)+\delta_l^i(\delta_a^d-P^d_a)+\delta_a^dP_l^i-\delta_a^iP_l^d\right)\delta\mathcal{A}_d^l(\vec{x},t),
    \end{split}\\
    \begin{split}
    \mathcal{O}^{\textrm{dual, vac}(1)}_{\delta\mathcal{E}^a_i,\{\mathcal{C}_I^\prime\}}(\vec{x},t)=&\delta\mathcal{E}^a_i(\vec{x},t)+\frac{2\beta}{\kappa}\partial^a\left(\delta G_i^{\textrm{geo}}\ast G^{\Delta}\right)(\vec{x},t)+\frac{1}{\kappa}\delta^a_i\left(\delta C^{\textrm{geo}}\ast G^\Delta\right)(\vec{x},t)\\
    &+\frac{\beta}{\kappa}\delta^{gj}\delta^a_k\epsilon_{ji}^{\:\:\:\:k}\left(\delta C_g^{\textrm{geo}}\ast G^\Delta\right)(\vec{x},t)\\
    =& P_d^a \delta\mathcal{E}^d_i(\vec{x},t)\\
    &+\left(\delta_i^a\delta^c_j\delta_k^d\epsilon_l^{\:\:jk}\partial_c+\delta_k^d\epsilon_{il}^{\:\:\:\:k}\partial^a+\delta^{cj}\delta_k^a\delta_l^d\epsilon_{ji}^{\:\:\:\:k}\partial_c+\delta^{dj}\delta^a_k\delta^c_l\epsilon_{ij}^{\:\:\:\:k}\partial_c\right)\\
    &\cdot\left(\delta\mathcal{A}_d^l\ast G^{\Delta}\right)(\vec{x},t).
    \end{split}
\end{align}
The expressions after the first equal signs arise from calculating the Poisson brackets of the linearised Ashtekar-Barbero variables with the reference fields, where we did not incorporate the exact expressions of the geometric constraints. This is done in the second steps, along with some rewriting utilising the previously introduced projectors. The combination of projectors
\begin{align}
    \frac{1}{2}\left(P_{la}P^{di}+\delta^d_l(\delta^i_a-P^i_a)+\delta_l^i(\delta_a^d-P^d_a)+\delta_a^dP_l^i-\delta_a^iP_l^d\right),
\end{align}
projects onto the 2 symmetric transverse traceless plus the 3 longitudinal degrees of freedom with respect to the spatial index a of the linearised Ashtekar-Barbero connection \cite{Dittrich:2006ee}. The linearised densitised triad is projected by the dual observable map onto the transverse subspace with respect to the spatial index, along with a term that is linear in the linearised Ashtekar-Barbero connection.
\subsection{Application of the dual observable map}\label{app: action dual map}
Next, we investigate the application of the dual observable map \eqref{eq:linearised dual observable map} onto $\delta\mathcal{A}_a^i,\:\delta\mathcal{E}^a_i,\:A_a,\:E^a$. The electromagnetic variables were already discussed in the main text. For the geometric variables we get the following
\begin{align}
    \begin{split}
    \mathcal{O}^{\textrm{dual}(1)}_{\delta\mathcal{A}_a^i,\{\mathcal{C}_I^\prime\}}(\vec{x},t)
    =&\delta\mathcal{A}_a^i(\vec{x},t)+\frac{1}{2\kappa}\epsilon_a^{\:\:cb}\delta_c^i\partial_b\left(\delta C^{\textrm{geo}}\ast G^\Delta\right)(\vec{x},t)\\
    &-\frac{\beta}{\kappa}\left(\delta_a ^g\delta_b^i\partial^b-\frac{1}{2}\delta_a^i\partial^g\right)\left(\delta C_g^{\textrm{geo}}\ast G^\Delta\right)(\vec{x},t)\\
    &-\frac{\beta}{2\kappa}\delta^i_b\partial^g\partial^b\partial_a\left(\delta C_g^{\textrm{geo}}\ast G^{\Delta\Delta}\right)(\vec{x},t)\\
    &+\frac{1}{2\kappa}\epsilon_a^{\:\:cb}\delta_c^i\partial_b\left(\delta C^{\textrm{ph}'}\ast G^\Delta\right)(\vec{x},t)\\
    &-\frac{\beta}{\kappa}\left(\delta_a ^g\delta_b^i\partial^b-\frac{1}{2}\delta_a^i\partial^g\right)\left(\delta C_g^{\textrm{ph}'}\ast G^\Delta\right)(\vec{x},t)\\
    &-\frac{\beta}{2\kappa}\delta^i_b\partial^g\partial^b\partial_a\left(\delta C_g^{\textrm{ph}'}\ast G^{\Delta\Delta}\right)(\vec{x},t)\\
    =&\frac{1}{2}\left(P_{la}P^{di}+\delta^d_l(\delta^i_a-P^i_a)+\delta_l^i(\delta_a^d-P^d_a)+\delta_a^dP_l^i-\delta_a^iP_l^d\right)\delta\mathcal{A}_d^l(\vec{x},t)\\
    &+\frac{1}{2}\epsilon_a^{\:\:cb}\delta_c^i\partial_b\left(T^{00}[-P_b^aE^b,A_a]\ast G^\Delta\right)(\vec{x},t)\\
    &-\beta\left(\delta_a ^g\delta_b^i\partial^b-\frac{1}{2}\delta_a^i\partial^g\right)\left(T_{0g}[-P_b^aE^b,A_a]\ast G^\Delta\right)(\vec{x},t)\\
    &-\frac{\beta}{2}\delta^i_b\partial^g\partial^b\partial_a\left(T_{0g}[-P_b^aE^b,A_a]\ast G^{\Delta\Delta}\right)(\vec{x},t),
    \end{split}\\
    \begin{split}
    \mathcal{O}^{\textrm{dual}(1)}_{\delta\mathcal{E}^a_i,\{\mathcal{C}_I^\prime\}}(\vec{x},t)=&\delta\mathcal{E}^a_i(\vec{x},t)+\frac{2\beta}{\kappa}\partial^a\left(\delta G_i^{\textrm{geo}}\ast G^{\Delta}\right)(\vec{x},t)+\frac{1}{\kappa}\delta^a_i\left(\delta C^{\textrm{geo}}\ast G^\Delta\right)(\vec{x},t)\\
    &+\frac{\beta}{\kappa}\delta^{gj}\delta^a_k\epsilon_{ji}^{\:\:\:\:k}\left(\delta C_g^{\textrm{geo}}\ast G^\Delta\right)(\vec{x},t)\\
    &+\frac{1}{\kappa}\delta^a_i\left(\delta C^{\textrm{ph}'}\ast G^\Delta\right)(\vec{x},t)+\frac{\beta}{\kappa}\delta^{gj}\delta^a_k\epsilon_{ji}^{\:\:\:\:k}\left(\delta C_g^{\textrm{ph}'}\ast G^\Delta\right)(\vec{x},t)\\
    =& P_d^a \delta\mathcal{E}^d_i(\vec{x},t)\\
    &+\left(\delta_i^a\delta^c_j\delta_k^d\epsilon_l^{\:\:jk}\partial_c+\delta_k^d\epsilon_{il}^{\:\:\:\:k}\partial^a+\delta^{cj}\delta_k^a\delta_l^d\epsilon_{ji}^{\:\:\:\:k}\partial_c+\delta^{dj}\delta^a_k\delta^c_l\epsilon_{ij}^{\:\:\:\:k}\partial_c\right)\\
    &\cdot\left(\delta\mathcal{A}_d^l\ast G^{\Delta}\right)(\vec{x},t)+\delta^a_i\left(T^{00}[-P_b^aE^b,A_a]\ast G^\Delta\right)(\vec{x},t)\\
    &+\beta\delta^{gj}\delta_k^a\epsilon_{ji}^{\:\:\:\:k}\left(T_{0g}[-P_b^aE^b,A_a]\ast G^\Delta\right)(\vec{x},t).
    \end{split}
\end{align}
Again, the expressions after the first equal signs arise from calculating the Poisson bracket of the linearised Ashtekar-Barbero variables with the reference fields, where we did not incorporate the exact expressions of the constraints and separated them into their geometric and electromagnetic components. In the following step, we inserted these expressions and simplified the results. For the geometric part we get the same result as for the vacuum dual observable map in the previous section. Furthermore, for both expressions, we obtain contributions depending on the components of the electromagnetic energy momentum tensor. These contributions are caused by the electromagnetic components of the Hamiltonian and diffeomorphism constraints.
\subsection{Application of the observable map}\label{app: action observable map}
Next we look at the application of the observable map \eqref{eq:linearised observable map} on $\delta\mathcal{A}_a^i,\:\delta\mathcal{E}_i^a,A_a,-E^a$, which is given by
\begin{align}
    \begin{split}
    \mathcal{O}_{A_a,\{\mathcal{G}^I\}}(\vec{x},t)=& A_a (\vec{x},t)+\partial_a\left(T^{\textrm{U(1)}}-\gamma\right)(\vec{x},t)+\kappa^2P_{da}\left(\left(\delta T-\tau\right) P^d_c E^c\right)(\vec{x},t)\\
    &+\kappa^2P^c_a\left(\left(\delta T^g-\sigma^g\right) A_{[g,c]}\right)(\vec{x},t)+\mathcal{O}^{(2)}_{A_a,\{\mathcal{G}^I\}}(\vec{x},t)\\
    =&P^b_a A_b(\vec{x},t)-\partial_a\gamma(\vec{x},t)+\kappa^2P_{da}\left(\left(\delta T-\tau\right) P^d_c E^c\right)(\vec{x},t)\\
    &+\kappa^2P^c_a\left(\left(\delta T^g-\sigma^g\right) A_{[g,c]}\right)(\vec{x},t)
    +\mathcal{O}^{(2)}_{A_a,\{\mathcal{G}^I\}}(\vec{x},t),
    \end{split}\\
    \begin{split}
    \mathcal{O}_{-E^a,\{\mathcal{G}^I\}}(\vec{x},t)=&-E^a (\vec{x},t)-\frac{\kappa^2}{2}\delta^{a[c}\partial^{b]}\left(\left(\delta T-\tau\right) A_{[c,b]}\right)(\vec{x},t)\\
    &-\kappa^2\delta^a_{[g}\partial_{c]}\left(\left(\delta T^g-\sigma^g\right) P^c_b E^b\right)(\vec{x},t)+\mathcal{O}^{(2)}_{-E^a,\{\mathcal{G}^I\}}(\vec{x},t),
    \end{split}\\
    \begin{split}
    \mathcal{O}_{\delta\mathcal{A}_a^i,\{\delta\mathcal{G}^I\}}(\vec{x},t)=&\delta\mathcal{A}_a^i(\vec{x},t)+\frac{\kappa}{2}\partial_a\left(\delta\Xi^i-\xi^i\right)(\vec{x},t)+\mathcal{O}^{(2)}_{\delta A_a^i,\{\mathcal{G}^I\}}(\vec{x},t)\\
    =& P^g_a \delta\mathcal{A}_g^i(\vec{x},t)-\frac{1}{2}\xi^i_{\:\:,a}(\vec{x},t) +\mathcal{O}^{(2)}_{\delta\mathcal{A}^i_a,\{\mathcal{G}^I\}}(\vec{x},t),
    \end{split}\\
    \begin{split}
    \mathcal{O}_{\delta\mathcal{E}^a_i,\{\delta\mathcal{G}^I\}}(\vec{x},t)= &\delta\mathcal{E}^a_i(\vec{x},t)+\frac{\kappa}{2}\epsilon_{ji}^{\:\:\:\:k}\delta_k^a\left(\delta\Xi^j-\xi^j\right)(\vec{x},t)-\beta\kappa\epsilon_{i}^{\:\:lk}\delta_l^d\delta_k^a\partial_d\left(\delta T-\tau\right)(\vec{x},t)\\
    &-\kappa\delta_{i}^d\delta^a_{[d}\partial_{g]}\left(\delta T^g-\sigma^g\right) (\vec{x},t)+\mathcal{O}^{(2)}_{\delta\mathcal{E}_i^a,\{\mathcal{G}^I\}}(\vec{x},t)\\
    =&\frac{1}{2}\left(P_{id}P^{al}+\delta_i^l(\delta_d^a-P_d^a)+\delta_i^a(\delta_d^l-P_d^l)+\delta_d^aP_i^l-\delta_d^lP^a_i\right)\delta\mathcal{E}_l^d(\vec{x},t)\\
    &+\left(\delta_l^d\delta_j^c\delta_k^a\epsilon_i^{\:\:jk}\partial_c+\delta_k^a\epsilon_{li}^{\:\:\:\:k}\partial^d+\delta_i^a\delta^{jc}\delta_k^d\epsilon_{jl}^{\:\:\:\:k}\partial_c+\delta_i^c\delta^{aj}\delta_k^d\epsilon_{lj}^{\:\:\:\:k}\partial_c\right)\left(\delta\mathcal{A}_d^l\ast G^\Delta\right)(\vec{x},t)\\
    &-\frac{1}{2}\epsilon_{ji}^{\:\:\:\:k}\delta_k^a\delta\xi^j(\vec{x},t)+\beta\epsilon_{i}^{\:\:lk}\delta_l^d\delta_k^a\partial_d \tau(\vec{x},t)+\delta_{i}^d\delta^a_{[d}\partial_{g]}\sigma^g (\vec{x},t)\\
    &+\mathcal{O}^{(2)}_{\delta\mathcal{E}^a_i,\{\mathcal{G}^I\}}(\vec{x},t).
    \end{split}
\end{align}
Again, we applied the observable map only up to first order, as we do not require the explicit form of the second order. Following the first equal sign, we calculated the Poisson bracket of the elementary variable with the first order primed constraints. Subsequently, we inserted the clocks and rewrote the term utilising the projectors introduced earlier. Regarding the electromagnetic variables, we did not insert the explicit form of the geometrical reference fields.
~\\
~\\
For $A_a$, the observable map projects onto the transverse subspace, with additional contributions that are linear in $\kappa$, along with the geometrical clocks multiplied by the transverse parts of $A_a$ and $E^a$. Regarding $-E^a$, the observable map acts trivially in the electromagnetic sector since $\{G^{\textrm{U(1)}},E^a\}=0$, but adds additional terms, in the linearised phase space, being linear in the reference fields multiplied by the transverse parts of $A_a,-E^a$.
~\\
~\\
Concerning the Ashtekar-Barbero connection, the observable map acts as the projector onto the transverse subspace, and an additional term given by the spatial derivative of the coordinate gauge-fixing term $\xi$, which in our case is chosen to be zero.
~\\
~\\
Regarding $\mathcal{O}^{\textrm{dual}(1)}_{\delta\mathcal{E}^a_i, \mathcal{G}^I}$, we obtain the projector
\begin{align}
    \frac{1}{2}\left(P_{id}P^{al}+\delta_i^l(\delta_d^a-P_d^a)+\delta_i^a(\delta_d^l-P_d^l)+\delta_d^aP_i^l-\delta_d^lP^a_i\right),
\end{align}
on the linearised densitised triad, which is similar to the projector for $\mathcal{O}^{\textrm{dual}(1)}_{\delta\mathcal{A}_a^i, \mathcal{G}^I}$, with only the indices being switched. Thus, it projects $\delta\mathcal{E}_i^a$ onto the 2 symmetric transverse traceless plus the 3 longitudinal degrees of freedom with respect to the spatial index $d$. Additionally, we obtain a term depending on the linearised Ashtekar-Barbero connection and terms depending on the coordinate gauge fixing parameters.
\section{Construction of Dirac observables}\label{app: More Details on the construction of the Dirac observables}
In this section we will give more details on the derivation of the Dirac observables introduced in the main text \eqref{eq: sets of Dirac observables}. As discussed we construct them by first applying the (vacuum) dual observable map and then the observable map onto $\delta\mathcal{A}_a^i,\:\delta\mathcal{E}^a_i,\:A_a,\:-E^a$.
\subsection{Dirac observables: electromagnetic sector}\label{app: Dirac observables: electromagnetic sector}
Since for both, set $S$ and set $S^\prime$, the electromagnetic Dirac observables $(A_a)^{\textrm{phys}},(-E^a)^{\textrm{phys}}$ are the same, we will start with them. Utilising the results of appendix \ref{app:action dual and observable map} they read
\begin{align}
    \begin{split}
        (A_a)^{\textrm{phys}}(\vec{x},t)\coloneqq &\mathcal{O}_{\mathcal{O}^{\textrm{dual}}_{A_a,\{C_I^\prime\}},\{\mathcal{G}^I\}}=\mathcal{O}_{A_a,\{\mathcal{G}^I\}}\\
        =&P^b_a A_b (\vec{x},t)+\kappa^2P_{ca}\left(\left(\delta T-\tau\right) E^c\right)(\vec{x},t)+\kappa^2P^c_a\left(\left(\delta T^g-\sigma^g\right) A_{[g,c]}\right)(\vec{x},t)\\
        &+\partial_a\gamma(\vec{x},t)+\mathcal{O}^{(2)}_{A,\{\mathcal{G}^I\}}(\vec{x},t),
    \end{split}\\
    \begin{split}
        (-E^a)^{\textrm{phys}}(\vec{x},t)\coloneqq &\mathcal{O}_{\mathcal{O}^{\textrm{dual}}_{-E^a,\{C_I^\prime\}},\{\mathcal{G}^I\}}=\mathcal{O}_{-P_b^aE^b,\{\mathcal{G}^I\}}\\
        =& -P_b^aE^b-\frac{\kappa^2}{2}P^{a[c}\partial^{b]}\left(\left(\delta T-\tau\right) A_{[c,b]}\right)(\vec{x},t)-\kappa^2 P^a_{[g}\partial_{c]}\left(\left(\delta T^g-\sigma^g\right) E^c\right)(\vec{x},t)\\
        &+\mathcal{O}^{(2)}_{-P^a_b,\{\mathcal{G}^I\}}(\vec{x},t).
    \end{split}
\end{align}
For the zeroth order we obtain the transverse degrees of freedom of $A_a,\:-E^a$. However, for the first order, we acquire a combination of $A_a, E^a$, the transverse projector $P^a_b$ and the geometrical reference fields, arising from the electromagnetic contributions of the Hamiltonian and diffeomorphism constraints.
\subsection{Dirac observables: vacuum dual observable map and observable map}\label{app: Dirac observables: vacuum dual observable map and observable map}
Next we want to discuss the geometric Dirac observables $(\delta\mathcal{A}^i_a)^{\textrm{phys}},\: (\delta\mathcal{E}_i^a)^{\textrm{phys}}$. As discussed in the main text they are given by first applying the first order vacuum dual observable map \eqref{eq: vacuum dual map first order} and then the observable map \eqref{eq:linearised observable map} onto $\delta\mathcal{A}_a^i,\:\delta\mathcal{E}^a_i$. We will start with $\delta\mathcal{A}^i_a$
\begin{align}
    \begin{split}
        (\delta\mathcal{A}^i_a)^{\textrm{phys}}(\vec{x},t)\coloneqq &\mathcal{O}_{\mathcal{O}^{\textrm{dual, vac}(1)}_{\delta\mathcal{A}_a^i,\{C_I^\prime\}},\{\mathcal{G}^I\}}\\
        =& P_{al}^{id}\delta\mathcal{A}^l_a 
        -\frac{\kappa}{4}\left(\delta^d_l\partial_d(\delta_a^i-P^i_a)+\delta_l^i\partial_a+P_l^i\partial_a\right)\xi^l(\vec{x},t)\\
        &+\mathcal{O}^{(2)}_{\mathcal{O}^{\textrm{dual, vac}(1)}_{\delta\mathcal{A}_a^i,\{C_I^\prime\}},\{\mathcal{G}^I\}}(\vec{x},t).
    \end{split}
\end{align}
Here, we utilised the results of appendices \ref{app: action vacuum dual map} and \ref{app: action observable map}, as well as the fact that the first order observable map commutes with derivatives and therefore also with the projectors. Consequently, it acts solely on $\delta\mathcal{A}_a^i$, in the result of the $\delta\mathcal{A}_a^i$ under the application of the vacuum dual observable map. Also we used again \eqref{eq:dual and observable on phase space function}. Up to first order, the result consists of the symmetric transverse traceless degrees of freedom of $\delta\mathcal{A}_a^i$, as well as spatial derivatives of $\xi^l$.
~\\
~\\
For the linearised densitised triad, the calculation is more involved, as the results of the vacuum dual observable map and observable map acting on the linearised densitised triad are linear in the linearised Ashtekar-Barbero connection (see appendices \ref{app: action vacuum dual map} and \ref{app: action observable map})
\begin{align}
    \begin{split}
        (\delta\mathcal{E}_i^a)^{\textrm{phys}}(\vec{x},t)\coloneqq &\mathcal{O}_{\mathcal{O}^{\textrm{dual, vac}(1)}_{\delta\mathcal{E}_a^i,\{C_I^\prime\}},\{\mathcal{G}^I\}}=P_d^a \mathcal{O}_{\delta\mathcal{E}^d_i,\{\delta\mathcal{G}^I\}}(\vec{x},t)\\&+\left(\delta_i^a\delta^c_j\delta_k^d\epsilon_l^{\:\:jk}\partial_c+\delta_k^d\epsilon_{il}^{\:\:\:\:k}\partial^a+\delta^{cj}\delta_k^a\delta_l^d\epsilon_{ji}^{\:\:\:\:k}\partial_c+\delta^{dj}\delta^a_k\delta^c_l\epsilon_{ij}^{\:\:\:\:k}\partial_c\right)\left(\mathcal{O}_{\delta\mathcal{A}_d^l,\{\mathcal{G}^I\}}\ast G^{\Delta}\right)(\vec{x},t)\\
        &+\mathcal{O}^{(2)}_{\mathcal{O}^{\textrm{dual, vac}(1)}_{\delta\mathcal{E}^a_i,\{C_I^\prime\}},\{\mathcal{G}^I\}}(\vec{x},t)\\
        =&P^{aj}_{ib}\delta\mathcal{E}_j^b+\Omega^{ag}_{il}\left(\delta\mathcal{A}_g^l\ast G^\Delta\right)(\vec{x},t)\\
        &+\left(\delta_l^g\delta_j^c\delta_k^d\epsilon_i^{\:\:jk}\partial_c+\delta_k^d\epsilon_{li}^{\:\:\:\:k}\partial^g+\delta_i^d\delta^{jc}\delta_k^g\epsilon_{jl}^{\:\:\:\:k}\partial_c+\delta_i^c\delta^{dj}\delta_k^g\epsilon_{lj}^{\:\:\:\:k}\partial_c\right)\partial_d\left(\xi^l\ast G^\Delta\right)(\vec{x},t)\\ \nonumber
        &-\frac{1}{2}\epsilon_{ji}^{\:\:\:\:k}\delta^g_kP^a_g\xi^j(\vec{x},t)+\beta\epsilon_i^{\:\:lk}\delta_l^d\delta_k^a\partial_d\tau(\vec{x},t)+\delta_i^b\delta^a_{[b}\partial_{c]}\sigma^c(\vec{x},t)\\
        &+\mathcal{O}^{(2)}_{\mathcal{O}^{\textrm{dual}(1)}_{\delta\mathcal{E}^a_i,\{C_I^\prime\}},\{\mathcal{G}^I\}}(\vec{x},t).
    \end{split}
\end{align}
We obtain a structurally mostly similar expression to that of the linearised Ashtekar-Barbero connection. $\delta\mathcal{E}^a_i$ is mapped mapped onto its symmetric transverse traceless degrees of freedom, together with combinations of spatial derivatives of the coordinate gauge fixing parameters. However, in contrast to $(\delta\mathcal{A}_a^i)^{\textrm{phys}}$, there exists a contribution linear in the linearised Ashtekar-Barbero connection, with a prefactor given by
\begin{align}\label{eq: extra term in physical linearised densitised triad}
    \begin{split}
        \Omega_{il}^{ag}\left(\delta\mathcal{A}_g^l\ast G^{\Delta}\right)(\vec{x},t)\coloneqq \Big(&P_d^a\left(\delta_l^g\delta_j^c\delta_k^d\epsilon_i^{\:\:jk}\partial_c+\delta_k^d\epsilon_{li}^{\:\:\:\:k}\partial^g+\delta_i^d\delta^{jc}\delta_k^g\epsilon_{jl}^{\:\:\:\:k}\partial_c+\delta_i^c\delta^{dj}\delta_k^g\epsilon_{lj}^{\:\:\:\:k}\partial_c\right)\\
        &+\left(\delta_i^a\delta^c_j\delta_k^d\epsilon_l^{\:\:jk}\partial_c+\delta_k^d\epsilon_{il}^{\:\:\:\:k}\partial^a+\delta^{cj}\delta_k^a\delta_l^d\epsilon_{ji}^{\:\:\:\:k}\partial_c+\delta^{dj}\delta^a_k\delta^c_l\epsilon_{ij}^{\:\:\:\:k}\partial_c\right)P^g_d\Big)\\
        &\cdot\left(\delta\mathcal{A}_g^l\ast G^{\Delta}\right)(\vec{x},t).
    \end{split}
\end{align}
Inserting the projectors $P^a_d = \delta_d^a+\partial_d\partial^a,\:\:P^g_d = \delta_d^g+\partial_d\partial^g$ and with some rewriting $\Omega_{il}^{ag}$ can be simplified to
\begin{align}\label{eq: rewritten further operator on physical densitised triad}
    \Omega_{il}^{ag}=\epsilon_{li}^{\:\:\:\:k}\delta_{[k}^{a}\delta_{c]}^{g}\partial^{c}+\epsilon_b^{\:\:ag}\delta_l^{[b}\delta_i^{c]}\partial_c.
\end{align}
One can show the following identities for this operator
\begin{align}\label{eq: identities for Omega operator}
    \begin{split}
    \Omega_{il}^{ag}=&-\delta_k^l\delta_g^b\Omega_{ib}^{ak}\\
    \Omega_{il}^{ag}\delta_g^l =& 0\\ \:\:\Omega_{il}^{ag}\epsilon_{g}^{\:\:cb}\delta_c^l=&0\:\:\forall b\,.
    \end{split}
\end{align}
We now use that we can decompose $\delta\mathcal{A}^l_g$ in the following way \cite{arfken2013mathematical}
\begin{align}\label{eq: decompostion of perturbed ashtekar variable in position space}
    \delta\mathcal{A}^l_g = \frac{1}{3}\delta\mathcal{A}_{\textrm{scalar}}\delta_g^l+\frac{1}{2}\delta_j^g\epsilon^l_{\:\:jk}c^k+\left(\delta\mathcal{A}^l_g\right)^{\textrm{ST}},
\end{align}
where $\delta\mathcal{A}_{\textrm{scalar}}$ is a scalar,
\begin{align}
    c^k = \epsilon^k_{\:\:lj}\delta^{gj}\delta\mathcal{A}_g^l,
\end{align}
and the symmetric traceless tensor
\begin{align}
    \left(\delta\mathcal{A}^l_g\right)^{\textrm{ST}} = \frac{1}{2}\left(\delta\mathcal{A}_g^l+\delta\mathcal{A}_l^g\right)-\frac{1}{3}\delta\mathcal{A}_{\textrm{scalar}}\delta^l_g.
\end{align}
If we insert \eqref{eq: decompostion of perturbed ashtekar variable in position space} into \eqref{eq: extra term in physical linearised densitised triad} and use that $\Omega_{il}^{ag}$ as well as $\ast$ are linear operators, we find
\begin{align}
   \begin{split} \Omega_{il}^{ag}\left(\delta\mathcal{A}_g^l\ast G^{\Delta}\right) =& \frac{1}{3}\Omega_{il}^{ag}\delta_g^l\left(\delta\mathcal{A}_{\textrm{scalar}}\ast G^\Delta\right)+\frac{1}{2}\Omega_{il}^{ag}\delta_j^g\epsilon^l_{\:\:jk}\left(c^k\ast G^\Delta\right)+\Omega_{il}^{ag}\left(\left(\delta\mathcal{A}^l_g\right)^{\textrm{ST}}\ast G^\Delta\right)\\
    =& \frac{1}{2}\left(\Omega_{il}^{ag}\left(\delta\mathcal{A}_g^l\ast G^\Delta\right)+\Omega_{il}^{ag}\left(\delta\mathcal{A}_l^g \ast G^\Delta\right)\right)-\frac{1}{3}\Omega_{il}^{ag} \delta_g^l\left(\delta\mathcal{A}_{\textrm{scalar}}\ast G^\Delta\right)\\
    =& \frac{1}{2}\left(\Omega_{il}^{ag}\left(\delta\mathcal{A}_g^l\ast G^\Delta\right)+\Omega_{ig}^{al}\left(\delta\mathcal{A}_g^l \ast G^\Delta\right)\right)\\
    =& \frac{1}{2}\left(\Omega_{il}^{ag}\left(\delta\mathcal{A}_g^l\ast G^\Delta\right)-\Omega_{il}^{ag}\left(\delta\mathcal{A}_g^l \ast G^\Delta\right)\right)\\
    =& 0,
    \end{split}
\end{align}
where we used the identities in \eqref{eq: identities for Omega operator}. Hence, overall we get
\begin{align}
    \begin{split}
        (\delta\mathcal{E}_i^a)^{\textrm{phys}}(\vec{x},t)
        =&P^{aj}_{ib}\delta\mathcal{E}_j^b+\beta\epsilon_i^{\:\:lk}\delta_l^d\delta_k^a\partial_d\tau(\vec{x},t)+\delta_i^b\delta^a_{[b}\partial_{c]}\sigma^c(\vec{x},t)\\
        &+\mathcal{O}^{(2)}_{\mathcal{O}^{\textrm{dual}(1)}_{\delta\mathcal{E}^a_i,\{C_I^\prime\}},\{\mathcal{G}^I\}}(\vec{x},t).
    \end{split}
\end{align}
\subsection{Dirac observables: dual observable map and observable map}\label{app: Dirac observables: (full) dual observable map and observable map}
Next, we want to construct the Dirac observables $(\delta\mathcal{A}_a^i)^{\textrm{phys}\prime},\:(\delta\mathcal{E}_i^a)^{\textrm{phys}\prime}$ of the set $S^\prime$. Here, as discussed in the main text, first the first order dual observable map \eqref{eq:linearised dual observable map} and then the observable map is applied to $\delta\mathcal{A}_a^i,\:\delta\mathcal{E}^a_i$. For this we will use the results of appendices \ref{app: action dual map} and \ref{app: action observable map}. From these results we see that we already calculated the geometric contributions in the last section. What is left, is to calculate the application of the observable map on the energy-momentum tensor components, arising from the application of the dual observable map onto $\delta\mathcal{A}_a^i,\:\delta\mathcal{E}^a_i$. These contributions look all similar to
\begin{align}
    \left(T^{00}[-P_b^aE^b,A_a]\ast G^\Delta\right) = \left(T^{00}[\mathcal{O}^{\textrm{dual}}_{-E^a,\{C^\prime\}},\mathcal{O}^{\textrm{dual}}_{A_a,\{C^\prime\}}]\ast G^\Delta\right).
\end{align}
The task is now to apply the observable map to these expressions. Using \eqref{eq:dual and observable on phase space function} this can be easily evaluated
\begin{align}
    \begin{split}
        \mathcal{O}_{\left(T^{00}[\mathcal{O}^{\textrm{dual}}_{-E^a,\{C^\prime\}},\mathcal{O}^{\textrm{dual}}_{A_a,\{C^\prime\}}]\ast G^\Delta\right),\{\mathcal{G}^I\}}=& \left(T^{00}[\mathcal{O}_{\mathcal{O}^{\textrm{dual}}_{-E^a,\{C^\prime\}},\{\mathcal{G}^I\}},\mathcal{O}_{\mathcal{O}^{\textrm{dual}}_{A_a,\{C^\prime\}},\mathcal{G}^I\}}]\ast G^\Delta\right)\\
        =&\left(T^{00}[(-E^a)^{\textrm{phys}},(A_a)^{\textrm{phys}}]\ast G^\Delta\right).
    \end{split}
\end{align}
Hence, we get the energy density depending on the electromagnetic Dirac observables. Using this and the result of the previous section we get
\begin{align}
    \begin{split}
        (\delta\mathcal{A}^i_a)^{\textrm{phys}\prime}(\vec{x},t)\coloneqq &\mathcal{O}_{\mathcal{O}^{\textrm{dual}(1)}_{\delta\mathcal{A}_a^i,\{C_I^\prime\}},\{\mathcal{G}^I\}}\\
        =& P_{al}^{id}\delta\mathcal{A}^l_a + \frac{1}{2}\epsilon_a^{\:\:cb}\delta_c^i\partial_b\left(T^{00}\left((-E^b)^{\textrm{phys}},(A_a)^{\textrm{phys}}\right)\ast G^\Delta\right)(\vec{x},t)\\
        &-\beta\left(\delta_a ^g\delta_b^i\partial^b-\frac{1}{2}\delta_a^i\partial^g\right)\left(T_{0g}\left((-E^b)^{\textrm{phys}},(A_a)^{\textrm{phys}}\right)\ast G^\Delta\right)(\vec{x},t)\\
        &-\frac{\beta}{2}\delta^i_b\partial^g\partial^b\partial_a\left(T_{0g}\left((-E^b)^{\textrm{phys}},(A_a)^{\textrm{phys}}\right)\ast G^{\Delta\Delta}\right)(\vec{x},t)\\
        &-\frac{1}{4}\left(\delta^d_l\partial_d(\delta_a^i-P^i_a)+\delta_l^i\partial_a+P_l^i\partial_a\right)\xi^l(\vec{x},t)+\mathcal{O}^{(2)}_{\mathcal{O}^{\textrm{dual}(1)}_{\delta\mathcal{A}_a^i,\{C_I^\prime\}},\{\mathcal{G}^I\}}(\vec{x},t).
    \end{split}\\
    \begin{split}
        (\delta\mathcal{E}_i^a)^{\textrm{phys}\prime}(\vec{x},t)\coloneqq &\mathcal{O}_{\mathcal{O}^{\textrm{dual}(1)}_{\delta\mathcal{E}_a^i,\{C_I^\prime\}},\{\mathcal{G}^I\}}\\
        =& P^{aj}_{ib}\delta\mathcal{E}_j^b
        +\delta^a_i\left(T^{00}\left((-E^b)^{\textrm{phys}},(A_a)^{\textrm{phys}}\right)\ast G^\Delta\right)(\vec{x},t)\\
        &+\beta\delta^{gj}\delta_k^a\epsilon_{ji}^{\:\:\:\:k}\left(T_{0g}\left((-E^b)^{\textrm{phys}},(A_a)^{\textrm{phys}}\right)\ast G^\Delta\right)(\vec{x},t)\\
        &+\left(\delta_l^g\delta_j^c\delta_k^d\epsilon_i^{\:\:jk}\partial_c+\delta_k^d\epsilon_{li}^{\:\:\:\:k}\partial^g+\delta_i^d\delta^{jc}\delta_k^g\epsilon_{jl}^{\:\:\:\:k}\partial_c+\delta_i^c\delta^{dj}\delta_k^g\epsilon_{lj}^{\:\:\:\:k}\partial_c\right)\\
        &\cdot\partial_d\left(\xi^l\ast G^\Delta\right)(\vec{x},t)\\
        &-\frac{1}{2}\epsilon_{ji}^{\:\:\:\:k}\delta^g_kP^a_g\xi^j(\vec{x},t)+\beta\epsilon_i^{\:\:lk}\delta_l^d\delta_k^a\partial_d\tau(\vec{x},t)+\delta_i^b\delta^a_{[b}\partial_{c]}\sigma^c(\vec{x},t)\\
        &+\mathcal{O}^{(2)}_{\mathcal{O}^{\textrm{dual}(1)}_{\delta\mathcal{E}^a_i,\{C_I^\prime\}},\{\mathcal{G}^I\}}(\vec{x},t).
    \end{split}
\end{align}
We get the symmetric transverse traceless degrees of freedom of $\delta\mathcal{A}_a^i,\:\delta\mathcal{E}^a_i$, along with the electromagnetic Pointing vector and energy density, both depending on the electromagnetic Dirac observables, as well as contributions depending on the gauge fixing parameters.
\section{Poisson algebra of the Dirac observables for the different sets}\label{app: More Details on the derivation of the Poisson algebra of the Dirac observables for the different sets}
In this section we give more details on the derivation of the Poisson algebra of the Dirac observables of the two sets \eqref{eq: sets of Dirac observables} defined in the main text. The Poisson brackets between $(A_a)^{\textrm{phys}}$ and $(-E^a)^{\textrm{phys}}$ were already discussed in the main text. Thus we will focus here on the geometric Dirac observables.
\subsection{Poisson algebra of the Dirac observables of set $S^\prime$}\label{app: derivation of Poisson algebra of full physical Variables}
We will start with the set $S^\prime$, that is the Poisson algebra of the Dirac observables \\
$(\delta\mathcal{A}_a^i)^{\textrm{phys}\prime},(\delta\mathcal{E}^a_i)^{\textrm{phys}\prime},(A_a)^{\textrm{phys}},(-E^a)^{\textrm{phys}}$. First we look at the following Poisson bracket
\begin{align}\label{Poisson bracket of full physical Ashtekar variables}
    \begin{split}
    \{(\delta\mathcal{A}_a^i)^{\textrm{phys}\prime}(\vec{x},t),(\delta\mathcal{E}^b_j)^{\textrm{phys}\prime}(\vec{y},t)\}=&\{\mathcal{O}_{\mathcal{O}^{\textrm{dual}(1)}_{\delta\mathcal{A}^i_a,\{C_I^\prime\}}(\vec{x},t),\{\mathcal{G}^I\}},\mathcal{O}_{\mathcal{O}^{\textrm{dual}(1)}_{\delta\mathcal{E}_j^b,\{C_I^\prime\}},\{\mathcal{G}^I\}}(\vec{y},t)\}\\
    =&\mathcal{O}_{\{\mathcal{O}^{\textrm{dual}(1)}_{\delta\mathcal{A}^i_a,\{C_I^\prime\}}(\vec{x},t), \mathcal{O}^{\textrm{dual}(1)}_{\delta\mathcal{E}^b_j,\{C_I^\prime\}}(\vec{y},t)\}^\ast,\{\mathcal{G}^I\}}\\
    =&\mathcal{O}_{\{\mathcal{O}^{\textrm{dual}(1)}_{\delta\mathcal{A}^i_a,\{C_I^\prime\}}(\vec{x},t), \mathcal{O}^{\textrm{dual}(1)}_{\delta\mathcal{E}^b_j,\{C_I^\prime\}}(\vec{y},t)\},\{\mathcal{G}^I\}},
    \end{split}
\end{align}
here, in the second line, we made use of identity \eqref{pb equals dirac bracket under observable map}. In the third line, we exploited the fact that by construction, both expressions in the Dirac bracket Poisson commute with all reference fields, thereby the Dirac bracket reduces to the Poisson bracket.
~\\
~\\
We could once again apply identity \eqref{pb equals dirac bracket under observable map} to the Poisson brackets of the dual observable map acting on the l´inearised Ashtekar-Barbero variables to obtain the dual observable map of the Dirac bracket of the linearised Ashtekar-Barbero variables. However, since they do not Poisson commute with all reference fields nor constraints, this would lead to the calculation of a very complicated Dirac bracket. On the other hand, we have already calculated the application of the dual observable map on the linearised Ashtekar-Barbero variables \eqref{eq: dual map on Ashtekar connection} and \eqref{eq: dual map on densifed triad}. Thus, we will first calculate the Poisson bracket of these expressions and then apply the observable map to the result. To do this, we will consider the electromagnetic and geometric degrees of freedom of $(\delta\mathcal{A}_a^i)^{\textrm{phys}\prime},\:(\delta\mathcal{E}^a_i)^{\textrm{phys}\prime}$ separately.
\subsubsection{Electromagnetic sector}
We begin with the electromagnetic sector. The Dirac observables $(\delta\mathcal{A}_a^i)^{\textrm{phys}\prime}$ and $(\delta\mathcal{E}^a_i)^{\textrm{phys}\prime}$ depend on the energy density $T^{00}(A_a,-P^a_bE^b)$ and the Pointing vector $T_{0g}(A_a,-P^a_bE^b)$. However, the linearised Ashtekar–Barbero variables $\delta\mathcal{A}^i_a,\:\delta\mathcal{E}_i^a$ Poisson commute with these quantities, as they also Poisson commute with $A_a$ and $-E^a$. Therefore, we only need to compute the Poisson brackets between the electromagnetic energy-momentum tensor components appearing in $\mathcal{O}^{\textrm{dual}(1)}_{\delta\mathcal{A}_a^i,\{C^\prime\}}$ and $\mathcal{O}^{\textrm{dual}(1)}_{\delta\mathcal{E}^a_i,\{C^\prime\}}$ (see appendix \ref{app: action dual map}). This can be done straightforwardly using the Poisson bracket between $A_a$ and $-E^a$
\begin{align}
    \begin{split}
    \Sigma_g(\vec{x},\vec{y},t)\coloneqq &\{\left(T^{00}\ast G^\Delta\right)(\vec{x},t), \left(T_{0g}\ast G^\Delta\right)(\vec{y},t)\}\\
    =&\int\limits_{\mathbb{R}^3} d^3z\Big(P_{a[f}\partial_{g]}^z\left(\left(P_d^a E^d P_c^f E^c\right)(\vec{z},t)G^\Delta (\vec{x}-\vec{z})G^\Delta(\vec{y}-\vec{z})\right)\\
    &+P^{f[a}\partial^{b]z}\left(\left(A_{[f,g]}A_{a,b}\right)(\vec{z},t)G^\Delta(\vec{x}-\vec{z})G^\Delta(\vec{y}-\vec{z})\right)\Big),
    \end{split}\\
    \begin{split}
    \Sigma_g^\prime(\vec{x},\vec{y},t)\coloneqq &\{\left(T^{00}\ast G^\Delta\right)(\vec{x},t), \left(T_{0g}\ast G^{\Delta\Delta}\right)(\vec{y},t)\}\\
    =&\int\limits_{\mathbb{R}^3} d^3z\Big(P_{a[f}\partial_{g]}^z\left(\left(P^a_d E^d P^f_c E^c\right)(\vec{z},t)G^{\Delta}(\vec{x}-\vec{z})G^{\Delta\Delta}(\vec{y}-\vec{z})\right)\\
    &+P^{f[a}\partial^{b]z}\left(\left(A_{[f,g]} A_{a,b}\right)(\vec{z},t)G^{\Delta}(\vec{x}-\vec{z})G^{\Delta\Delta}(\vec{y}-\vec{z})\right)\Big).
    \end{split}
\end{align}
Here, the energy-momentum tensor components are to be understood as functions of $\mathcal{O}^{\textrm{dual}}_{A_a,\{C_I\}}$ and $\mathcal{O}^{\textrm{dual}}_{-E^a,\{C_I\}}$. The Poisson brackets of $T^{00}$ with itself and of $T_{0a}$ with itself vanish. If we apply the observable map to this expression, we can again use identity \eqref{eq:dual and observable on phase space function} and obtain an expression depending on the physical electromagnetic variables. Using this and the linearity of the observable map, we can simplify \eqref{Poisson bracket of full physical Ashtekar variables} to
\begin{align}\label{eq: app Poisson algebra full physical Ashtekar variables em part}
    \begin{split}
    \{(\delta\mathcal{A}_a^i)^{\textrm{phys}\prime}(\vec{x},t),(\delta\mathcal{E}^b_j)^{\textrm{phys}\prime}(\vec{y},t)\}=&\{\mathcal{O}_{\mathcal{O}^{\textrm{dual,vac}(1)}_{\delta\mathcal{A}^i_a,\{C_I^\prime\}}(\vec{x},t),\{\mathcal{G}^I\}},\mathcal{O}_{\mathcal{O}^{\textrm{dual,vac}(1)}_{\delta\mathcal{E}_j^b,\{C_I^\prime\}}(\vec{y},t),\{\mathcal{G}^I\}}\}\\
    &+\left(\frac{1}{2}\epsilon_a^{\:\:cd}\delta_c^i\epsilon^{feb}\delta_{ej}\partial_d^x+\beta\left(\delta_a^g\delta_c^i\partial^c_x-\frac{1}{2}\delta_a^i\partial^g_x\right)\delta_j^b\right)\\
    &\cdot\Sigma_g\left((A_a)^{\textrm{phys}},(-E^a)^{\textrm{phys}}\right)(\vec{x},\vec{y},t)\\
    &+\frac{\beta}{2}\delta_c^i\partial^c_x\partial^g_x\partial_a^x\delta_j^b\Sigma_g^\prime\left((A_a)^{\textrm{phys}},(-E^a)^{\textrm{phys}}\right)(\vec{x},\vec{y},t).
    \end{split}
\end{align}
Here, we used that when the electromagnetic energy-momentum contributions are factored out, the dual observable map acts on the geometric degrees of freedom as the vacuum dual observable map. The electromagnetic energy-momentum tensor components generate contributions by applying the Poisson bracket, which depend on the phase space functions $\Sigma_g$ and $\Sigma_g^\prime$ with a prefactor, depending solely on $(A_a)^{\textrm{phys}},\:(-E^a)^{\textrm{phys}}$. These terms are non-zero and highly non-trivial. As already discussed in the main text, these contributions represent the self-interaction for the Dirac observables of $S^\prime$.
~\\
~\\
The Poisson brackets of $(\delta\mathcal{A}_a^i)^{\textrm{phys}\prime},(\delta\mathcal{E}^a_i)^{\textrm{phys}\prime}$ with themselves can be calculated analogously, and the results also depend non-trivially on the electromagnetic Dirac observables
\begin{align}
    \label{eq: app Poisson algebra full physical ashtekar connection em part}
    \begin{split}
    \{(\delta\mathcal{A}_a^i)^{\textrm{phys}\prime}(\vec{x},t),(\delta\mathcal{A}_a^i)^{\textrm{phys}\prime}(\vec{y},t)\}=&\{\mathcal{O}_{\mathcal{O}^{\textrm{dual,vac}(1)}_{\delta\mathcal{A}^i_a,\{C_I^\prime\}}(\vec{x},t),\{\mathcal{G}^I\}},\mathcal{O}_{\mathcal{O}^{\textrm{dual,vac}(1)}_{\delta\mathcal{A}^j_b,\{C_I^\prime\}}(\vec{y},t),\{\mathcal{G}^I\}}\}\\
    &+\frac{\beta}{2}\left(\epsilon_b^{\:\:ed}\delta_e^j\partial_d^y\left(\delta^g_a\delta_c^i\partial^c_x-\frac{1}{2}\delta^i_a\partial^g_x\right)-\epsilon_a^{\:\:ed}\delta_e^i\partial_d^x\left(\delta^g_b\delta_c^j\partial^c_y-\frac{1}{2}\delta^j_b\partial^g_y\right)\right)\\
    &\cdot\Sigma_g\left((A_a)^{\textrm{phys}},(-E^a)^{\textrm{phys}}\right)(\vec{x},\vec{y},t)\\
    &+\frac{\beta}{4}\left(\epsilon_b^{\:\:ed}\delta_e^j\delta_c^i\partial_d^y\partial^c_x\partial^g_x\partial_a^x-\epsilon_a^{\:\:ed}\delta_e^i\delta_c^j\partial_d^x\partial^c_y\partial^g_y\partial_b^y\right)\\
    &\cdot\Sigma_g^\prime\left((A_a)^{\textrm{phys}},(-E^a)^{\textrm{phys}}\right)(\vec{x},\vec{y},t),
    \end{split}\\
    \label{eq: app Poisson algebra full physical densitised triad em part}
    \begin{split}
    \{(\delta\mathcal{E}^a_i)^{\textrm{phys}\prime}(\vec{x},t),(\delta\mathcal{E}^b_j)^{\textrm{phys}\prime}(\vec{y},t)\}=&\{\mathcal{O}_{\mathcal{O}^{\textrm{dual,vac}(1)}_{\delta\mathcal{E}_i^a,\{C_I^\prime\}}(\vec{x},t),\{\mathcal{G}^I\}},\mathcal{O}_{\mathcal{O}^{\textrm{dual,vac}(1)}_{\delta\mathcal{E}_j^b,\{C_I^\prime\}}(\vec{y},t),\{\mathcal{G}^I\}}\}\\
    &+\beta\left(\delta_i^a\delta_{cj}\epsilon^{gcb}-\delta_j^b\delta_{ci}\epsilon^{gca}\right)\Sigma_g\left((A_a)^{\textrm{phys}},(-E^a)^{\textrm{phys}}\right)(\vec{x},\vec{y},t).
    \end{split}
\end{align}
~\\
~\\
For the Poisson brackets of $(\delta\mathcal{A}_a^i)^{\textrm{phys}\prime},(\delta\mathcal{E}^a_i)^{\textrm{phys}\prime}$ with $(A_a)^{\textrm{phys}},(-E^a)^{\textrm{phys}}$, the arguments are very similar, where here only the Poisson brackets of the energy-momentum tensor components depending on $-P_a^bE^a$ and $A_a$ with $-P_a^bE^a$ and $A_a$ are needed. Upon applying the observable map, the results are again non-trivial functions of the electromagnetic Dirac observables $(A_a)^{\textrm{phys}},\:(-E^a)^{\textrm{phys}}$:
\begin{align}
    \begin{split}
    \{(\delta\mathcal{A}_a^i)^{\textrm{phys}\prime}(\vec{x},t),(A_f)^{\textrm{phys}}(\vec{y},t)\}=&
    \frac{1}{2}\epsilon_a^{\:\:cb}\delta_c^i\partial_b^x P_{df}\left(G^\Delta(\vec{x}-\vec{y})(E^d)^{\textrm{phys}}(\vec{y},t)\right)\\
    &-\beta\left(\delta_a^g\delta_b^i\partial_x^b-\frac{1}{2}\delta_a^i\partial^g_x\right)P^c_f\left(G^\Delta(\vec{x}-\vec{y})(A_{[c,g]})^{\textrm{phys}}(\vec{y},t)\right)\\
    &-\frac{\beta}{2}\delta_b^i\partial_x^g\partial_x^b\partial_a^x P_f^c\left(G^{\Delta\Delta}(\vec{x}-\vec{y})(A_{[c,g]})^{\textrm{phys}}(\vec{y},t)\right),
    \end{split}\\
    \begin{split}
    \{(\delta\mathcal{A}_a^i)^{\textrm{phys}\prime}(\vec{x},t),(-E^f)^{\textrm{phys}}(\vec{y},t)\}=&
    -\frac{1}{2}\epsilon_a^{\:\:cb}\delta_c^i\partial_b^x P^{f[d}\partial^{e]}_y\left(G^\Delta(\vec{x}-\vec{y})(A_{d,e})^{\textrm{phys}}(\vec{y},t)\right)\\
    &+\beta\left(\delta_a^g\delta_b^i\partial_x^b-\frac{1}{2}\delta_a^i\partial^g_x\right)P^f_{[d}\partial_{g]}^y\left(G^\Delta(\vec{x}-\vec{y})(E^d)^{\textrm{phys}}(\vec{y},t)\right)\\
    &+\frac{\beta}{2}\delta_b^i\partial_x^g\partial_x^b\partial_a^x P^f_{[d}\partial^y_{g]}\left(G^{\Delta\Delta}(\vec{x}-\vec{y})(E^d)^{\textrm{phys}}(\vec{y},t)\right),
    \end{split}\\
    \begin{split}
    \{(\delta\mathcal{E}^a_i)^{\textrm{phys}\prime}(\vec{x},t),(A_f)^{\textrm{phys}}(\vec{y},t)\}=&
    \delta^a_iP_{df}\left(G^\Delta(\vec{x}-\vec{y})(E^d)^{\textrm{phys}}(\vec{y},t)\right)\\
    &+\beta\epsilon^{gca}\delta_i^c P_f^d\left(G^\Delta(\vec{x}-\vec{y})(A_{[d,g]})^{\textrm{phys}}(\vec{y},t)\right),
    \end{split}\\
    \begin{split}
    \{(\delta\mathcal{E}^a_i)^{\textrm{phys}\prime}(\vec{x},t),(-E^f)^{\textrm{phys}}(\vec{y},t)\}=&
    -\delta^a_iP^{f[d}\partial^{b]}_y\left(G^\Delta(\vec{x}-\vec{y})(A_{d,b})^{\textrm{phys}}(\vec{y},t)\right)\\
    &-\beta\epsilon^{gca}\delta_i^c P^f_{[b}\partial^y_{g]}\left(G^\Delta(\vec{x}-\vec{y})(E^b)^{\textrm{phys}}(\vec{y},t)\right).
    \end{split}
\end{align}
\subsubsection{Geometric sector}\label{app: geometric Dirac observables of S Poisson algebra calculation}
In this subsection we will compute the geometric sector of \eqref{eq: app Poisson algebra full physical Ashtekar variables em part}, \eqref{eq: app Poisson algebra full physical ashtekar connection em part} and \eqref{eq: app Poisson algebra full physical densitised triad em part}. We will start with \eqref{eq: app Poisson algebra full physical Ashtekar variables em part}, where the geometric part is given by 
\begin{align}
    \begin{split}
    \{\mathcal{O}_{\mathcal{O}^{\textrm{dual,vac}(1)}_{\delta\mathcal{A}^i_a,\{C_I^\prime\}}(\vec{x},t),\{\mathcal{G}^I\}},\mathcal{O}_{\mathcal{O}^{\textrm{dual,vac}(1)}_{\delta\mathcal{E}_j^b,\{C_I^\prime\}}(\vec{y},t),\{\mathcal{G}^I\}}\}
    =&\mathcal{O}_{\{\mathcal{O}^{\textrm{dual,vac}(1)}_{\delta\mathcal{A}^i_a,\{C_I^\prime\}}(\vec{x},t), \mathcal{O}^{\textrm{dual,vac}(1)}_{\delta\mathcal{E}^b_j,\{ C_I^\prime\}}(\vec{y},t)\}^\ast,\{\mathcal{G}^I\}}\\
    =&\mathcal{O}_{\{\mathcal{O}^{\textrm{dual,vac}(1)}_{\delta\mathcal{A}^i_a,\{C_I^\prime\}}(\vec{x},t), \mathcal{O}^{\textrm{dual,vac}(1)}_{\delta\mathcal{E}^b_j,\{C_I^\prime\}}(\vec{y},t)\},\{\mathcal{G}^I\}}.
    \end{split}
\end{align}
Here, in the first line, we used identity \eqref{pb equals dirac bracket under observable map}. In the second line, we exploited the fact that, as discussed in the main text, both arguments in the Dirac bracket Poisson commute with all reference fields. Thus, the Dirac bracket reduces to the Poisson bracket. We could once again apply identity \eqref{pb equals dirac bracket under observable map} to the Poisson brackets of the dual observable map acting on the linearised Ashtekar-Barbero variables to obtain the dual observable map of the Dirac bracket of $\delta\mathcal{A}^i_a$ and $\delta\mathcal{E}^a_i$. But the Dirac bracket would not reduce to the Poisson bracket since $\delta\mathcal{A}^i_a$ and $\delta\mathcal{E}^a_i$ are neither invariant under the application of all constraints nor Poisson commute with the clocks. Hence, we first calculate the Poisson bracket $\{\mathcal{O}^{\textrm{dual,vac}(1)}_{\delta\mathcal{A}^i_a,\{\delta C_I^\prime\}}(\vec{x},t), \mathcal{O}^{\textrm{dual,vac}(1)}_{\delta\mathcal{E}^b_j,\{\delta C_I^\prime\}}(\vec{y},t)\}$ and then apply the observable map onto the result. Using the results of appendix \ref{app: action vacuum dual map} this leads to
\begin{align}\label{eq: app calculation of geometric variables poisson bracket under vacuum dual map mixed}
    \begin{split}
    \{\mathcal{O}^{\textrm{dual,vac}(1)}_{\delta\mathcal{A}^i_a,\{C_I^\prime\}}(\vec{x},t), \mathcal{O}^{\textrm{dual,vac}(1)}_{\delta\mathcal{E}^b_j,\{ C_I^\prime\}}(\vec{y},t)\} =& \frac{1}{2}\prescript{y}{}{P}_g^b\left(\prescript{x}{}{P}_{la}\prescript{x}{}{P}^{di}+\delta^d_l(\delta^i_a-\prescript{x}{}{P}^i_a)+\delta_l^i(\delta_a^d-\prescript{x}{}{P}^d_a)+\delta_a^d\prescript{x}{}{P}_l^i-\delta_a^i\prescript{x}{}{P}_l^d\right)\\
    &\cdot\{\delta\mathcal{A}_d^l(\vec{x},t),\delta\mathcal{E}_j^g(\vec{y},t)\}\\
    =& \frac{1}{2}\prescript{y}{}{P}_g^b\left(\prescript{x}{}{P}_{la}\prescript{x}{}{P}^{di}+\delta^d_l(\delta^i_a-\prescript{x}{}{P}^i_a)+\delta_l^i(\delta_a^d-\prescript{x}{}{P}^d_a)+\delta_a^d\prescript{x}{}{P}_l^i-\delta_a^i\prescript{x}{}{P}_l^d\right)\\
    &\cdot\frac{\beta}{\kappa}\delta_j^l\delta_d^g\delta(\vec{x}-\vec{y})\\
    =&\frac{\beta}{\kappa}\prescript{x}{}{P}^{ib}_{aj}\delta(\vec{x}-\vec{y}),
    \end{split}
\end{align}
where $\prescript{x}{}{P}^{a}_b=\delta_b^a+\partial_b^x\partial^a_x$. In the first line, we used that $\{\delta\mathcal{A}_d^l(\vec{x},t),\delta\mathcal{A}_g^m(\vec{y},t)\}=0$. In the second, we used $\{\delta\mathcal{A}_d^l(\vec{x},t),\delta\mathcal{E}^g_j(\vec{y},t)\}=\frac{\beta}{\kappa}\delta_d^g\delta_j^l\delta(\vec{x}-\vec{y})$. In the last, we used the properties of the projectors given in appendix \ref{app: projectors}, as well as $\prescript{y}{}{P}^{a}_b\delta(\vec{x}-\vec{y})=\prescript{x}{}{P}^{a}_b\delta(\vec{x}-\vec{y})$. This result has no degrees of freedom in the geometric nor electromagnetic sector and therefore trivially Poisson commutes with all constraints. Therefore, the observable map acts trivially, and we obtain the following Poisson bracket
\begin{align}\label{eq: app calculation of geometric variables poisson bracket under vacuum dual map ashtekar connection}
    \begin{split}
    \mathcal{O}_{\{\mathcal{O}^{\textrm{dual,vac}(1)}_{\delta\mathcal{A}^i_a,\{C_I^\prime\}}(\vec{x},t), \mathcal{O}^{\textrm{dual,vac}(1)}_{\delta\mathcal{E}^b_j,\{C_I^\prime\}}(\vec{y},t)\},\{\mathcal{G}^I\}}
    =& \mathcal{O}_{\frac{\beta}{\kappa}\prescript{x}{}{P}^{ib}_{aj}\delta(\vec{x}-\vec{y}),\{\mathcal{G}^I\}}\\
    =&\frac{\beta}{\kappa}\prescript{x}{}{P}^{ib}_{aj}\delta(\vec{x}-\vec{y}).
    \end{split}
\end{align}
Note that the observable map is here applied up to second order.
~\\
~\\
Next we calculate, using again the application of the first order vacuum dual observable map on the linearised Ashtekar-Barbero variables, the following
\begin{align}\label{eq: app self Poisson bracket of asktekar connection under first order dual vac map}
    \begin{split}
    \{\mathcal{O}^{\textrm{dual,vac}(1)}_{\delta\mathcal{A}^i_a,\{C_I^\prime\}}(\vec{x},t), \mathcal{O}^{\textrm{dual,vac}(1)}_{\delta\mathcal{A}_b^j,\{C_I^\prime\}}(\vec{y},t)\} =& 0,
    \end{split}
\end{align}
where we used that $\{\delta\mathcal{A}_d^l(\vec{x},t),\delta\mathcal{A}_g^m(\vec{y},t)\}=0$. Using this, we can, as argued above, immediately follow
\begin{align}
    \begin{split}
    \mathcal{O}_{\{\mathcal{O}^{\textrm{dual,vac}(1)}_{\delta\mathcal{A}^i_a,\{C_I^\prime\}}(\vec{x},t), \mathcal{O}^{\textrm{dual,vac}(1)}_{\delta\mathcal{A}_b^j,\{C_I^\prime\}}(\vec{y},t)\},\{\mathcal{G}^I\}}
    =& \mathcal{O}_{0,\{\mathcal{G}^I\}}=0.
    \end{split}
\end{align}
~\\
~\\
The next Poisson bracket we need to calculate is the one of $\mathcal{O}^{\textrm{dual,vac}(1)}_{\delta\mathcal{E}_i^a,\{C_I^\prime\}}(\vec{x},t)$ with itself. As can be seen in appendix \ref{app: action vacuum dual map}, $\mathcal{O}^{\textrm{dual,vac}(1)}_{\delta\mathcal{E}^i_a,\{\delta C_I^\prime\}}$ has degrees of freedom linear in $\delta\mathcal{A}_a^i$ and $\delta\mathcal{E}^a_i$, therefore, it appears at first sight to have a non-trivial Poisson bracket with itself. For simplicity, we will use in the following an abuse of notation by writing Levi-Civita symbols, derivatives, the projectors as well as the Kronecker deltas with mixed Lie(SU(2)) and spatial indices in the sense that for example $\epsilon^{ai}_{\:\:\:\:b}\coloneqq\delta_c^i\epsilon^{ac}_{\:\:\:\:b}$. Using this we get
\begin{align}
    \begin{split}
    \{\mathcal{O}^{\textrm{dual,vac}(1)}_{\delta\mathcal{E}_i^a,\{C_I^\prime\}}(\vec{x},t), \mathcal{O}^{\textrm{dual,vac}(1)}_{\delta\mathcal{E}^b_j,\{C_I^\prime\}}(\vec{y},t)\} =\int\limits_{\mathbb{R}^3} d^3z\Big(&G^\Delta(\vec{y}-\vec{z})\prescript{x}{}{P}_d^a\left(\delta_j^b\epsilon_n^{\:\:cg}\partial_c^z+\epsilon_{jn}^{\:\:\:\:g}\partial^b_z+\delta_n^g\epsilon^{c\:\:\:b}_{\:\:j}\partial_c^z+\epsilon_j^{\:\:gb}\partial_n^z\right)\\
    &\cdot\{\delta\mathcal{E}_i^d(\vec{x},t),\delta\mathcal{A}_g^n(\vec{z},t)\}\\
    &+G^\Delta(\vec{x}-\vec{z})\prescript{y}{}{P}_g^b\left(\delta_i^a\epsilon_l^{\:\:cd}\partial_c^z+\epsilon_{il}^{\:\:\:\:d}\partial^a_z+\delta_l^d\epsilon^{c\:\:\:a}_{\:\:i}\partial_c^z+\epsilon_i^{\:\:da}\partial_l^z\right)\\
    &\cdot\{\delta\mathcal{A}_d^l(\vec{z},t),\delta\mathcal{E}_j^g(\vec{y},t)\}\\
    =\int\limits_{\mathbb{R}^3} d^3z\Big(&G^\Delta(\vec{y}-\vec{z})\prescript{x}{}{P}_d^a\left(\delta_j^b\epsilon_i^{\:\:cd}\partial_c^z+\epsilon_{ji}^{\:\:\:\:d}\partial^b_z+\delta_i^d\epsilon^{c\:\:\:b}_{\:\:j}\partial_c^z+\epsilon_j^{\:\:db}\partial_i^z\right)\\
    &\cdot\frac{\beta}{\kappa}\delta(\vec{x}-\vec{z})(-1)\\
    &+G^\Delta(\vec{x}-\vec{z})\prescript{y}{}{P}_d^b\left(\delta_i^a\epsilon_j^{\:\:cd}\partial_c^z+\epsilon_{ij}^{\:\:\:\:d}\partial^a_z+\delta_j^d\epsilon^{c\:\:\:a}_{\:\:i}\partial_c^z+\epsilon_i^{\:\:da}\partial_j^z\right)\\
    &\cdot\frac{\beta}{\kappa}\delta(\vec{z}-\vec{y}).
    \end{split}
\end{align}
To evaluate the delta distribution, we apply partial integration, which gives us a boundary term that vanishes since we assume compact support for all involved functions. Now, all derivatives act on $G^\Delta(\vec{x}-\vec{y})$. We want to have only derivatives with respect to $\vec{x}$ so that we can apply identities for the projector. To do this, we use that $\prescript{y}{}{P_g^b}G^\Delta(\vec{x}-\vec{y})=\prescript{x}{}{P_g^b}G^\Delta(\vec{x}-\vec{y})$, as well as that $\partial^x_c G^\Delta(\vec{y}-\vec{x})=-\partial^x_c G^\Delta(\vec{x}-\vec{y}),\:\:\partial^y_c G^\Delta(\vec{x}-\vec{y})=-\partial^x_c G^\Delta(\vec{x}-\vec{y)}$. This leads to
\begin{align}
    \begin{split}
    \{\mathcal{O}^{\textrm{dual,vac}(1)}_{\delta\mathcal{E}_i^a,\{C_I^\prime\}}(\vec{x},t), \mathcal{O}^{\textrm{dual,vac}(1)}_{\delta\mathcal{E}^b_j,\{ C_I^\prime\}}(\vec{y},t)\}
    =& \frac{\beta}{\kappa}\Big(\prescript{x}{}{P}_d^b \left(\delta_i^a\epsilon_j^{\:\:\:\:cd}\partial_c^x+\epsilon_{ij}^{\:\:\:\:d}\partial^a_x+\delta_j^d\epsilon^{c\:\:a}_{\:\:i}\partial_c^x+\epsilon_i^{\:\:da}\partial_j^x\right)\\
    &+\prescript{x}{}{P}_d^a \left(\delta_j^b\epsilon_i^{\:\:\:\:cd}\partial_c^x+\epsilon_{ji}^{\:\:\:\:\:d}\partial^b_x+\delta_i^d\epsilon^{c\:\:b}_{\:\:j}\partial_c^x+\epsilon_j^{\:\:db}\partial_i^x\right)\Big)\\
    &\cdot G^\Delta(\vec{x}-\vec{y}).
    \end{split}
\end{align}
The next step is now to write out the projectors \eqref{eq: projector transverse subspace} and simplify. After a rather long but tedious calculation, we end up with
\begin{align}
    \begin{split}
    \{\mathcal{O}^{\textrm{dual,vac}(1)}_{\delta\mathcal{E}_i^a,\{C_I^\prime\}}(\vec{x},t), \mathcal{O}^{\textrm{dual,vac}(1)}_{\delta\mathcal{E}^b_j,\{C_I^\prime\}}(\vec{y},t)\}
    =& \frac{\beta}{\kappa}\Big(\epsilon_{ij}^{\:\:\:\:b}\partial_x^a+\epsilon_i^{\:\:ba}\partial_j^x+\epsilon_{ji}^{\:\:\:\:a}\partial^b_x+\epsilon_j^{\:\:ab}\partial_i^x\Big)G^\Delta(\vec{x}-\vec{y}).
    \end{split}
\end{align}
In fact, this term is always zero. To see this, one needs to reconsider that $a,\:b,\:i,\:j\in\{1,2,3\}$, this means that, in the expression, at least two of the four indices must always be the same. We will start by checking the case for $a=b$
\begin{align}\label{eq: app calculation of geometric variables poisson bracket under vacuum dual map densitised triad}
    \begin{split}
    \{\mathcal{O}^{\textrm{dual,vac}(1)}_{\delta\mathcal{E}_i^a,\{C_I^\prime\}}(\vec{x},t), \mathcal{O}^{\textrm{dual,vac}(1)}_{\delta\mathcal{E}^a_j,\{C_I^\prime\}}(\vec{y},t)\}
    =& \frac{\beta}{\kappa}\Big(\epsilon_{ij}^{\:\:\:\:a}\partial_x^a+\epsilon_i^{\:\:aa}\partial_j^x+\epsilon_{ji}^{\:\:\:\:a}\partial^a_x+\epsilon_j^{\:\:aa}\partial_i^x\Big)G^\Delta(\vec{x}-\vec{y})\\
    =& \frac{\beta}{\kappa}\Big(\epsilon_{ij}^{\:\:\:\:a}\partial_x^a+\epsilon_{ji}^{\:\:\:\:a}\partial^a_x\Big)G^\Delta(\vec{x}-\vec{y})\\
    =& 0,
    \end{split}
\end{align}
and analogously one can show that the term vanishes for the cases $a=i,\:a=j,\:b=i,\:b=j,\:i=j$. Therefore, we can conclude that the Poisson bracket vanishes. From this follows immediately that also the application of the observable map onto it vanishes
\begin{align}
    \begin{split}
    \mathcal{O}_{\{\mathcal{O}^{\textrm{dual,vac}(1)}_{\delta\mathcal{E}_i^a,\{\delta C_I^\prime\}}(\vec{x},t), \mathcal{O}^{\textrm{dual,vac}(1)}_{\delta\mathcal{E}^b_j,\{\delta C_I^\prime\}}(\vec{y},t)\},\{\mathcal{G}^I\}}
    =& \mathcal{O}_{0,\{\mathcal{G}^I\}}=0.
    \end{split}
\end{align}
\subsection{Poisson algebra of the Dirac observables of set S}\label{app: PB of ashtekar and photon physical variables}
In this section we give more details on the calculation of the Poisson algebra of\\
$(\delta\mathcal{A}_a^i)^{\textrm{phys}},(\delta\mathcal{E}^a_i)^{\textrm{phys}},(A_a)^{\textrm{phys}},(-E^a)^{\textrm{phys}}$, which are the Dirac observables we chose to quantise in this work. We will start with $(\delta\mathcal{A}_a^i)^{\textrm{phys}},(\delta\mathcal{E}^a_i)^{\textrm{phys}}$. As already discussed in the main text, these observables can be produced by applying first the first order vacuum  dual observable map \eqref{eq: vacuum dual map first order} and then the observable map \eqref{eq:linearised observable map} onto the elementary geometric variables $\delta\mathcal{A}_a^i,\:\delta\mathcal{E}^a_i$
\begin{align}\label{eq: app geometric physical variables with the use of vacuum dual map}
    \begin{split}
    \left(\delta\mathcal{A}_a^i\right)^{\textrm{phys}}(\vec{x},t)=& \mathcal{O}_{\mathcal{O}^{\textrm{dual,vac}(1)}_{\delta\mathcal{A}^i_a,\{C_I^\prime\}},\{\mathcal{G}^I\}}(\vec{x},t) ,\\
    (\delta\mathcal{E}_i^a)^{\textrm{phys}}(\vec{x},t)=& \mathcal{O}_{\mathcal{O}^{\textrm{dual,vac}(1)}_{\delta\mathcal{E}_i^a,\{C_I^\prime\}},\{\mathcal{G}^I\}}(\vec{x},t).
    \end{split}
\end{align}
We will use this to calculate their Poisson brackets. We will start with the Poisson bracket of $(\delta\mathcal{A}_a^i)^{\textrm{phys}}$ with $(\delta\mathcal{E}^b_j)^{\textrm{phys}}$
\begin{align}
    \begin{split}
    \{(\delta\mathcal{A}_a^i)^{\textrm{phys}}(\vec{x},t),(\delta\mathcal{E}^b_j)^{\textrm{phys}}(\vec{y},t)\}
    =&\{\mathcal{O}_{\mathcal{O}^{\textrm{dual,vac}(1)}_{\delta\mathcal{A}^i_a,\{C_I^\prime\}},\{\mathcal{G}^I\}}(\vec{x},t), \mathcal{O}_{\mathcal{O}^{\textrm{dual,vac}(1)}_{\delta\mathcal{E}_j^b,\{C_I^\prime\}},\{\mathcal{G}^I\}}(\vec{y},t)\},\\
    =&\mathcal{O}_{\{\mathcal{O}^{\textrm{dual,vac}(1)}_{\delta\mathcal{A}^i_a,\{C_I^\prime\}}(\vec{x},t), \mathcal{O}^{\textrm{dual,vac}(1)}_{\delta\mathcal{E}^b_j,\{C_I^\prime\}}(\vec{y},t)\}^\ast,\{\mathcal{G}^I\}}\\
    =&\mathcal{O}_{\{\mathcal{O}^{\textrm{dual,vac}(1)}_{\delta\mathcal{A}^i_a,\{C_I^\prime\}}(\vec{x},t), \mathcal{O}^{\textrm{dual,vac}(1)}_{\delta\mathcal{E}^b_j,\{C_I^\prime\}}(\vec{y},t)\},\{\mathcal{G}^I\}}\\
    =& \mathcal{O}_{\frac{\beta}{\kappa}\prescript{x}{}{P}^{ib}_{aj}\delta(\vec{x}-\vec{y}),\{\mathcal{G}^I\}}\\
    =&\frac{\beta}{\kappa}\prescript{x}{}{P}^{ib}_{aj}\delta(\vec{x}-\vec{y}).
    \end{split}
\end{align}
In the first line we inserted \eqref{eq: app geometric physical variables with the use of vacuum dual map}. In the second we applied identity \eqref{pb equals dirac bracket under observable map}. In the third we used that, as already discussed in the main text, the expressions $\mathcal{O}^{\textrm{dual,vac}(1)}_{\delta\mathcal{A}_a^j,\{C_I^\prime\}},\:\mathcal{O}^{\textrm{dual,vac}(1)}_{\delta\mathcal{E}_j^b,\{C_I^\prime\}}$ Poisson commute by construction with all reference fields. Hence, the Dirac bracket reduces to the Poisson bracket. In the fourth line we used the result for this Poisson bracket we already calculated above in \eqref{eq: app calculation of geometric variables poisson bracket under vacuum dual map mixed} (see section \ref{app: geometric Dirac observables of S Poisson algebra calculation} for the exact derivation).
~\\
~\\
With exactly the same arguments and the use of \eqref{eq: app calculation of geometric variables poisson bracket under vacuum dual map densitised triad} and \eqref{eq: app self Poisson bracket of asktekar connection under first order dual vac map}, the self Poisson brackets for $(\delta\mathcal{A}^i_a)^{\textrm{phys}},\:(\delta\mathcal{E}_i^a)^{\textrm{phys}}$ can be calculated to
\begin{align}
    \begin{split}
    \{(\delta\mathcal{A}_a^i)^{\textrm{phys}}(\vec{x},t),(\delta\mathcal{A}_b^j)^{\textrm{phys}}(\vec{y},t)\}
    =&\mathcal{O}_{\{\mathcal{O}^{\textrm{dual,vac}(1)}_{\delta\mathcal{A}^i_a,\{ C_I^\prime\}}(\vec{x},t), \mathcal{O}^{\textrm{dual,vac}(1)}_{\delta\mathcal{A}_b^j,\{C_I^\prime\}}(\vec{y},t)\},\{\mathcal{G}^I\}}\\
    =& \mathcal{O}_{0,\{\mathcal{G}^I\}}\\
    =&0,\\
    \{(\delta\mathcal{E}^a_i)^{\textrm{phys}}(\vec{x},t),(\delta\mathcal{E}^b_j)^{\textrm{phys}}(\vec{y},t)\}
    =&\mathcal{O}_{\{\mathcal{O}^{\textrm{dual,vac}(1)}_{\delta\mathcal{E}_i^a,\{ C_I^\prime\}}(\vec{x},t), \mathcal{O}^{\textrm{dual,vac}(1)}_{\delta\mathcal{E}^b_j,\{C_I^\prime\}}(\vec{y},t)\},\{\mathcal{G}^I\}}\\
    =& \mathcal{O}_{0,\{\mathcal{G}^I\}}\\
    =&0.
    \end{split}
\end{align}
~\\
~\\
Next, we look at the Poisson bracket of the geometric Dirac observables with the electromagnetic ones. For this, we again apply identity \eqref{pb equals dirac bracket under observable map}, note that by construction, all expressions after the dual observable map has been applied for a given phase space function, Poisson commute with all clocks. Therefore, we can use a similar approach as for the geometric observables. This means that we first will calculate the following Poisson brackets
\begin{align}
    &\{\mathcal{O}^{\textrm{dual,vac}(1)}_{\delta\mathcal{A}^i_a,\{ C_I^\prime\}}(\vec{x},t), \mathcal{O}^{\textrm{dual}}_{A_b,\{C_I^\prime\}}(\vec{y},t)\},\\
    &\{\mathcal{O}^{\textrm{dual,vac}(1)}_{\delta\mathcal{A}^i_a,\{C_I^\prime\}}(\vec{x},t), \mathcal{O}^{\textrm{dual}}_{-E^b,\{C_I^\prime\}}(\vec{y},t)\},\\
    &\{\mathcal{O}^{\textrm{dual,vac}(1)}_{\delta\mathcal{E}_i^a,\{C_I^\prime\}}(\vec{x},t), \mathcal{O}^{\textrm{dual}(1)}_{A_b,\{C_I^\prime\}}(\vec{y},t)\},\\
    &\{\mathcal{O}^{\textrm{dual,vac}(1)}_{\delta\mathcal{E}_i^a,\{C_I^\prime\}}(\vec{x},t), \mathcal{O}^{\textrm{dual}}_{-E^b,\{C_I^\prime\}}(\vec{y},t)\}.
\end{align}
To do this, we use the already calculated form of the dual observable map onto $A_a$ and $-E^a$ given by
\begin{align}
    \begin{split}
    \mathcal{O}^{\textrm{dual}}_{A_a,\{G^{\textrm{U(1)}}\}}(\vec{x},t)
    =& A_a(\vec{x},t)
    \end{split},\\
    \begin{split}
    \mathcal{O}^{\textrm{dual}}_{-E^a,\{G^{\textrm{U(1)}}\}}(\vec{x},t)
    =&-P_b^a E^b(\vec{x},t)
    \end{split}.
\end{align}
These expressions only have electromagnetic degrees of freedom. On the other hand, $\mathcal{O}^{\textrm{dual,vac}(1)}_{\delta\mathcal{A}^i_a,\{ C_I^\prime\}}$ and $\mathcal{O}^{\textrm{dual,vac}(1)}_{\delta\mathcal{E}_i^a,\{C_I^\prime\}}$ have only geometric ones. From this, we can immediately follow that all Poisson brackets vanish. Therefore, we get for the Dirac observables also vanishing Poisson brackets
\begin{align}
    &\{(\delta\mathcal{A}_a^i)^{\textrm{phys}}(\vec{x},t),(A_b)^{\textrm{phys}}(\vec{y},t)\}
    =\mathcal{O}_{0,\{\mathcal{G}^I\}}=0,\\
    &\{(\delta\mathcal{A}_a^i)^{\textrm{phys}}(\vec{x},t),(-E^b)^{\textrm{phys}}(\vec{y},t)\}
    =\mathcal{O}_{0,\{\mathcal{G}^I\}}=0,\\
    &\{(\delta\mathcal{E}^a_i)^{\textrm{phys}}(\vec{x},t),(A_b)^{\textrm{phys}}(\vec{y},t)\}
    =\mathcal{O}_{0,\{\mathcal{G}^I\}}=0,\\
    &\{(\delta\mathcal{E}^a_i)^{\textrm{phys}}(\vec{x},t),(-E^b)^{\textrm{phys}}(\vec{y},t)\}
    =\mathcal{O}_{0,\{\mathcal{G}^I\}}=0.
\end{align}
\section{Second order contribution of the dual observable map}\label{app: More details on the inclusion of the second order of the dual map into the construction}
Until now, we have only considered the first order application of the dual observable map onto $\delta\mathcal{A}_a^i,\:\delta\mathcal{E}^a_i$. In this section, we discuss what happens when the second order is also included. In our case, for linear clocks, the application of the dual observable map up to second order is given by 
\begin{align}
    \begin{split}
    \mathcal{O}^{\textrm{dual}}_{F,\{C_I^\prime\}}\coloneqq&
    \mathcal{O}^{\textrm{dual}}_{F,\{G^{\textrm{U(1)}}\}}-\kappa\int\limits_{\mathbb{R}^3} d^3y \delta C_I^\prime(\vec{y})\{\delta \mathcal{G}^I(\vec{y}),\delta f\}\\
    &-\kappa\int\limits_{\mathbb{R}^3} d^3y \delta^2 C_I^\prime(\vec{y})\{\delta \mathcal{G}^I(\vec{y}),\delta f\}
    -\kappa\int\limits_{\mathbb{R}^3} d^3y \delta {C^I}^\prime(\vec{y})\{\delta \mathcal{G}^I(\vec{y}),\delta^2 f\}\\
    &+\frac{\kappa^2}{2}\int\limits_{\mathbb{R}^3} d^3 y\int\limits_{\mathbb{R}^3} d^3 z \delta C_I^\prime(\vec{y})\delta C_J^\prime(\vec{z})\{\delta \mathcal{G}^J(\vec{z}),\{\delta \mathcal{G}^I(\vec{y}),\delta^2 f\}\}\\
    &+O(\delta^3,\kappa^3),
    \end{split}
\end{align}
for a phase space function $F[A_a,E^a,\delta\mathcal{A}_a^i,\delta\mathcal{E}^a_i]=f[A_a,E^a]+\delta f[A_a,E^a,\delta\mathcal{A}_a^i,\delta\mathcal{E}^a_i]+\delta^2 f[A_a,E^a,\delta\mathcal{A}_a^i,\delta\mathcal{E}^a_i]$.
Since the geometric reference fields consist solely of geometric degrees of freedom, the application of the dual observable map onto $(A_a)^{\textrm{phys}},\:(-E^a)^{\textrm{phys}}$ is not altered by including the second order, since all involved Poisson brackets with the reference fields vanish for any order anyway. As discussed in the main text, we choose as Dirac observables the ones which are produced by applying first the first order vacuum dual observable map and then the observable map onto $\delta\mathcal{A}_a^i,\:\delta\mathcal{E}^a_i$. Thus, we will also look here at the vacuum dual observable map applied onto a first order function $F[\delta\mathcal{A}_a^i,\delta\mathcal{E}^a_i]=\delta f[\delta\mathcal{A}_a^i,\delta\mathcal{E}^a_i]$ which is, including the second order, given by
\begin{align}
    \begin{split}
    \mathcal{O}^{\textrm{dual, vac}}_{F,\{C_I^\prime\}}\coloneqq&
    -\kappa\int\limits_{\mathbb{R}^3} d^3y \delta C_I^{\textrm{geo}}(\vec{y})\{\delta \mathcal{G}^I(\vec{y}),\delta f\}-\kappa\int\limits_{\mathbb{R}^3} d^3y \delta^2 C_I(\vec{y})^{\textrm{geo}\prime}\{\delta \mathcal{G}^I(\vec{y}),\delta f\}\\
    &+O(\delta^3,\kappa^3)\\
    &\coloneqq \mathcal{O}^{\textrm{dual,vac} (1)}_{F,\{C_I^\prime\}}+\mathcal{O}^{\textrm{dual,vac} (2)}_{F,\{C_I^\prime\}}+O(\delta^3,\kappa^3).
    \end{split}
\end{align}
The difference to the construction in the main text is that now also the primed second order geometric constraints $\delta^2 C_I^{\textrm{geo}\prime}$ are considered. Note that since $\delta^2 C_I^{\textrm{geo}\prime}$ Poisson commutes with all clocks, the further contributions will also commute with all clocks, in the perturbation order we are interest in. We will refuse from explicitly calculating the resulting further expressions, if we act on $\delta\mathcal{A}_a^i,\:\delta\mathcal{E}^a_i$, since we are still mainly interested in the linear space. But in general these second order contributions need to be taken into account in the Poisson brackets such as $\{\mathcal{O}_{\delta\mathcal{A}_a^i,\{C_I^\prime\}}^{\textrm{dual, vac}}(\vec{x},t), \mathcal{O}_{\delta\mathcal{E}^b_j,\{C_I^\prime\}}^{\textrm{dual, vac}}(\vec{y},t)\}$, this will be investigated in the following.
~\\
~\\
For the further derivation we will define the Dirac observables 
\begin{align}
    \label{eq app: Dirac observables with second order vacuum dual observable map}
    \begin{split}
    \left(\delta\mathcal{A}_a^i\right)^{\textrm{phys}\prime\prime}(\vec{x},t)\coloneqq& \mathcal{O}_{\mathcal{O}^{\textrm{dual, vac}}_{\delta\mathcal{A}^i_a,\{C_I^\prime\}},\{\mathcal{G}^I\}}(\vec{x},t),\\
    (\delta\mathcal{E}_i^a)^{\textrm{phys}\prime\prime}(\vec{x},t)\coloneqq& \mathcal{O}^{\textrm{vac}}_{\mathcal{O}^{\textrm{dual, vac}}_{\delta\mathcal{E}_i^a,\{C_I^\prime\}},\{\mathcal{G}^I\}}(\vec{x},t).
    \end{split}
\end{align}
First we look at the Poisson bracket of $(\delta\mathcal{A}_a^i)^{\textrm{phys}\prime\prime}$ and $(\delta\mathcal{E}_i^a)^{\textrm{phys}\prime\prime}$. As a first step, we write out the second order expansion of the dual observable map for the geometric variables in their Poisson bracket, up to second order
\begin{align}\label{eq: second order dual map expansion on Poisson bracket 1}
    \begin{split}
    &\{(\delta\mathcal{A}_a^i)^{\textrm{phys}\prime\prime}(\vec{x},t),(\delta\mathcal{E}^b_j)^{\textrm{phys}\prime\prime}(\vec{y},t)\}\\
    &=\mathcal{O}_{\left(\{\mathcal{O}^{\textrm{dual,vac}(1)}_{\delta\mathcal{A}^i_a,\{C_I^\prime\}}(\vec{x},t), \mathcal{O}^{\textrm{dual,vac}(1)}_{\delta\mathcal{E}^b_j,\{ C_I^\prime\}}(\vec{y},t)\}+\{\mathcal{O}^{\textrm{dual,vac}(1)}_{\delta\mathcal{A}^i_a,\{ C_I^\prime\}}(\vec{x},t), \mathcal{O}^{\textrm{dual,vac}(2)}_{\delta\mathcal{E}^b_j,\{C_I^\prime\}}(\vec{y},t)\}+\{\mathcal{O}^{\textrm{dual,vac}(2)}_{\delta\mathcal{A}^i_a,\{ C_I^\prime\}}(\vec{x},t), \mathcal{O}^{\textrm{dual,vac}(1)}_{\delta\mathcal{E}^b_j,\{C_I^\prime\}}(\vec{y},t)\}\right),\{\mathcal{G}^I\}}\\
    &\hspace{0.2in}+O(\kappa^2,\delta^2),
    \end{split}
\end{align}
where we used identity \eqref{pb equals dirac bracket under observable map}. We have already calculated the first Poisson bracket in this expression (see appendix \ref{app: PB of ashtekar and photon physical variables}). To evaluate the second and third, we need the second order vacuum dual observable map action on $\delta\mathcal{A}_a^i,\:\delta\mathcal{E}^a_i$, which is given by the following expressions
\begin{align}\label{eq: second order dual expansion}
    \begin{split}
        \mathcal{O}^{\textrm{dual,vac}(2)}_{\delta\mathcal{A}_a^i,\{ C_I^\prime\}}(\vec{x},t)=&\frac{1}{2\kappa}\epsilon_a^{\:\:ic}\partial_c\left(\delta^2 C^{\textrm{geo}\prime}\ast G^\Delta\right)(\vec{x},t)-\frac{\beta}{\kappa}\left(\delta_a^g\partial^i-\frac{1}{2}\delta_a^i\partial^g\right)\left(\delta^2 C_g^{\textrm{geo}\prime}\ast G^\Delta\right)(\vec{x},t)\\
        &-\frac{\beta}{2\kappa}\partial^g\partial^i\partial_a\left(\delta^2 C_g^{\textrm{geo}\prime}\ast G^{\Delta\Delta}\right)(\vec{x},t),
    \end{split}\\
    \begin{split}
        \mathcal{O}^{\textrm{dual,vac}(2)}_{\delta\mathcal{E}^b_j,\{C_I^\prime\}}(\vec{x},t)=&\frac{1}{\kappa}\delta_j^b\left(\delta^2 C^{\textrm{geo}\prime}\ast G^\Delta\right)(\vec{x},t)+\frac{2\beta}{\kappa}\partial^b\left(\delta^2 G_j^{\textrm{geo}\prime}\ast G^\Delta\right)(\vec{x},t)\\
        &-\frac{\beta}{\kappa}\epsilon_j^{\:\:gb}\left(\delta^2 C_g^{\textrm{geo}\prime}\ast G^{\Delta}\right)(\vec{x},t).
    \end{split}
\end{align}
For later convenience, we have not explicitly written out the geometric primed second order constraints. Note that both expressions are, by construction, depend only on geometric degrees of freedom. We can now insert this into \eqref{eq: second order dual map expansion on Poisson bracket 1} and use that the observable map is linear, as well as the integrals involved. Hence, we get
\begin{align}\label{eq: Poisson bracket of physical geometric variables second order 1}
    \begin{split}
        \{(\delta\mathcal{A}_a^i)^{\textrm{phys}\prime\prime}(\vec{x},t),(\delta\mathcal{E}^b_j)^{\textrm{phys}\prime\prime}(\vec{y},t)\}&=\frac{\beta}{\kappa}P^{ib}_{aj}\delta(\vec{x}-\vec{y})\\
        +\int d^3z\Big[&\frac{1}{\kappa}\delta_j^b G^{\Delta}(\vec{y}-\vec{z})\mathcal{O}_{\{\mathcal{O}^{\textrm{dual,vac}(1)}_{\delta\mathcal{A}_a^i,\{ C_I^\prime\}}(\vec{x},t),\delta^2 C^{\textrm{geo}\prime}(\vec{z},t)\},\{\mathcal{G}^I\}}\\
        &+\frac{2\beta}{\kappa}\partial^b_y G^{\Delta}(\vec{y}-\vec{z})\mathcal{O}_{\{\mathcal{O}^{\textrm{dual,vac}(1)}_{\delta\mathcal{A}_a^i,\{C_I^\prime\}}(\vec{x},t),\delta^2 G_j^{\textrm{geo}\prime}(\vec{z},t)\},\{\mathcal{G}^I\}}\\
        &-\frac{\beta}{\kappa}\epsilon_j^{\:\:gb} G^{\Delta}(\vec{y}-\vec{z})\mathcal{O}_{\{\mathcal{O}^{\textrm{dual,vac}(1)}_{\delta\mathcal{A}_a^i,\{ C_I^\prime\}}(\vec{x},t),\delta^2 C_g^{\textrm{geo}\prime}(\vec{z},t)\},\{\mathcal{G}^I\}}\\
        &+\frac{1}{2\kappa}\epsilon_a^{\:\:ic}\partial_c^x G^{\Delta}(\vec{x}-\vec{z})\mathcal{O}_{\{\delta^2 C^{\textrm{geo}\prime}(\vec{z},t),\mathcal{O}^{\textrm{dual,vac}(1)}_{\delta\mathcal{E}_j^b,\{C_I^\prime\}}(\vec{y},t)\},\{\mathcal{G}^I\}}\\
        &-\frac{\beta}{\kappa}\left(\delta_a^g\partial_x^i-\frac{1}{2}\delta_a^i\partial^g_x\right) G^{\Delta}(\vec{x}-\vec{z})\mathcal{O}_{\{\delta^2 C^{\textrm{geo}\prime}_g(\vec{z},t),\mathcal{O}^{\textrm{dual,vac}(1)}_{\delta\mathcal{E}_j^b,\{ C_I^\prime\}}(\vec{y},t)\},\{\mathcal{G}^I\}}\\
        &-\frac{\beta}{2\kappa}\partial^g_x\partial^i_x\partial_a^x G^{\Delta\Delta}(\vec{x}-\vec{z})\mathcal{O}_{\{\delta^2 C^{\textrm{geo}\prime}_g(\vec{z},t),\mathcal{O}^{\textrm{dual,vac}(1)}_{\delta\mathcal{E}_j^b,\{ C_I^\prime\}}(\vec{y},t)\},\{\mathcal{G}^I\}}\Big].
    \end{split}
\end{align}
In all Poisson brackets both contributions by construction Poisson commute with all clocks, thus $\{.,.\}=\{.,.\}^\ast$ and by that we can use identity \eqref{pb equals dirac bracket under observable map} to plug in the observable map into the Poisson bracket. We know that $\left(\delta\mathcal{A}_a^i\right)^{\textrm{phys}}=\mathcal{O}_{\mathcal{O}^{\textrm{dual,vac} (1)}_{\delta\mathcal{A}_i^a,\{C_I^\prime\}},\{\delta\mathcal{G}^J\}}$ and similar for the linearised densitised triad. Since the other expression in the Poisson bracket above is always a second order function, we only need to assume the first order of the observable map. Also by using \eqref{eq:dual and observable on phase space function}, as well as \eqref{eq: transformed second order constraints}, we can easily see that the second order constraints in the brackets only depend on the geometric Dirac observables and since the constraints are second order, only on the first order of them. Thus, we can write \eqref{eq: Poisson bracket of physical geometric variables second order 1} as
\begin{align}\label{eq: Poisson bracket of physical geometric variables second order 2}
    \begin{split}
        \{(\delta\mathcal{A}_a^i)^{\textrm{phys}\prime\prime}(\vec{x},t),(\delta\mathcal{E}^b_j)^{\textrm{phys}\prime\prime}(\vec{y},t)\}&=\frac{\beta}{\kappa}P^{ib}_{aj}\delta(\vec{x}-\vec{y})\\
        +\int d^3z\Big[&\frac{1}{\kappa}\delta_j^b G^{\Delta}(\vec{y}-\vec{z})\{\prescript{x}{}{P}^{id}_{al}\delta\mathcal{A}_d^l(\vec{x},t),\delta^2_p C^{\textrm{geo}\prime}(\vec{z},t)\}\\
        &+\frac{2\beta}{\kappa}\partial^b_y G^{\Delta}(\vec{y}-\vec{z})\{\prescript{x}{}{P}^{id}_{al}\delta\mathcal{A}_d^l(\vec{x},t),\delta^2_p G_j^{\textrm{geo}\prime}(\vec{z},t)\}\\
        &-\frac{\beta}{\kappa}\epsilon_j^{\:\:gb} G^{\Delta}(\vec{y}-\vec{z})\{\prescript{x}{}{P}^{id}_{al}\delta\mathcal{A}_d^l(\vec{x},t),\delta^2_p C^{\textrm{geo}\prime}_g(\vec{z},t)\}\\
        &+\frac{1}{2\kappa}\epsilon_a^{\:\:ic}\partial_c^x G^{\Delta}(\vec{x}-\vec{z})\{\delta^2_p C^{\textrm{geo}\prime}(\vec{z},t),\prescript{y}{}{P}^{bl}_{jd}\delta\mathcal{E}_l^d(\vec{y},t)\}\\
        &-\frac{\beta}{\kappa}\left(\delta_a^g\partial_x^i-\frac{1}{2}\delta_a^i\partial^g_x\right) G^{\Delta}(\vec{x}-\vec{z})\{\delta^2_p C^{\textrm{geo}\prime}_g(\vec{z},t),\prescript{y}{}{P}^{bl}_{jd}\delta\mathcal{E}_l^d(\vec{y},t)\}\\
        &-\frac{\beta}{2\kappa}\partial^g_x\partial^i_x\partial_a^x G^{\Delta\Delta}(\vec{x}-\vec{z})\{\delta^2_p C^{\textrm{geo}\prime}_g(\vec{z},t),\prescript{y}{}{P}^{bl}_{jd}\delta\mathcal{E}_l^d(\vec{y},t)\}\Big].
    \end{split}
\end{align}
Here, we defined the geometric primed second order constraints depending on the symmetric transverse traceless degrees of freedom of $\delta\mathcal{A}_a^i,\:\delta\mathcal{E}^a_i$
\begin{align}
    \begin{split}
        \delta^2_p G_j^{\textrm{geo}\prime}=&\frac{\kappa^2}{2\beta}\epsilon_{jl}^{\:\:\:k}P^{lg}_{cn}\delta\mathcal{A}_g^n P_{kd}^{cm}\delta\mathcal{E}^d_m,
    \end{split}\\
    \begin{split}
        \delta^2_p C^{\textrm{geo}\prime} =& \frac{\kappa^2}{2\beta^2}\Big(P^{bd}_{nl}\delta\mathcal{A}_d^lP^{nf}_{bk}\delta\mathcal{A}_f^k-(\beta^2+1)P^{jl}_{bh}\delta\mathcal{E}_l^h\Delta P^{bk}_{jd}\delta\mathcal{E}^d_k\\
        &+2\epsilon_c^{\:\:bg}P^{nf}_{bl}\delta\mathcal{A}_f^l\partial_g P^{ck}_{nd}\delta\mathcal{E}^d_k\Big),
    \end{split}\\
    \begin{split}
        \delta^2_p C_h^{\textrm{geo}\prime} =& \frac{\kappa^2}{\beta}\left(P^{ld}_{gm}\delta\mathcal{A}_{d,h}^mP^{gn}_{lc}\delta\mathcal{E}^c_n-P^{ld}_{hm}\delta\mathcal{A}^m_{d,g}P^{gn}_{lc}\delta\mathcal{E}^c_n\right).
    \end{split}
\end{align}
To calculate the Poisson brackets involved in \eqref{eq: Poisson bracket of physical geometric variables second order 2}, we use the following
\begin{align}
    \begin{split}
    \{\prescript{x}{}{P}^{ic}_{ak}\delta\mathcal{A}^k_c(\vec{x},t),\prescript{y}{}{P}^{bl}_{jd}\delta\mathcal{E}^d_l(\vec{x},t)\}=&\frac{\beta}{\kappa}\prescript{x}{}{P}^{ic}_{ak}\prescript{y}{}{P}^{bl}_{jd}\delta^k_l\delta_c^d\delta(\vec{x}-\vec{y})\\
    =&\frac{\beta}{\kappa}\prescript{x}{}{P}^{ib}_{aj}\delta(\vec{x}-\vec{y}),
    \end{split}\\
    \begin{split}
    \{\prescript{x}{}{P}^{ic}_{ak}\delta\mathcal{A}^k_c(\vec{x},t),\prescript{y}{}{P}^{jd}_{bl}\delta\mathcal{A}^l_d(\vec{y},t)\}=& 0,
    \end{split}\\
    \begin{split}
    \{\prescript{x}{}{P}^{ak}_{ic}\delta\mathcal{E}^c_k(\vec{x},t),\prescript{y}{}{P}^{bl}_{jd}\delta\mathcal{E}_l^d(\vec{y},t)\}=& 0,
    \end{split}
\end{align}
where we used the Poisson algebra of $\delta\mathcal{A}_a^i$ and $\delta\mathcal{E}^a_i$ and that we can pull the projector out of the Poisson bracket. Using this, we get the following expression for the Poisson bracket \eqref{eq: Poisson bracket of physical geometric variables second order 2}
\begin{align}
    \begin{split}
        &\{(\delta\mathcal{A}_a^i)^{\textrm{phys}\prime\prime}(\vec{x},t),(\delta\mathcal{E}^b_j)^{\textrm{phys}\prime\prime}(\vec{y},t)\}=\frac{\beta}{\kappa}\prescript{x}{}{P}^{ib}_{aj}\delta(\vec{x}-\vec{y})\\
        &+\beta\Big[\left(\prescript{y}{}{P}^{kb}_{gj}\partial_h^y-\prescript{y}{}{P}_{hj}^{kb}\partial_g^y\right)\left(\left(\left(\delta^a_h\partial_y^i-\frac{1}{2}\delta_a^i\partial^h_y\right)G^\Delta(\vec{x}-\vec{y})+\frac{1}{2}\partial_a^y\partial^h_y\partial_y^iG^\Delta(\vec{x}-\vec{y})\right)\prescript{y}{}{P}^{gl}_{kd}\delta\mathcal{E}^d_l(\vec{y},t)\right)\\
        &-\frac{1}{2}\delta_j^b\left(\prescript{x}{}{P}^{ik}_{ac}\left(G^\Delta(\vec{x}-\vec{y})\prescript{x}{}{P}^{cl}_{kd}\Delta^x\delta\mathcal{E}^d_l(\vec{x},t)\right)+\prescript{x}{}{P}^{ic}_{ak}\Delta^x\left(G^\Delta(\vec{x}-\vec{y})\prescript{x}{}{P}^{kl}_{cd}\delta\mathcal{E}_l^d(\vec{x},t)\right)\right)\\
        &-\epsilon_{jn}^{\:\:\:\:k}\prescript{x}{}{P}^{ic}_{ak}\left(\partial^b_xG^{\Delta}(\vec{x}-\vec{y})\prescript{x}{}{P}^{nd}_{cl}\delta\mathcal{A}_d^l(\vec{x},t)\right)\\
        &+\epsilon^{h\:\:\:\:b}_{\:\:j}\prescript{x}{}{P}^{ig}_{ak}\left(G^\Delta(\vec{x}-\vec{y})\left(\prescript{x}{}{P}^{kd}_{gl}\delta\mathcal{A}^l_{d,h}(\vec{x},t)-\prescript{x}{}{P}^{kd}_{hl}\delta\mathcal{A}^l_{d,g}(\vec{x},t)\right)\right)\Big]\\
        &+\frac{1}{\beta}\Big[\frac{1}{2}\epsilon_a^{\:\:ih}\prescript{y}{}{P}^{nb}_{cj}\left(\partial_h^yG^\Delta(\vec{x}-\vec{y})\prescript{y}{}{P}^{cd}_{nl}\delta\mathcal{A}^l_d(\vec{y},t)\right)-\delta_j^b\epsilon_c^{\:\:hg}\prescript{x}{}{P}^{ic}_{an}\partial_g^x\left(G^\Delta(\vec{y}-\vec{x})\prescript{x}{}{P}_{hl}^{nd}\delta\mathcal{A}_d^l(\vec{x},t)\right)\\
        &+\frac{1}{2}\epsilon_a^{\:\:ih}\epsilon_c^{\:\:fg}\prescript{y}{}{P}^{nb}_{fj}\left(\partial_h^yG^\Delta(\vec{x}-\vec{y})\prescript{y}{}{P}^{cl}_{nd}\delta\mathcal{E}^d_{l,g}(\vec{y},t)\right)\\
        &-\frac{1}{2}\delta_j^b\prescript{x}{}{P}^{ik}_{ac}\left(G^\Delta(\vec{x}-\vec{y})\prescript{x}{}{P}^{cl}_{kd}\Delta^x\delta\mathcal{E}^d_l(\vec{x},t)\right)-\frac{1}{2}\delta_j^b\prescript{x}{}{P}^{ic}_{ak}\Delta^x\left(G^\Delta(\vec{x}-\vec{y})\prescript{x}{}{P}^{kl}_{cd}\delta\mathcal{E}_l^d(\vec{x},t)\right)\Big].
    \end{split}
\end{align}
This result is in general non-zero, highly non-trivial, and linear in $P^{id}_{al}\delta\mathcal{A}d^l,\:\:P^{bl}_{jd}\delta\mathcal{E}^d_l$. The derivation for the self Poisson brackets of $(\delta\mathcal{A}_a^i)^{\textrm{phys}\prime\prime},\:(\delta\mathcal{E}_i^a)^{\textrm{phys}\prime\prime}$ can be calculated analogously and gives similar expressions. The Poisson algebra with the electromagnetic Dirac observables $(A_a)^{\textrm{phys}},\:(-E^a)^{\textrm{phys}}$ does not change, since $(\delta\mathcal{A}_a^i)^{\textrm{phys}\prime\prime},\:(\delta\mathcal{E}^a_i)^{\textrm{phys}\prime\prime}$ have no electromagnetic degrees of freedom.
~\\
~\\
Note that the Dirac observables $(\delta\mathcal{A}_a^i)^{\textrm{phys}\prime\prime},\:(\delta\mathcal{E}^a_i)^{\textrm{phys}\prime\prime}$, by construction, Poisson commute with all constraints and clocks of \eqref{eq:FinalSet}. Thus, together with $(A_a)^{\textrm{phys}},\:(-E^a)^{\textrm{phys}}$, they form another set $S^{\prime\prime}$ of Dirac observables, with the desired properties we discussed in the main text, like $S$ and $S^\prime$. But, similar to $S^\prime$, the Poisson algebra of this set is quite complicated and quite challenging to quantise.
\section{Physical Hamiltonian}\label{app: more details on the derivation of the physical Hamiltonian}
In this section we give more details on the derivation of the physical Hamiltonian, for both the Dirac observables of set $S$ and $S^\prime$. As discussed in the main text, the starting point is to apply the observable map onto $\frac{1}{\kappa} h$ (see equation \eqref{eq: physical Hamiltonian construction deparametrised system argument}) and integrate over all spatial coordinates. Using once again equation \eqref{eq:dual and observable on phase space function} this leads to
\begin{align}\label{eq app: full physical Hamiltonian in position space}
    \begin{split}
    H^{\textrm{phys}\prime}\coloneqq &\int\limits_{\mathbb{R}^3} d^3 x \mathcal{O}_{\frac{1}{\kappa} h,\{\mathcal{G}^I\}}\\
    =&\int\limits_{\mathbb{R}^3} d^3 x\Big[ T^{00}\left(\mathcal{O}_{\mathcal{O}^{\textrm{dual}}_{A_a,\{C_I^\prime\}},\{\mathcal{G}^I\}}, \mathcal{O}_{\mathcal{O}^{\textrm{dual}}_{-E^a,\{C_I^\prime\}},\{\mathcal{G}^I\}}\right)\\
    &-\frac{\kappa}{2}W_{abc}^i\mathcal{O}_{\mathcal{O}^{\textrm{dual}(1)}_{\delta\mathcal{E}_i^c,\{C_I^\prime\}},\{\mathcal{G}^I\}}T^{ab}\left(\mathcal{O}_{\mathcal{O}^{\textrm{dual}}_{A_a,\{C_I^\prime\}},\{\mathcal{G}^I\}},\mathcal{O}_{\mathcal{O}^{\textrm{dual}}_{-E^a,\{C_I^\prime\}},\{\mathcal{G}^I\}}\right)\\
    &+\frac{1}{\kappa}\delta^2 C^{\textrm{geo}}\left(\mathcal{O}_{\mathcal{O}^{\textrm{dual}(1)}_{\delta\mathcal{A}_a^i,\{ C_I^\prime\}},\{\mathcal{G}^I\}}, \mathcal{O}_{\mathcal{O}^{\textrm{dual}(1)}_{\delta\mathcal{E}^a_i ,\{ C_I^\prime\}},\{\mathcal{G}^I\}}\right)\Big]+O(\delta^3, \kappa^2)\\
    =& \int\limits_{\mathbb{R}^3} d^3 x\Big[T^{00}((A_a)^{\textrm{phys}},(-E^a)^{\textrm{phys}})\\
    &-\frac{\kappa}{2}W_{abc}^i\big(\delta\mathcal{E}_i^c)^{\textrm{phys}\prime} T^{ab}((A_a)^{\textrm{phys}},\:(-E^a)^{\textrm{phys}})\\
    &+\frac{1}{\kappa}\delta^2 C^{\textrm{geo}}\big((\delta\mathcal{E}^a_i)^{\textrm{phys}\prime},(\delta\mathcal{A}_a^i)^{\textrm{phys}\prime})\Big)\Big]+O(\delta^3,\kappa^2),
    \end{split}
\end{align}
where the detailed expression of the last contribution is given in the main text (see equation \eqref{eq: physical Hamiltonian for S prime}). As discussed, this is the physical Hamiltonian for the Dirac observables of the set $S^\prime$, whereas we are interested in the one corresponding to the Dirac observables of the set $S$. Thus, the task is now to calculate the different contributions of \eqref{eq app: full physical Hamiltonian in position space} as functions of the chosen Dirac observables. The first line is already solved since we just have to plug $(A_a)^{\textrm{phys}},\:(-E^a)^{\textrm{phys}}$ into the 00-component of the energy-momentum tensor, which is in zeroth order just the ordinary Hamiltonian for Maxwell theory in flat spacetime. 
~\\
~\\
Next, we come to the second contribution of \eqref{eq app: full physical Hamiltonian in position space}. We want the expression to depend on our chosen Dirac observables. Therefore, we insert \eqref{eq: transformation between sets of physical variables} leading to
\begin{align}\label{eq app: second term physical hamiltonian vcalculation}
    \begin{split}
        -\frac{\kappa}{2}W_{abc}^i\big(\delta\mathcal{E}_i^c)^{\textrm{phys}\prime} T^{ab}((A_a)^{\textrm{phys}},\:(-E^a)^{\textrm{phys}})=&-\int\limits_{\mathbb{R}^3} d^3x \frac{\kappa}{2}W_{abc}^i\Big( T^{ab}\left((-E^a)^{\textrm{phys}},(A_a)^{\textrm{phys}}\right)\Big[P^{cj}_{id}\delta\mathcal{E}_j^d\\
        &+\delta^c_i\left(T^{00}\left((-E^a)^{\textrm{phys}},(A_a)^{\textrm{phys}}\right)\ast G^\Delta\right)\\
        &+\beta\delta^{gj}\delta_k^c\epsilon_{ji}^{\:\:\:\:k}\left(T_{0g}\left((-E^a)^{\textrm{phys}},(A_a)^{\textrm{phys}}\right)\ast G^\Delta\right)\\
        &+\beta\epsilon_i^{\:\:lk}\delta_l^d\delta_k^c\partial_d\tau+\delta_i^g\delta^c_{[g}\partial_{d]}\sigma^d\Big]\Big)(\vec{x},t).
    \end{split}
\end{align}
The next step is to calculate the application of $ W_{abc}^i:= -\delta_b^i\delta_{ac}-\delta_a^i\delta_{bc}+\delta_{ab}\delta_c^i$ onto the different contributions, where we will leave the spatial energy-momentum tensor as it is
\begin{align}
    W_{abc}^iP^{cj}_{id}\delta\mathcal{E}_j^d=&-2\delta_b^i\delta_{ac}P^{cj}_{id}\delta\mathcal{E}_j^d,\\
    W_{abc}^i\delta^c_i\left(T^{00}\left((-E^a)^{\textrm{phys}},(A_a)^{\textrm{phys}}\right)\ast G^\Delta\right)=&\delta_{ab}\left(T^{00}\left((-E^a)^{\textrm{phys}},(A_a)^{\textrm{phys}}\right)\ast G^\Delta\right),\\
    W_{abc}^i\delta^{gj}\delta_k^c\epsilon_{ij}^{\:\:\:\:k}\left(T_{0g}\left((-E^a)^{\textrm{phys}},(A_a)^{\textrm{phys}}\right)\ast G^\Delta\right)=&0,\\
    W_{abc}^i\epsilon_i^{\:\:lk}\delta_l^d\delta_k^c\partial_d\tau=0,\\
    W_{abc}^i\delta_i^g\delta^c_{[g}\partial_{d]}\sigma^d=\left(\delta_{ad}\partial_b+\delta_{bd}\partial_a\right)\sigma^d,
\end{align}
in the first line we used that $P^{cj}_{cd}=0$ (see appendix \ref{app: projectors}). With this, \eqref{eq app: second term physical hamiltonian vcalculation} reads
\begin{align}
    \begin{split}
        &\kappa\int\limits_{\mathbb{R}^3} d^3x\Big(T^{ab}\left((-E^a)^{\textrm{phys}},(A_a)^{\textrm{phys}}\right)\delta_b^i\delta_{ac}P^{cj}_{id}\delta\mathcal{E}_j^d\\
        &-\frac{1}{2}T^{ab}\left((-E^a)^{\textrm{phys}},(A_a)^{\textrm{phys}}\right)\delta_{ab}\left(T^{00}\left((-E^a)^{\textrm{phys}},(A_a)^{\textrm{phys}}\right)\ast G^\Delta\right)\\
        &- T^{ab}\left((-E^a)^{\textrm{phys}},(A_a)^{\textrm{phys}}\right)\delta_{ad}\partial_b\sigma^d\Big)(\vec{x},t),
    \end{split}
\end{align}
where we made use of the symmetry of the energy-momentum tensor in the last line. In the special case of Maxwell theory we have that the trace of the energy-momentum tensor is zero ($T_\mu^{\:\:\mu}=0$) in Minkowski spacetime. Thus, $T^{ab}\delta_{ab}=T^{00}$. On the other hand, for our choice of $\sigma^d=x^d$ we can explicitly calculate the derivative $\partial_bx^d=\delta_b^d$ and therefore \eqref{eq app: second term physical hamiltonian vcalculation} simplifies to
\begin{align}\label{eq app: second part of physical Hamiltonian}
    \begin{split}
        &\kappa\int\limits_{\mathbb{R}^3} d^3x\Big( -T^{ab}\left((E^a)^{\textrm{phys}},(A_a)^{\textrm{phys}}\right)\delta_b^i\delta_{ac}P^{cj}_{id}\delta\mathcal{E}_j^d\\
        &-\frac{1}{2}T^{00}\left((-E^a)^{\textrm{phys}},(A_a)^{\textrm{phys}}\right)\left(T^{00}\left((-E^a)^{\textrm{phys}},(A_a)^{\textrm{phys}}\right)\ast G^\Delta\right)\\
        &-T^{00}\left((-E^a)^{\textrm{phys}},(A_a)^{\textrm{phys}}\right)\Big)(\vec{x},t).
    \end{split}
\end{align}
The expression consists of two interactions. First the interaction between linearised gravity and electromagnetism, given by the contraction of the spatial, electromagnetic energy-momentum tensor and the symmetric, transverse, traceless part of the linearised densitised triad. On the other hand, we have a gravitationally induced electromagnetic self-interaction, which arises from the contraction of the spatial energy-momentum tensor with the electromagnetic energy density.
~\\
~\\
The next step is to calculate the third contribution of \eqref{eq app: full physical Hamiltonian in position space}, which is the application of the observable map onto the geometric primed second order Hamiltonian constraint. We again use \eqref{eq:dual and observable on phase space function} to get the following expression
\begin{align}
    \begin{split}
    \int\limits_{\mathbb{R}^3} d^3 x\frac{1}{\kappa}\delta^2 C^{\textrm{geo}}\big((\delta\mathcal{E}^a_i)^{\textrm{phys}\prime},(\delta\mathcal{A}_a^i)^{\textrm{phys}\prime})\Big)=\frac{\kappa}{2\beta^2}\int\limits_{\mathbb{R}^3} d^3 x\delta^a_{[n}\delta^b_{m]}\Big[&(\delta\mathcal{A}^m_a)^{\textrm{phys}\prime}(\delta\mathcal{A}^n_b)^{\textrm{phys}\prime}\\
    &+(\beta^2+1)(\delta\Gamma^m_a)^{\textrm{phys}\prime}(\delta\Gamma^n_b)^{\textrm{phys}\prime}\\
    &-2(\delta\Gamma^m_a)^{\textrm{phys}\prime}(\delta\mathcal{A}^n_b)^{\textrm{phys}\prime}\Big],
    \end{split}
\end{align}
where we inserted the exact expression of the geometric second order Hamiltonian constraint \cite{Fahn:2022zql}. Furthermore, we defined $(\delta\Gamma^m_b)^{\textrm{phys}\prime}$, the linearised spin-connection \eqref{LinearisedSpinconnectionintermsoftriads1} under the application of first the dual observable map and then the observable map. It is given by (here just the first order is relevant)
\begin{align}
    \begin{split}
    (\delta\Gamma^m_b)^{\textrm{phys}\prime}\coloneqq & \epsilon^{m\:\:l}_{\:\:\:j}\delta^b_l\partial_b P^{jk}_{ad}\delta\mathcal{E}_k^d-\beta\delta_d^m\partial^d\partial_a\tau-\frac{1}{2}\epsilon^{c\:\:\:b}_{\:\:a}\delta^m_c\partial_b\left(T^{00}\left((-E^a)^{\textrm{phys}},(A_a)^{\textrm{phys}}\right)\ast G^\Delta\right)\\
    &+\beta\delta^{mg}\partial_a\left(T_{0g}\left((-E^a)^{\textrm{phys}},(A_a)^{\textrm{phys}}\right)\ast G^\Delta\right).
    \end{split}
\end{align}
This result is valid for arbitrary $\tau$ and $\sigma$. In our case $\tau=t$, thus the term involving $\tau$ vanishes under the application of the spatial derivatives. The next step is to get this expression in terms of $(\delta\mathcal{A}_a^i)^{\textrm{phys}},\:(\delta\mathcal{E}_i^a)^{\textrm{phys}}$, for which we will use again \eqref{eq: transformation between sets of physical variables}. The calculation is rather long but tedious, overall we get
\begin{align}\label{eq app: third part of physical Hamiltonian}
    \begin{split}
        \kappa\int\limits_{\mathbb{R}^3} d^3x\Big[&\frac{1}{2\beta^2}\Big(P^{bd}_{nl}\delta\mathcal{A}_d^l P^{nf}_{bk}\delta\mathcal{A}_f^k-(\beta^2+1)P^{jl}_{bh}\delta\mathcal{E}_l^h\Delta P^{bk}_{jd}\delta\mathcal{E}_k^d\\
        &+2\epsilon_c^{\:\:bg}P^{nf}_{bl}\delta\mathcal{A}^l_f\partial_g P_{nd}^{ck}\delta\mathcal{E}_k^d+\kappa\beta^2T_{0f}\partial^f\tau\Big)\\
        &+\left(T_{0a}\ast G^\Delta\right)\frac{\partial^a\partial^b}{4}\left(T_{0b}\ast G^\Delta\right)+T_{0a}\delta^{ab}\left(T_{0b}\ast G^\Delta\right)-\frac{1}{4}T^{00}\left(T^{00}\ast G^\Delta\right)\Big],
    \end{split}
\end{align}
where all contributions of the energy-momentum tensor are to be understood as functions of the Dirac observables $(A_a)^{\textrm{phys}},\:(-E^a)^{\textrm{phys}}$. The first line is purely geometric and gives the free physical Hamiltonian of $\left(\delta\mathcal{A}^i_a\right)^{\textrm{phys}},\: \left(\delta\mathcal{E}_i^a\right)^{\textrm{phys}}$. The second term vanishes if we insert $\tau = t$. The third part is again a self-interaction term of the electromagnetic Dirac observables $(A_a)^{\textrm{phys}},\:(-E^a)^{\textrm{phys}}$. 
~\\
~\\
Inserting \eqref{eq app: second part of physical Hamiltonian} and \eqref{eq app: third part of physical Hamiltonian} into \eqref{eq app: full physical Hamiltonian in position space} and performing some rewriting to separate the different types of couplings and evolutions, leads to the physical Hamiltonian given in \eqref{eq: physical Hamiltonian for set S} for the Dirac observables of the set $S$.
\end{appendices}
\newpage
\bibliographystyle{unsrtnat}
\bibliography{PhotonDec.bib}
\end{document}